\documentclass[runningheads]{llncs}
\usepackage{graphicx}
\usepackage{comment}
\usepackage{amsmath,amssymb} 
\usepackage{color}

\usepackage[width=122mm,left=12mm,paperwidth=146mm,height=193mm,top=12mm,paperheight=217mm]{geometry}

\usepackage{tango}
\usepackage[caption=false]{subfig}
\usepackage{xspace}
\usepackage{cite}
\usepackage[colorlinks=true]{hyperref}
\usepackage{amsmath}
\usepackage{amssymb}
\usepackage{tabulary,multirow}

\usepackage[title]{appendix}

\newcommand{\figref}[1]{Fig.~\ref{fig:#1}}
\newcommand{\refsec}[1]{Section~\ref{sec:#1}}
\newcommand{\reffig}[1]{Figure~\ref{fig:#1}}
\newcommand{\tabref}[1]{Table~\ref{tab:#1}}
\newcommand{\eqnref}[1]{Equation~\eqref{eq:#1}}

\newcommand{\lblsec}[1]{\label{sec:#1}}

\newcommand{\x}{\times}
\newcommand{\etal}{\textit{et~al.}\xspace}
\newcommand{\rbr}[1]{\left(#1\right)}
\newcommand{\sbr}[1]{\left[#1\right]}
\newcommand{\cbr}[1]{\left\{#1\right\}}

\newcolumntype{x}[1]{>{\centering\arraybackslash}p{#1pt}}
\newlength\savewidth\newcommand\shline{\noalign{\global\savewidth\arrayrulewidth
  \global\arrayrulewidth 1pt}\hline\noalign{\global\arrayrulewidth\savewidth}}
\newcommand{\tablestyle}[2]{\setlength{\tabcolsep}{#1}\renewcommand{\arraystretch}{#2}\centering\footnotesize}
\makeatletter\renewcommand\paragraph{\@startsection{paragraph}{4}{\z@}
  {.5em \@plus1ex \@minus.2ex}{-.5em}{\normalfont\normalsize\bfseries}}\makeatother
\definecolor{demphcolor}{RGB}{100,100,100}
\newcommand{\demph}[1]{\textcolor{demphcolor}{#1}}

\newcommand{\app}{\raise.17ex\hbox{$\scriptstyle\sim$}}

\newcommand{\percent}[1]{{\tiny\demph{#1}}}

\newcommand{\toptwo}{\raisebox{-0.5ex}{\protect\includegraphics[width=1.5em]{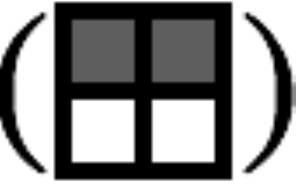}}}
\newcommand{\topleft}{\raisebox{-0.5ex}{\protect\includegraphics[width=1.5em]{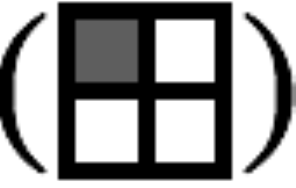}}}

\newcommand{\topleftb}{\raisebox{-0.42ex}{\protect\includegraphics[height=1.0em]{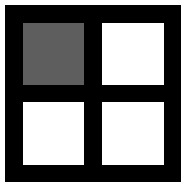}}}
\newcommand{\toprightb}{\raisebox{-0.42ex}{\protect\includegraphics[height=1.0em]{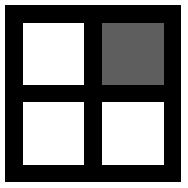}}}
\newcommand{\bottomleftb}{\raisebox{-0.42ex}{\protect\includegraphics[height=1.0em]{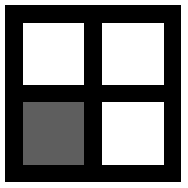}}}
\newcommand{\bottomrightb}{\raisebox{-0.42ex}{\protect\includegraphics[height=1.0em]{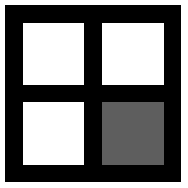}}}
\usepackage{etoolbox}
\patchcmd{\paragraph}{\itshape}{\bfseries\boldmath}{}{}

\begin{document}
\pagestyle{headings}
\mainmatter

\title{Lossless Image Compression through Super-Resolution} 

\author{Sheng Cao, Chao-Yuan Wu, Philipp Kr\"ahenb\"uhl}
\institute{The University of Texas at Austin}

\maketitle

\begin{abstract}

We introduce a simple and efficient lossless image compression algorithm.
We store a low resolution version of an image as raw pixels, followed by several iterations of lossless super-resolution.
For lossless super-resolution, we predict the probability of a high-resolution image, conditioned on the low-resolution input, and use entropy coding to compress this super-resolution operator.
Super-Resolution based Compression (SReC) is able to achieve state-of-the-art compression rates with practical runtimes on large datasets.
Code is available online.\footnote{\url{https://github.com/caoscott/SReC}}
\end{abstract}

\section{Introduction}
Mankind captures and shares a collective trillion of new photos annually~\cite{marymeeker}.
Images capture all aspects of our beautifully complex and diverse visual world.
Yet, not every arrangement of pixel color values forms a real image.
Most are illegible noise.
This is the main insight behind lossless image compression.
In fact, the Shannon source coding theorem directly links the likelihood of a group of pixels to be a real image, and our ability to compress that image~\cite{shannon}.
The main challenge is designing an effective probabilistic model of pixel values.

\begin{figure*}[t]
\centering
\includegraphics[width=1.0\linewidth]{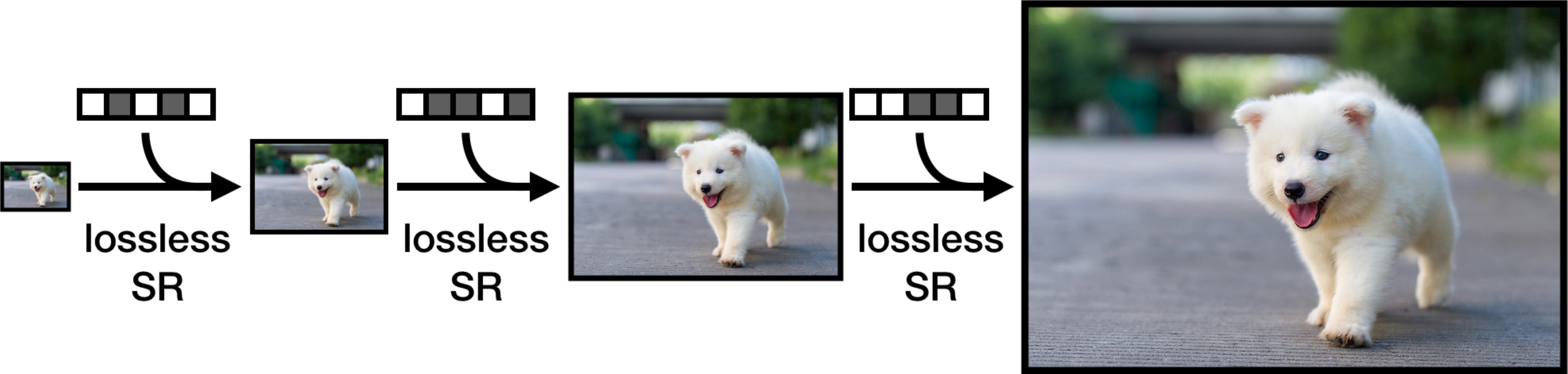}
\vspace{-2em}
\caption{\textbf{Model Overview.} We propose lossless image compression through super resolution (SR).
Our method first encodes a low-resolution image efficiently,
and then leverages SR models to efficiently entropy-code the high-resolution images.}
\vspace{-1em}
\label{fig:concept}
\end{figure*}

In this paper, we propose lossless image compression through \mbox{\emph{super-resolution}} (SR).
Unlike standard super-resolution, which predicts one single output image, we predict a \emph{distribution} over all possible super-resolved images.
Each pixel in a low-resolution image induces an autoregressive distribution over four high-resolution output pixels.
This distribution is then entropy-coded using arithmetic coding (AC), yielding a losslessly compressed super-resolution operator.
Our overall compression algorithm stores a low-resolution version of the image as raw pixels, and then applies three iterations of losslessly-compressing super-resolution operator, as shown in \reffig{concept}.
We train the losslessly compressed super-resolution operator to maximize the log-likelihood of a high-resolution image, conditioned on its downsampled version on standard image datasets.

Compression through super-resolution shares components with existing deep lossless image compression methods~\cite{l3c,idf}, yet enjoys several additional advantages.
Our neural network is lightweight and efficient.
Each set of four output pixels is independent of all other outputs at the same level; hence super-resolution is easily performed in parallel.
Furthermore, we show that simple constraints imposed by the super-resolution process, allow
25\% of the pixels to be reconstructed ``for free".
Finally, we show how the SR setting strictly limits the range of the probability distribution, further reducing the bitrate used in entropy coding.

We evaluate our algorithm on the ImageNet64~\cite{imagenet64,deng2009imagenet} and Open Images datasets~\cite{openimages}.
Our experiments show that our super-resolution-based image compression algorithm outperforms state-of-the-art lossless image compression algorithms across a varying set of image resolutions and data sources.
The runtime of our method is comparable to the fastest prior work, and our models are as small as the most compact prior lossless deep image compression methods.

\section{Related Work}

Most lossless image compression algorithms rely on entropy coding of commonly repeating image patterns.
The two signals most commonly exploited are inter image similarities and natural image statistics.

\paragraph{Hand-designed codecs} encompass some of the most popular lossless compression methods.
PNG~\cite{png} compresses the raw bits of a color image using the DEFLATE algorithm~\cite{deflate}.
It exploits bit-level repetitions in the image but often fails to capture large structural similarities or common image statistics.
JPEG2000~\cite{jpeg2000} builds its lossless compression in a wavelet transform, which captures some local image statistics.
WebP~\cite{webp} combines multiple image transformations before entropy coding.
Currently, the best performing hand-designed codec is FLIF~\cite{flif}.
It builds on the MANIAC entropy coding algorithm and captures repeating local image patterns in entropy coding.
Hand-designed codecs efficiently exploit the local structure of image formation but only capture simple image statistics that are hand-specified.

\paragraph{Entropy coding,} such as arithmetic coding (AC)~\cite{arithmetic} or asymmetric numeral systems (ANS)~\cite{duda2009asymmetric}, is able to convert generative models of image formation into a compression algorithm, as long as the model provides a probability estimate of the current image.
However, not all generative models are equally efficient.

\paragraph{Compression with Autoregressive Models.}
Autoregressive models predict a distribution over natural images as a distribution over color values, conditioned on previously predicted colors.
For example, Van Oord~\etal's PixelCNN~\cite{pixelcnn} predicts probabilities for subpixels\footnote{1 pixel = 3 subpixels: R, G, and B.} in a raster scan and RGB order.
Each subpixel probability is conditioned on previously seen subpixels.
Salimans~\etal~\cite{pixelcnnpp} uses a discrete logistic mixture to model the joint distribution of a pixel.
Reed~\etal~\cite{mspixelcnn} speeds up these methods by predicting probabilities of multiple pixels at once.
Kolesnikov and Lampert~\cite{kolesnikov2017} use grayscale or downscaled versions of the image as auxiliary variables to PixelCNN to improve model performance.
PixelCNN and their variants generally perform well in terms of log-likelihood, but are impractical for compression due to the long runtime.
For images of size $W \x H$, the original PixelCNN requires $\mathcal{O}(W+H)$ distinct network evaluations, each predicting a diagonal slice of the image.
Reed~\etal~\cite{mspixelcnn} improves this to $\mathcal{O}(\log(W+H))$ using a hierarchical encoding scheme.
However, they use shallow PixelCNNs~\cite{pixelcnnpp} to model dependency between blocks of pixels.
In practice, this is still too slow for lossless compression; see \cite{l3c} for detailed analysis.
Our model uses a similar hierarchical structure with a few important differences:
We use a simpler factorization of the pixel-wise probabilities, allow different hierarchical level to share a feature embedding, and use a more efficient architecture.
Our final compression model is $\app$60$\x$ faster than Reed~\etal~\cite{mspixelcnn} when used for image compression and comparable to the fastest learned image compression techniques~\cite{l3c}.

\paragraph{Latent Vector Models.}
An efficient alternative to condition unseen pixels on previously seen pixels is through a latent vector.
Mentzer~\etal~\cite{l3c} encode an image $x$ into smaller latent vectors $z_1, \ldots, z_3$, and entropy code $x$ using $P(x \mid z_1, \ldots, z_3)$ estimated by a network.
The latent vectors are discretized such that they are also efficient to store.
Integer Discrete Flow (IDF)~\cite{idf} uses a flow-based deep generative model~\cite{rezende2015variational}
 to invertibly transform the input image into a latent vector.
It factorizes probabilities of the latent vectors and compress the latent vectors.
During decompression, the latent vectors are decompressed and inverted to obtain the image.
IDF has high performance on ImageNet32 and ImageNet64~\cite{imagenet64},
as it is able to optimize factorized log-likelihood directly.
However, it struggles to learn higher resolution models.
It learns a discretely parametrized flow, which leads to large approximation errors when many layers are stacked.
In addition, current implementation of flow-based methods are relatively inefficient.
Our method is about 55$\x$ faster on high-resolution images.

\paragraph{Dataset Compression.}
Bits-back methods~\cite{bitsback} are a family of methods that compress continuous latent vectors at fine discretization levels instead of discrete latent vectors.
Bits Back with ANS~\cite{bbans}, Bit-Swap~\cite{bitswap}, and Hierarchical Latent Lossless Compression~\cite{hilloc} build on variational auto-encoders~\cite{kingma2013auto}, while Local Bits-Back~\cite{lbb} is a flow-based method.
Bits-back methods yield the best performance on ImageNet32 and ImageNet64~\cite{imagenet64} when compressing the entire test set into a \emph{single} vector.
However, they are designed as dataset compression algorithms rather than single image compression algorithms.
They are currently unable to compress single images efficiently.
For example, Local Bits-Back (LBB)~\cite{lbb} requires an initial bit-buffer of 52 bits per subpixel (bpsp) at a dataset compression rate of 3.63 bpsp.
This initial investment is amortized when compressing a large dataset of images.
When used for single image compression,
LBB would compress a single images at 55 bpsp, which is much worse than the uncompressed BMP at 8 bpsp.
Our algorithm on the other hand does not rely on a bit-buffer, but instead entropy codes each image independently at a bitrate close to the best bits-back approaches.

\paragraph{Lossy Compression.}
Lossy image/video compression, on the other hand, allows for some distortion in the decompressed data in exchange for reduced storage size.
Recent deep learning based methods typically directly predict the decompressed output~\cite{toderici2015variable,toderici2017full,rippel2017real,johnston2018improved,balle2017end,li2018learning,agustsson2019generative,wu2018video,lu2019dvc,rippel2019learned},
as opposed to predicting a distribution of outputs, as in a lossless compression method.

\paragraph{Super-Resolution.}
Super-resolution (SR) is a task to construct a high-resolution image given a low-resolution image~\cite{dong2014cnnsr,resnet_sr,vggsr,gansr,prsr,edsr,li2019feedback,tong2017image,tai2017image,wang2019edvr,zhou2019kernel,rad2019srobb,cai2019toward}.
Recent works have advanced the state-of-the-art performance with the advances in CNN architecture~\cite{dong2014cnnsr},
image generation~\cite{gansr}, and the likelihood based methods~\cite{prsr}.
Our method leverages recent advances in SR to predict likely high-resolution images for compression.
However, unlike standard super-resolution, our algorithm predicts a probability for each high-resolution image, in order to entropy code the image in a lossless manner.

\section{Preliminaries}
\lblsec{prelim}
Lossless Compression methods encode an image $x$ into a bitstream $b_x$ using an invertible transformation.
The goal of a compression algorithm is to minimize the expected code length $L := \mathbb{E}_{x \sim P}\sbr{|b_x|}$ of the bitstream over a distribution $x \sim P$ of natural images.
The entropy $H_P = \mathbb{E}_{x \sim P}\sbr{-\log_2 P(x)}$ bounds the expected code length $L$ from below following Shannon's Source Coding Theorem~\cite{shannon,info_theory}.

\paragraph{Arithmetic coding (AC)~\cite{arithmetic}}
is a form of entropy encoding that reaches the theoretical lower-bound of the code length within a few bits if it is given access to the distribution of images $P(x)$.
For infinite precision numerical computation, AC obtains a code of length $L \le H_P + 1$.
For finite precision implementations, a few bits are wasted due to rounding.
AC maps the entire image into an interval within a range $[0, 1]$, where the size of the interval is equivalent to the probability $P(x)$ of that image.
The image is then encoded as the shortest integer number in that interval.
Intuitively, frequently used images are mapped to a larger interval, and thus require fewer bits to encode.

In our work, we learn a distribution $P_\theta(x)$ over natural images, and then use this distribution for arithmetic coding.
The code length induced by our distribution $P_\theta$ is bound by the cross entropy between the true natural image distribution $P$ and the learned distribution $P_\theta$: $L \le \mathbb{E}_{x \sim P}[-\log_2 P_\theta(x)]+1$.
Thus, minimizing the bit-length is equivalent to minimizing the cross-entropy, or negative log-likelihood of the model $P_\theta$ under our data distribution $P$.

\section{Method}
\lblsec{method}

Let $x^{(0)} \in \cbr{0, \ldots, 255}^{W \times H \times 3}$ be a 3-channel input image with width $W$ and height $H$.
Let $y^{(1)} = \mathrm{avgpool}_2\rbr{x^{(0)}} \in \mathbb{R}^{\left\lceil\frac{W}{2}\right\rceil \times \left\lceil\frac{H}{2}\right\rceil \times 3}$ be a downsampled version of the input,
where $\mathrm{avgpool}_2$ denotes average pooling of size 2 and stride 2: four neighboring pixels are averaged into a single output value. 
Finally, let $x^{(1)} \in \cbr{0, \ldots, 255}^{\left\lceil\frac{W}{2}\right\rceil \times \left\lceil\frac{H}{2}\right\rceil \times 3}$ be a rounded version of $y^{(1)}$.
Any further low-resolution image is then defined recursively $y^{\rbr{l+1}} = \mathrm{avgpool}_2\rbr{x^{(l)}}$ and $x^{(l)} = \mathrm{round}\rbr{y^{\rbr{l}}}$.

Our compression algorithm stores the low resolution image $x^{(3)}$ in its raw form.
It also stores the rounding values $r^{(l)} = y^{(l)} - x^{(l)} \in \cbr{-\frac{1}{4}, 0, \frac{1}{4}, \frac{1}{2}}$, for $l = 1, 2, 3$ raw using two bits per pixel and channel.
Rounding is close to uniformly random and contains little compressible information.
The super-resolved pixels on the other hand are highly compressible.
Our algorithm conditionally encodes the higher resolution image $x^{(l)}$ given a lower resolution image $y^{(l+1)}$ using AC based on probability estimated by a super-resolution network.
\refsec{network} describes the network structure and training objective, while \refsec{details} covers the exact architectural details.
\refsec{enc} and \refsec{dec} describe the encoding and decoding schemes respectively.

\subsection{Autoregressive Super-Resolution Network}
\lblsec{network}

The goal of the super-resolution network is to predict a distribution over $x^{(l)}$ given $y^{(l+1)}$,
so that we can efficiently entropy code $x^{(l)}$ based on $P\rbr{x^{(l)} \mid y^{(l+1)}}$.
In the following section, we describe our network for one level of super-resolution and omit the superscript for simplicity.
Note that since we define $y_{i,j}$ to be the average of 4 pixels, $x_{2i, 2j}$, $x_{2i, 2j+1}$, $x_{2i+1, 2j}$, and $x_{2i+1, 2j+1}$,
to super-resolve each pixel $y_{i,j}$, we only need to predict a distribution over three pixel values $P(x_{2i,2j}, x_{2i,2j+1}, x_{2i+1,2j} \mid Y_{i,j})$, where $Y_{i,j}$ is a local image region (receptive field) around pixel $y_{i,j}$.
We do not encode the fourth pixel $x_{2i+1,2j+1}$ as it is reconstructed for free using
$x_{2i+1,2j+1} = 4 y_{i,j} - x_{2i,2j}-x_{2i,2j+1}-x_{2i+1,2j}$.

\begin{figure*}[t]
\includegraphics[height=5cm,page=1]{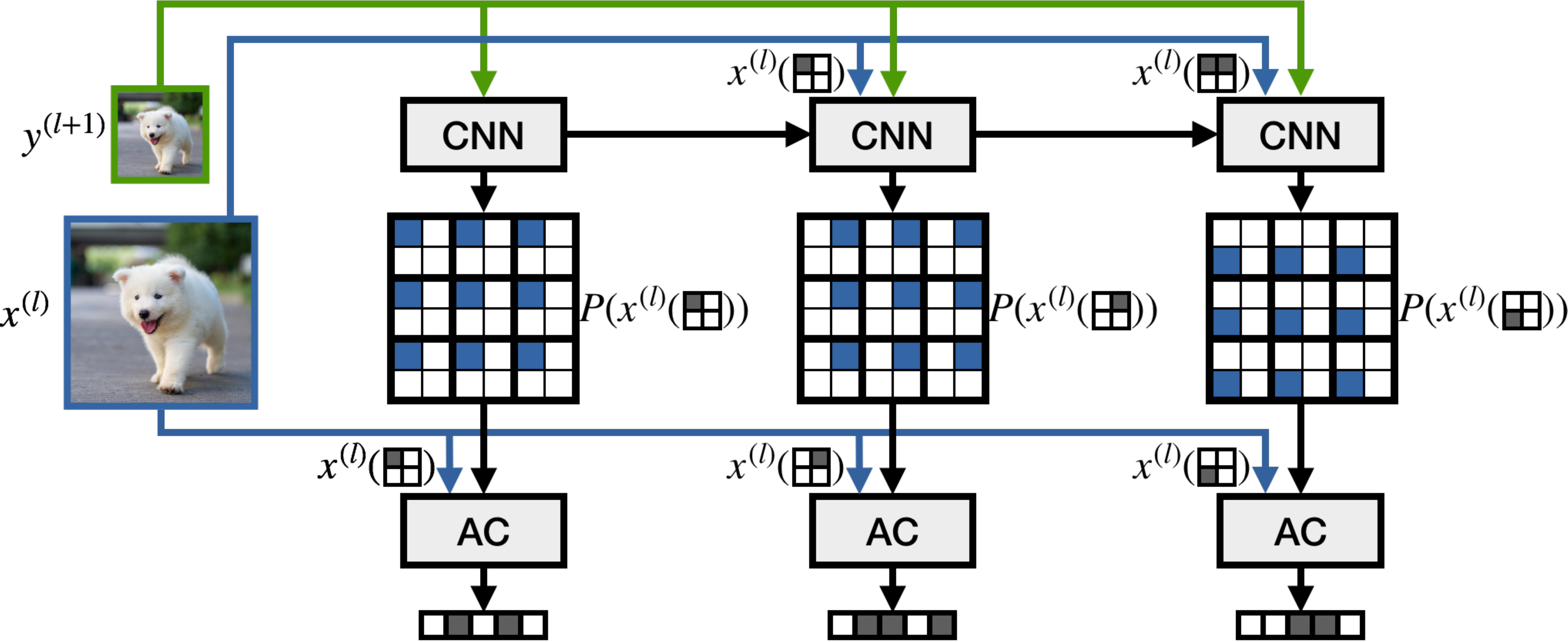}
\caption{\textbf{Encoding $x^{(l)}$ conditioned on a downsampled image $y^{(l+1)}$.}
Our autoregressive network predicts the probability distribution over the first three pixels in an upsampled block sequentially.
An arithmetic coder (AC) then entropy codes each of the pixels based on the estimated probability.
(The fourth pixel could be computed given previously decoded pixels, so we do not need to encode it.)
We use a block in parentheses to denote indexing. For example,
``\topleft" denotes indexing the top-left pixel (i.e., $x_{2i,2j}$) in each block.
``\toptwo" denotes indexing the top two pixels (i.e., $x_{2i,2j}$ and $x_{2i,2j+1}$) in each block.
}
\label{fig:compress}
\end{figure*}

To leverage the correlation among the four pixels,
we factorize their probabilities autoregressively:
\begin{align}
&P\rbr{x_{2i,2j}, x_{2i,2j+1}, x_{2i+1,2j} \mid Y_{i,j}}\nonumber\\
=&P\rbr{x_{2i,2j} \mid Y_{i,j}}
P\rbr{x_{2i,2j+1} \mid Y_{i,j}, x_{2i,2j}}
P\rbr{x_{2i+1,2j} \mid Y_{i,j}, x_{2i,2j}, x_{2i,2j+1}}.\label{eq:blockfactorization}
\end{align}
Each term in the factorization is estimated by a convolutional neural network (CNN), as in \figref{compress}.
The features from earlier networks are fed into later networks through skip connections.
Each network is further provided the true pixel values of previously coded pixels.

We similarly factorize the probability of a single pixel into three autoregressive terms for its three channels $x^R, x^G, x^B$ (omitted in \figref{compress} for simplicity):
\begin{align}
P\rbr{x_{ij} \mid \mathbf{Z}_{ij}}
=&P\rbr{x_{ij}^R, x_{ij}^{G}, x_{ij}^{B} \mid \mathbf{Z}}\nonumber\\
=&P\rbr{x_{ij}^{R} \mid \mathbf{Z}_{ij}}
P\rbr{x_{ij}^{G} \mid \mathbf{Z}_{ij}, x_{ij}^{R}}
P\rbr{x_{ij}^{B} \mid \mathbf{Z}_{ij}, x_{ij}^{R}, x_{ij}^{G}}.\label{eq:rgbfactorization}
\end{align}
$\mathbf{Z}$ denotes the conditioning variables of $x$ introduced by \eqnref{blockfactorization}.
We follow PixelCNN++~\cite{pixelcnnpp} and parametrize this probability as a mixture of logistic functions
$$
P\rbr{x_{ij}^{R} \mid \mathbf{Z}_{ij}} = \sum_{k=1}^K w_k \mathrm{logistic}(x_{ij}^{R} | \mu_{ijk}, s_{ijk}),
$$
where the logistic function $\mathrm{logistic}(x | \mu, s) = \sigma\left(\frac{x-\mu+0.5}{s}\right) - \sigma\left(\frac{x-\mu-0.5}{s}\right)$ is the difference of two sigmoid functions.
The distributions for $x_{ij}^{G}$ and $x_{ij}^{B}$ are defined analogously.
We use a total of $K=10$ mixture components for each.
Our deep network produces the mixture weights $w_{ijk}$, mean $\mu_{ijk}$, and standard deviation $s_{ijk}$ parameters.
For the green and blue color values, the mean $\mu_{ijk}$ and weight $w_{ijk}$ are a linear function of the previously decoded color values.
The linear functions allow for a weak form of conditioning, while keeping inference time low, as all distributional parameters are produced by a \emph{single} network forward pass.
See PixelCNN++~\cite{pixelcnnpp} for details.

Our overall super-resolution network contains three levels of super-resolution with skip-connections from lower-resolution to higher-resolution layers, see \refsec{details} for details.

\paragraph{Training Objective.}
We train our network to minimize the cross entropy between the predicted model probability $P_\theta$ and a data distribution $P$ given by samples from an image dataset.
As the model contains skip connections between levels, we train all three super-resolution levels jointly:
\begin{align}
\ell = \sum_{l=0}^2  -\mathbb{E}\sbr{\log P_\theta\rbr{x^{(l)} \mid y^{(l+1)}}}.
\end{align}
This objective tightly bounds to the expected bit length $\ell-1 \le L \le \ell$ (see \refsec{prelim} for more discussions).

Both training and evaluation are straightforward and only depend on known quantities, e.g.\ down-sampled versions of the original image $x^{(l)}$ and $y^{(l)}$.
They contain no interdependencies and are performed fully convolutionally in parallel.
However, encoding and decoding contain several dependencies, e.g.\ the probability of the green pixel is not known before the red pixel is decoded.
In the next two sections, we highlight how the structure of our model still allows a massively parallel encoding and decoding of the image.

\subsection{Encoding}
\lblsec{enc}
Arithmetic coding has one major drawback.
Encoding and decoding are inherently sequential and follow the same fixed order.
For encoding this is not a major limitation, as all probability estimates are known ahead of time.
However, at decoding time the super-resolution network contains many dependencies that constrain an efficiently parallel probability estimate.
In this section, we describe an encoding order which allows for massively parallel decoding in the next section.

Our algorithm first encodes the lowest-resolution level $x^{(3)}$ as raw pixels.
It then stores the rounding bits to reconstruct $y^{(3)}$, and it subsequently encodes the arithmetic codes of the super-resolution network and rounding bits for all consecutive levels $x^{(2)}$, $y^{(2)}$, $x^{(1)}$, $y^{(1)}$ and $x^{(0)}$.
The algorithm encodes rounding bits  as two bits per color channel corresponding to the four rounding values: $\cbr{-\frac{1}{4}, 0, \frac{1}{4}, \frac{1}{2}}$.

For each super-resolved image $x^{(l)}$,
we encode all blocks in raster scan order.
Let $\topleftb$ be the first of four pixels to be super-resolved, $\toprightb$ the second, $\bottomleftb$ the third.
Our algorithm first encodes all red values of the first super-resolved pixel $\topleftb R$ for all lower-resolution pixels.
All other channels then follow in order: \topleftb R$\rightarrow$\topleftb G$\rightarrow$\topleftb B, followed by \mbox{\toprightb R$\rightarrow$\toprightb G$\rightarrow$\toprightb B}, and finally \bottomleftb R$\rightarrow$\bottomleftb G$\rightarrow$\bottomleftb B.
All values of each channel are then arithmetically encoded.
\reffig{compress} illustrates the process.
Since we can predict probabilities for all blocks in parallel and arithmetic coding is computationally light-weight,
the whole process is efficient.

\subsection{Decoding}
\lblsec{dec}

\begin{figure*}[t]
\centering
\includegraphics[height=5cm,page=1]{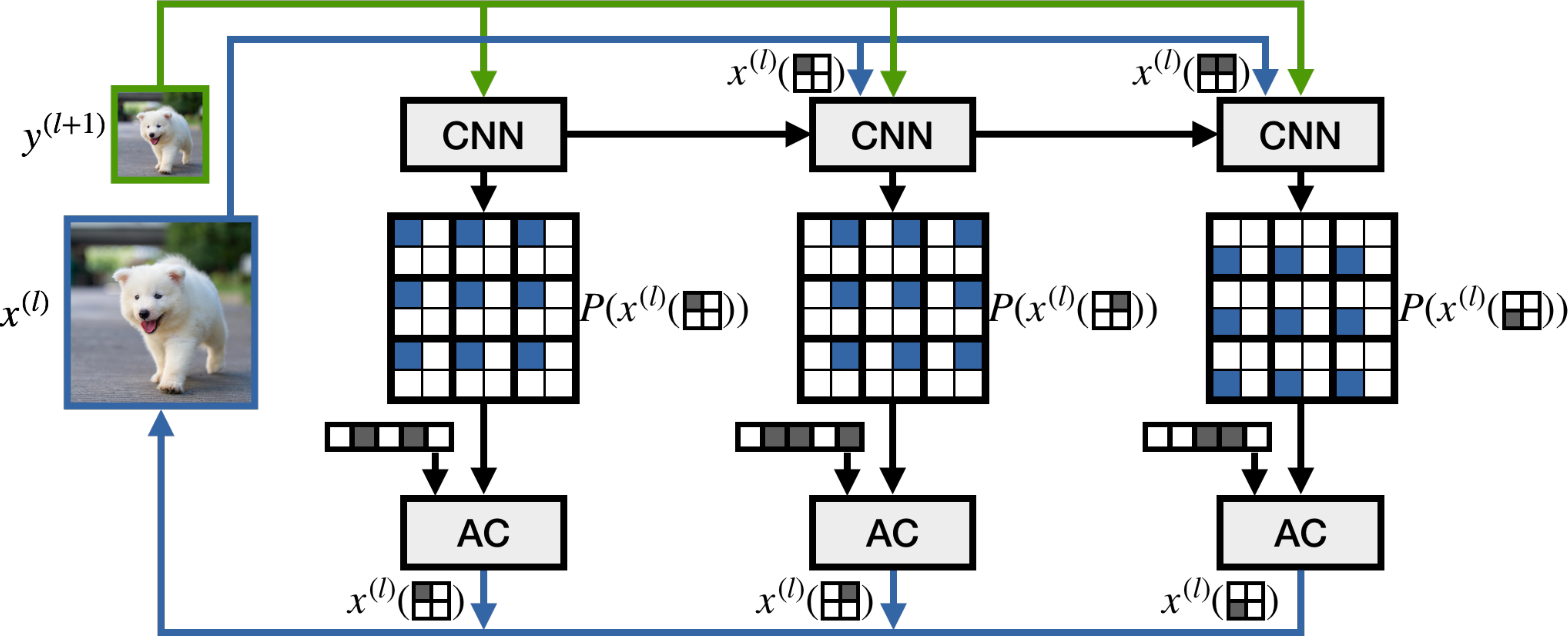}
\caption{\textbf{Decoding $x^{(l)}$ given encoded bits and a downsampled image $y^{(l+1)}$.}
Our autoregressive network predicts the probability distribution over the pixels in $x^{(l)}$ sequentially given $y^{(l+1)}$.
The arithmetic coder (AC) then decode the encoded bits based on the estimated probability.
We use a block in parentheses to denote indexing, similar to \figref{compress}.
}
\label{fig:decompress}
\end{figure*}

Both the network architecture and encoding procedure are chosen as to make decoding as efficient as possible, as shown in \reffig{decompress}.
Both the low-resolution image $x^{(3)}$ and all rounding parameters are stored raw and can be read directly from disk.
The compressed super-resolution operator depends on arithmetic coding and proceeds in three steps using three distinct network passes.
The super-resolution network first produces the mixture parameters of the RGB color values of the first pixel $\topleftb$.
These mixture parameters only depend on the low-resolution input image and can all be computed in parallel.
Arithmetic coding then decodes color values one at a time: \topleftb R$\rightarrow$\topleftb G$\rightarrow$\topleftb B.
Note that the mixture components of green and blue depend on the previously decoded values and need to be estimated in that order.
However, this can again happen in parallel once an entire color plane is decoded.

Once the first pixel $\topleftb$ is decoded, a second network pass produces the mixture parameters of the second pixel $\toprightb$, which is decoded analogous to the first. A final network pass then produces the third pixel value $\bottomleftb$.
The final pixel $\bottomrightb$ is reconstructed in closed form: $x_{2i+1,2j+1} = 4 y_{i,j} - x_{2i,2j}-x_{2i,2j+1}-x_{2i+1,2j}$.
All network passes and parameter computations are performed in parallel over the entire image, allowing for efficient decoding on a GPU.
However, decoding is marginally slower than encoding as it forces several GPU synchronizations caused by arithmetic coding.

\section{Implementation Details}
\lblsec{details}

\begin{figure*}[t]
\centering
\includegraphics[width=\linewidth]{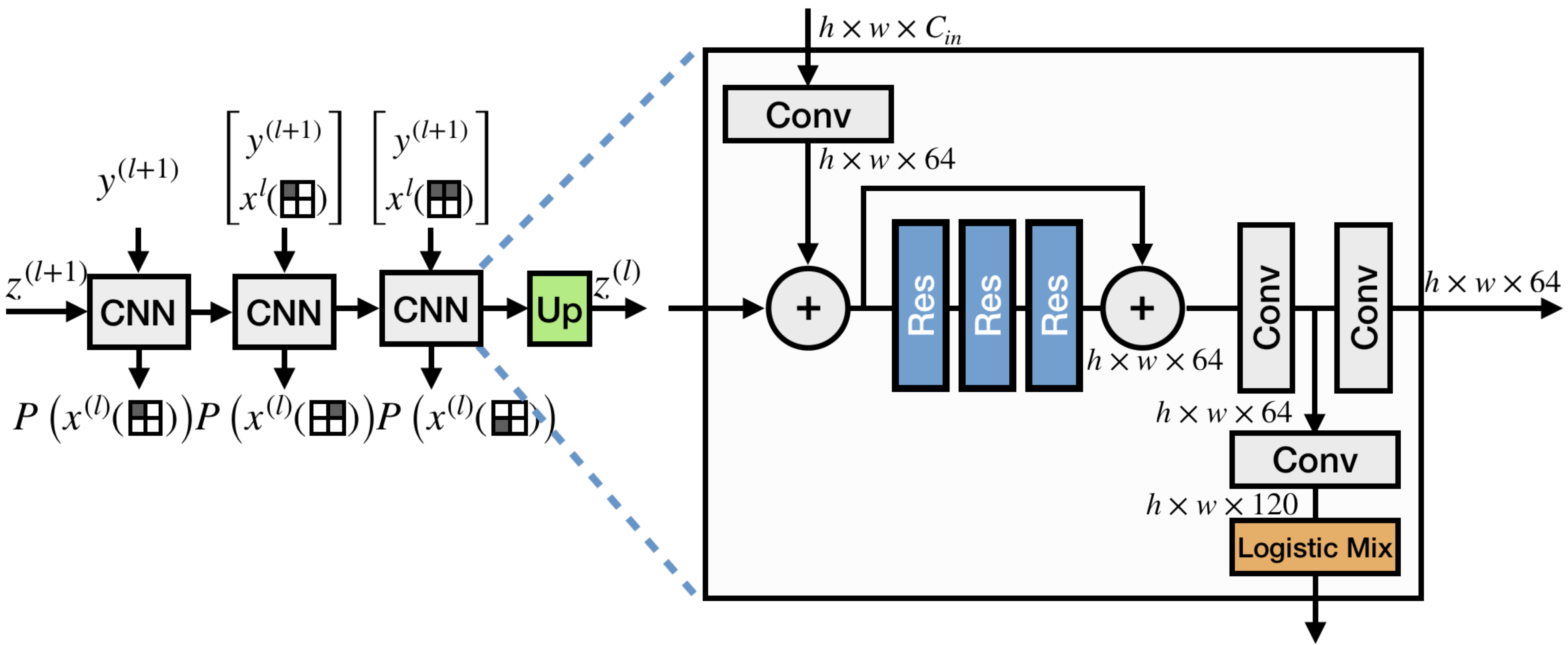}
\caption{\textbf{Network Architecture.}
We use a simple building block for each of the autoregressive super-resolution steps.
Right hand side shows the detailed architecture.
It is composed of simple convolutional layers with residual connections to model visual patterns
and a discretized-mixture-of-logistic output layer.
`Res' denotes a residual block; see the Appendix for details.
`Up' denotes an upsampling layer implemented as a PixelShuffle operator~\cite{shi2016real}.
It upsamples the output feature map such that it matches the resolution of the next super-resolution level.
Note the autoregressive structure: an output feature map and the pixel value of one step is passed as input to the next step.
}
\label{fig:architecture}
\end{figure*}

\paragraph{Network Architecture.}
\figref{architecture} shows the architecture of our autoregressive super-resolution network.
It has simple ResNet-like~\cite{resnet} network design, similar to a typical super-resolution architecture~\cite{edsr}.
Following Salimans~\etal~\cite{pixelcnnpp}, we use discretized mixture of logistics for the pixel probability distribution,
and adopt the same RGB pixel conditioning scheme.
See the Appendix for the exact architecture specification and details.

\paragraph{Handling General Image Sizes.}
In cases where an image's width/height is not divisible by 2$^3$,
we use repeat-padding at right-most column (or bottom row) when downsampling,
which is equivalent to pooling only 2 or 1 pixels at the border during down-sampling.
When the image has odd number of columns (or rows),
at decompression time, we discard the right-most column (or bottom row) to keep the original size.
Padded values do not need to be stored during encoding.

\paragraph{Super-Resolution Constraints.}
One important advantage of SR-based compression is that it naturally imposes range constraints in each block,
making predicting the values easier.
Specifically, given the definition of average pooling
$\frac{1}{4}\rbr{\sum_{i' = 2i}^{2i+1}\sum_{j' = 2j}^{2j+1}x^{(l)}_{i',j'}}=y_{i,j}^{(l+1)}$,
and the range constraint of a pixel $x_{i,j} \in \cbr{0, 1, \ldots, 255}, \forall i,j$, we have
\begin{align}
  4y_{i,j} - 3 \cdot 255 \le& x_{2i,2j} &\hspace{-7mm}\le& 4y_{i,j} \nonumber\\
  4y_{i,j} - x_{2i,2j} - 2 \cdot 255 \le& x_{2i,2j+1} &\hspace{-7mm}\le& 4y_{i,j} - x_{2i,2j} \label{eq:constraints}\\
  4y_{i,j} - x_{2i,2j} - x_{2i,2j+1} - 1 \cdot 255 \le& x_{2i+1, 2j}& \hspace{-7mm}\le& 4y_{i,j} - x_{2i,2j}  - x_{2i,2j+1}\nonumber
\end{align}
We can thus exclude impossible ranges during entropy coding, making our probability estimation more accurate.
For simple non-factorized models, these constraints significantly improve the performance.
For our autoregressive model, these constraints can be easily learned, so there is no need to explicitly impose them, as shown in experiments.

\section{Experiments}
We present detailed ablation study results in \refsec{ablation}, qualitative analysis in \refsec{qualitative},
and qualitative evaluation compared with other state-of-the-art methods in \refsec{compare}.

\paragraph{Datasets and Protocol.}
Our evaluation protocol follows Mentzer~\etal~\cite{l3c}.
We evaluate SReC on two datasets, ImageNet64~\cite{imagenet64,deng2009imagenet} and Open Images~\cite{openimages}.
ImageNet64 consists of downsampled 64$\x$64 images of ImageNet~\cite{deng2009imagenet}.
It contains $\app$1.28m training and 50k validation images.
Open Images~\cite{openimages} consists of high-resolution images.
We use 2 different versions of Open Images, JPEG
\footnote{In their published results L3C uses JPEG compressed images (with compression artifacts) for Open Images. This mistake was later discovered and fixed after publication. Here, we report results on both the JPEG compressed images, and lossless PNG images for a fair comparison. Note that JPEG compression is much easier to learn for all algorithms.}
and PNG.
For fair comparison, we follow the preprocessing steps of Mentzer~\etal~\cite{l3c}:
We downscale the images to 768 pixels on the longer side to reduce artifacts from prior compression.
We discard small ($<$1.25$\x$ downsampling) or high-saturation images.
For PNG images, we apply random downscaling while keeping the shorter side $\geq$ 512 pixels.
We use Lanczos~\cite{turkowski1990filters} interpolation instead of bilinear for downscaling.
We pick the same set of validation images as Mentzer~\etal~\cite{l3c}.
This process results in $\app$336k training and 500 validation images.

We measure compression rate by bits per subpixel (bpsp).
We measure runtime of all methods on the same machine
with AMD Ryzen 5 1600 and NVIDIA GTX 1060.
All runtime measurements use a single image (batch size of $1$).

\paragraph{Training Details.}
We use Adam~\cite{adam} with a batch size of 32 and no weight decay.
For regularization, we apply gradient norm clipping at 0.5.
We train our model for 10 epochs on ImageNet64~\cite{imagenet64} and 50 epochs on Open Images~\cite{openimages}.
We use a learning rate of 10$^{-4}$, which is then decreased by a factor of 0.75 every epoch for ImageNet64~\cite{imagenet64} and every 5 epochs for Open Images~\cite{openimages}.

We apply random horizontal flipping for training.
We train with the same crop sizes as L3C for fair comparison, which is 64$\x$64 on ImageNet64 and 128$\x$128 on Open Images.

\subsection{Ablation Experiments}
\lblsec{ablation}

We use Open Images~\cite{openimages} (PNG) for extensive ablation studies.
\begin{table}[t]
\vspace{-3mm}
\hfill
  \subfloat[\textbf{Network Design.}\label{tab:ablation:net}]{
\tablestyle{2.5pt}{1.12}\begin{tabular}{@{}lx{24}@{}}
& bpsp \\
\shline
SR& 3.36\\
SR + constraints&3.03\\
\textbf{SR + factorization}& \textbf{2.69}\\
SR + factorization + constraints& {2.69}\\
\end{tabular}}\hfill
  \subfloat[\textbf{Compression Scheme.}\label{tab:ablation:levels}]{
\tablestyle{2.5pt}{1.12}\begin{tabular}{@{}lx{24}@{}}
  & bpsp\\
\shline
1-level&3.87\\
2-level&2.90\\
\textbf{3-level}\quad\quad\quad\quad\quad\quad\quad\quad&\textbf{2.69}\\
4-level&{2.68}
\end{tabular}}\hfill
\vspace{2mm}
\caption{\textbf{Ablation Study.}
We perform ablations on the Open Images dataset~\cite{openimages}.
\tabref{ablation:net} validates that a super-resolution-based method benefits from the natural
constraints imposed by the setting, and our factorized method is able to easily leverage the constraints
and performs better than baselines.
\tabref{ablation:levels} demonstrates that each super-resolution level improves our modeling power,
leading to stronger compression.\vspace{-6mm}
}\label{tab:ablations}
\end{table}
\begin{table}[h!]
\subfloat[\textbf{Speed.}\label{tab:ablation:speed}]{
\tablestyle{2.5pt}{1.12}\begin{tabular}{@{}lx{32}r@{}}
& time (s) & {\tiny\demph{\%}}\\
\shline
$x^{(3)}$ & 0.0002 &\percent{0.01}\\
$x^{(2)}$ & 0.077 &\percent{6.7}\\
$x^{(1)}$ & 0.230 &\percent{20.0}\\
$x^{(0)}$ & 0.842 &\percent{73.3}\\
\hline
\demph{Total} & \demph{1.149}\\
\\
\\
\end{tabular}}\hfill
\subfloat[\textbf{Compression Rate.}\label{tab:ablation:size}]{
\tablestyle{2.5pt}{1.12}\begin{tabular}{@{}lx{24}r@{}}
  & bpsp & {\tiny\demph{\%}}\\
\shline
rounding bits & 0.656 & \percent{24.3}\\
metadata & 0.002 & \percent{0.1}\\
$x^{(3)}$ (raw image) & 0.125 & \percent{4.6}\\
$x^{(2)}$ & 0.126 & \percent{4.7}\\
$x^{(1)}$ & 0.432 & \percent{16.0}\\
$x^{(0)}$ & 1.360 & \percent{50.3}\\
\hline
\demph{Total} & \demph{2.701} \\
\end{tabular}}\hfill
\subfloat[\textbf{Scalability.}\label{tab:ablation:scale}]{
\tablestyle{2.5pt}{1.12}\begin{tabular}{@{}lx{32}x{32}@{}}
W$\x$H & enc (s) & dec (s) \\
\shline
32$^2$ & 0.036 & 0.049\\
64$^2$ & 0.045 & 0.085 \\
320$^2$ & 0.277 & 0.327\\
640$^2$ & 0.977 & 1.101\\
720$^2$ & 1.166 & 1.549\\
960$^2$ & 2.148 & 2.373\\
\\
\end{tabular}}
\vspace{2mm}
\caption{\textbf{Detailed Analysis}.
We present speed and compression-rate analysis on Open Images~\cite{openimages} in \tabref{ablation:speed} and \tabref{ablation:size} respectively.
We demonstrate that our method scales to high-resolution (up to 960$\x$960) images in \tabref{ablation:scale}.\vspace{-4mm}
}\label{tab:ablations}
\end{table}

\paragraph{Network Design.}
We first ablate our network design in \tabref{ablation:net}.
We evaluate different designs using log-likelihood in bits per subpixels (bpsp).
We start with a baseline which simply replaces the latent factors of a latent factor model by downsampled images (denoted `SR' in table).
This model is equivalent to L3C~\cite{l3c} with RGB latent factors, and intermediate supervision on those latent factors.
When we impose the range constraints of \eqnref{constraints} (denoted `SR + constraints' in table),
we can see a large improvement in bpsp (3.36 $\rightarrow$ \textbf{3.03}).
In fact, this latent factor RGB model with constraints performs as well as the full fledged L3C~\cite{l3c} model.
This result suggests that by framing lossless compression as super-resolution,
we can indeed benefit from the natural constraints, resulting in more accurate predictions.
If we use a factorized model (`SR + factorization') to model image structures,
we see even better results (3.03 $\rightarrow$ \textbf{2.69}).
Note that the factorized model can easily learn these simple range constraints, and is most likely learning an even more complex prior in the color distribution.
In fact, adding range constraints to a factorized model does not further improve performance (denoted `SR + factorization + constraints').
In the following, we thus use our factorized variant without explicit constraints as our default model.

\paragraph{Compression Scheme.}
\tabref{ablation:levels} compares log-likelihoods of different numbers of super-resolution levels.
Each additional level improves the performance,
saturating at three levels.
We thus use a 3-level design as our default choice.

\paragraph{Speed.}
\tabref{ablation:speed} presents detailed decompression runtime analysis with AC.
Decompressing $x^{(3)}$ is very efficient (0.01\% of total runtime) as they are simply stored as raw pixels.
Higher-resolution super-resolution consumes most of the runtime.
Overall we observe runtime of each level roughly linear to the number of pixels in the level.

\paragraph{Compression Rate.}
\tabref{ablation:size} shows detailed analysis of compression rate with AC.
We see that while $x^{(3)}$ is simply stored as raw images, it contributes to only a small fraction of the overall bpsp.
Finer-level of super-resolution requires more bpsp as expected, as finer-grained details are harder to predict.
Rounding bits also have less structure to exploit.
We store them uncompressed, taking $\app$24\% of the overall bpsp.
Metadata to store width and height is negligible.
The rounding in AC causes 0.01 bpsp increase compared to raw log-likelihoods.

\paragraph{Scalability.}
\tabref{ablation:scale} presents the runtime with AC when scaling to high-resolution images.
Images $\geq$640$^2$ come from the higher-resolution DIV2K dataset~\cite{agustsson2017ntire}.
We see that even for high-resolution $960^2$ images,
both the encoding and decoding time stay practical.
Encoding is more efficient than decoding, because it requires fewer CPU-GPU synchronizations.

\begin{figure}[t]
\setlength{\tabcolsep}{0.1em}
\center
  \begin{tabular}{@{}cccccccc@{}}
  $x^{(0)}$ (full) & $x^{(0)}$ & $x^{(3)}$ & sample 1 & sample 2 & sample 3 & sample 4 & sample 5\\
\frame{\includegraphics[height=0.12\textwidth]{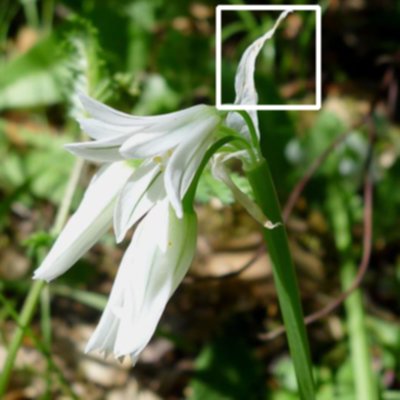}}&
\frame{\includegraphics[height=0.12\textwidth]{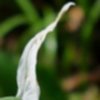}}&
\frame{\includegraphics[height=0.12\textwidth]{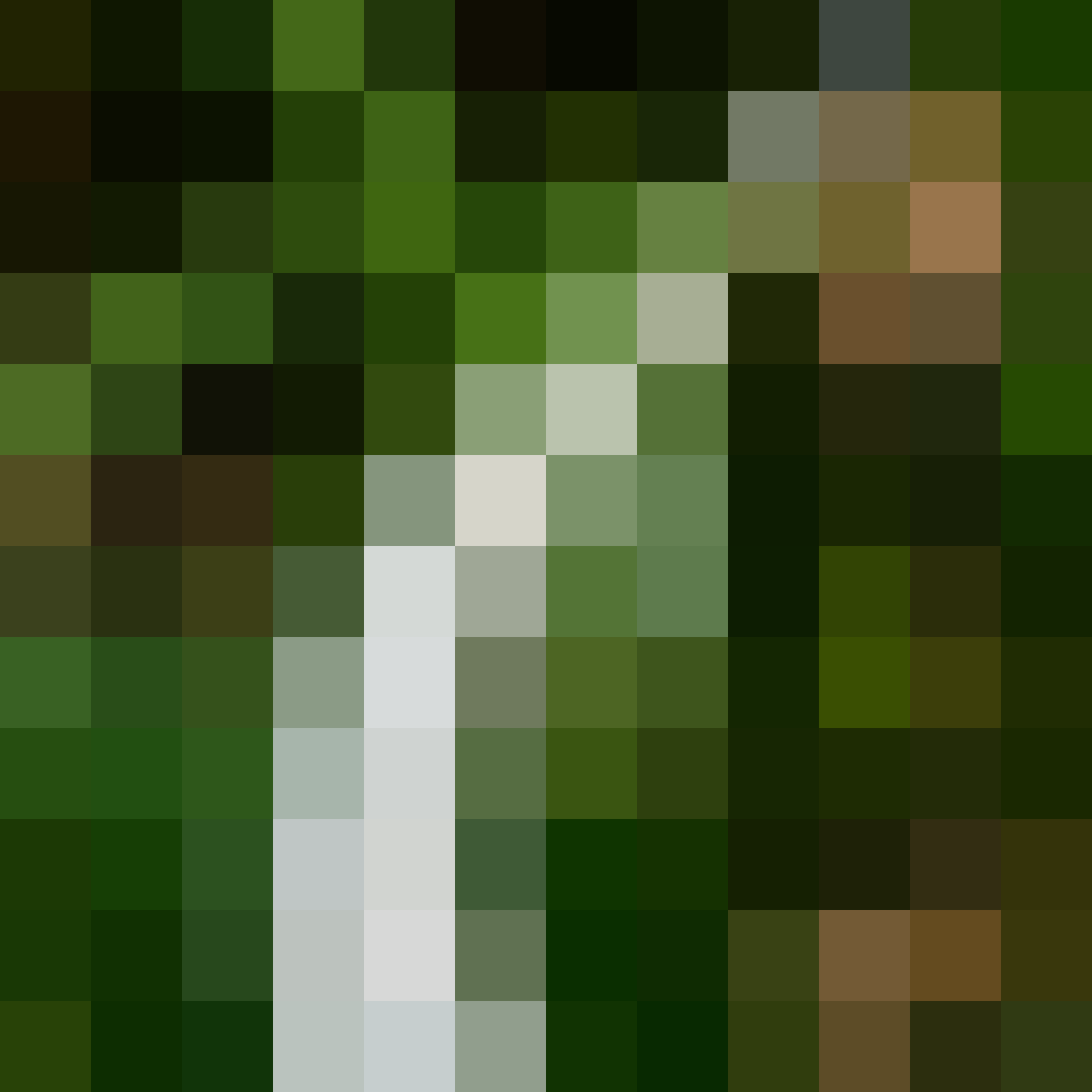}}&
\frame{\includegraphics[height=0.12\textwidth]{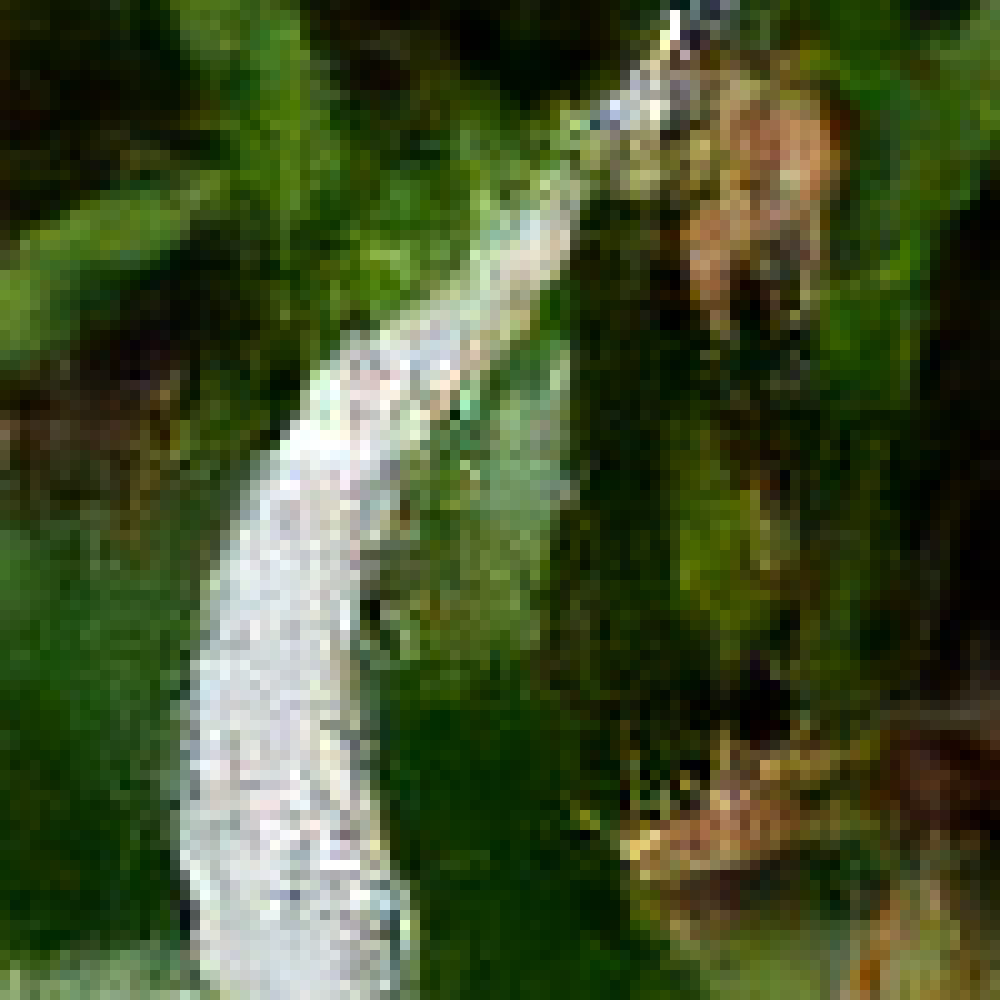}}&
\frame{\includegraphics[height=0.12\textwidth]{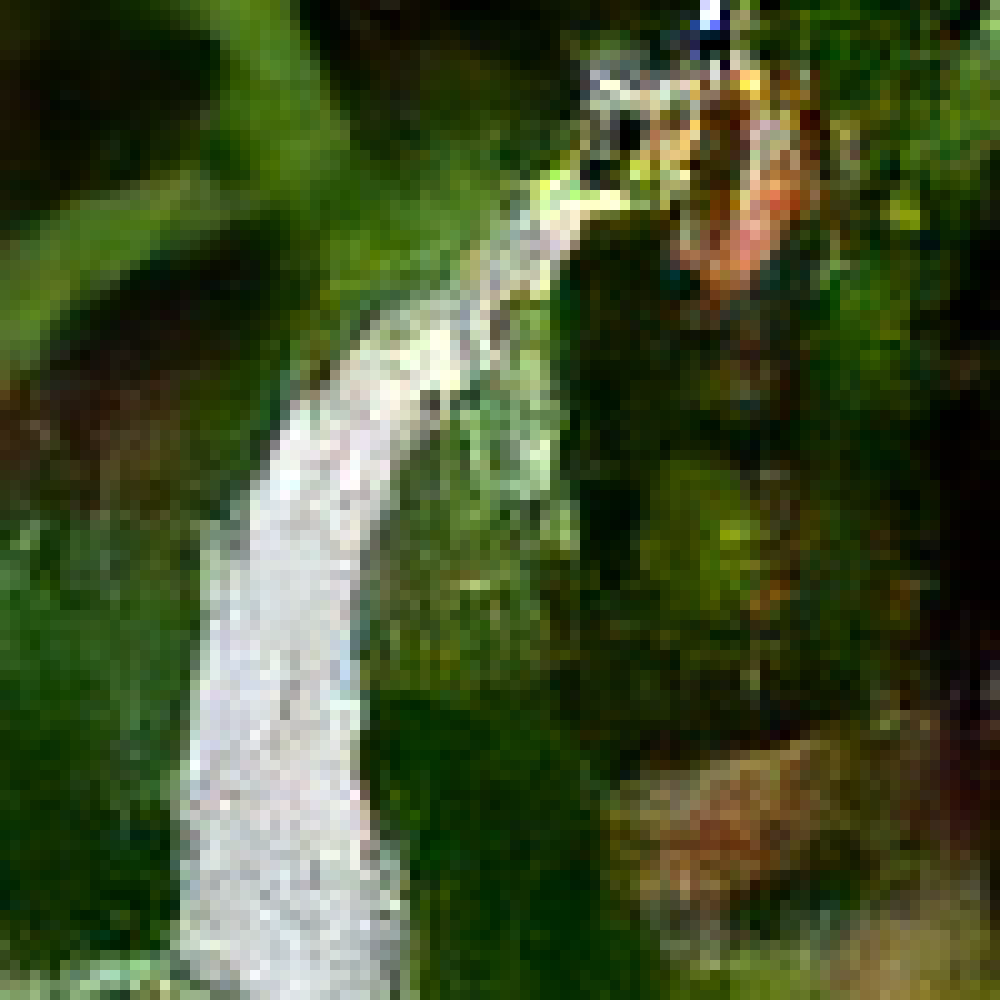}}&
\frame{\includegraphics[height=0.12\textwidth]{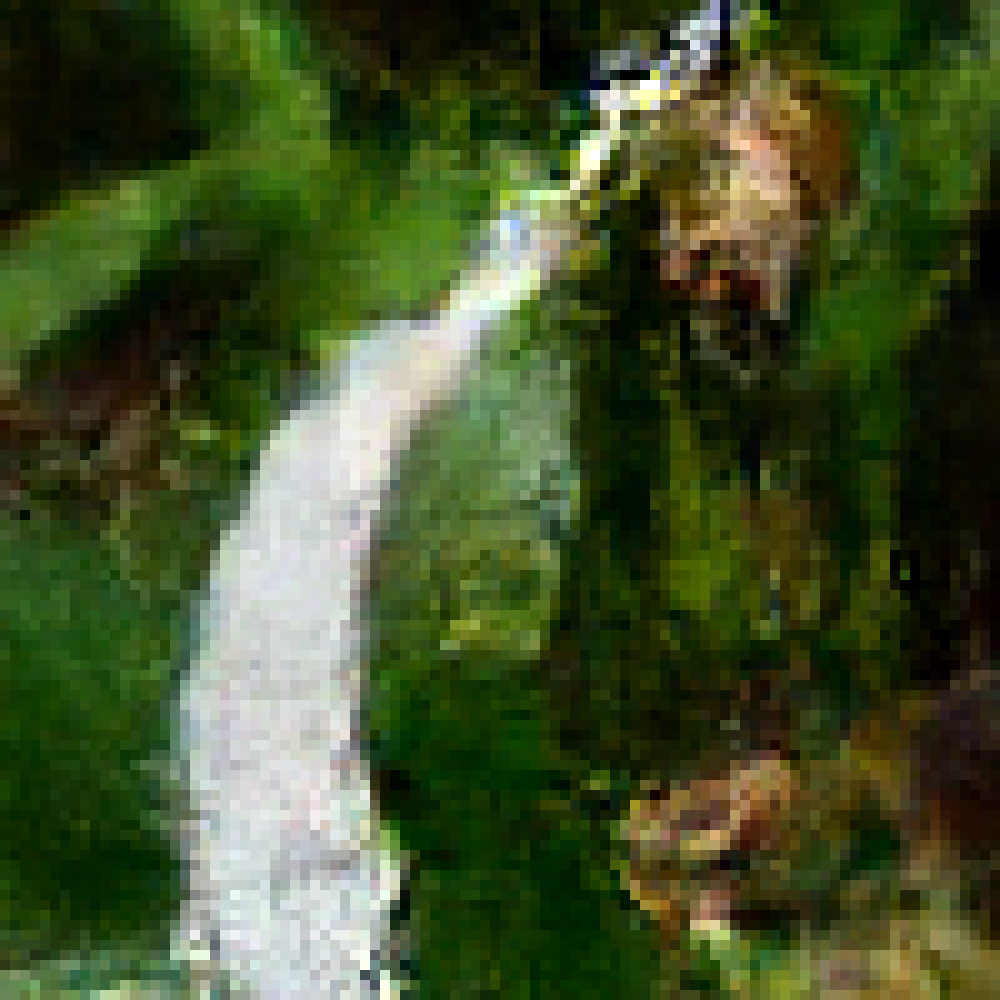}}&
\frame{\includegraphics[height=0.12\textwidth]{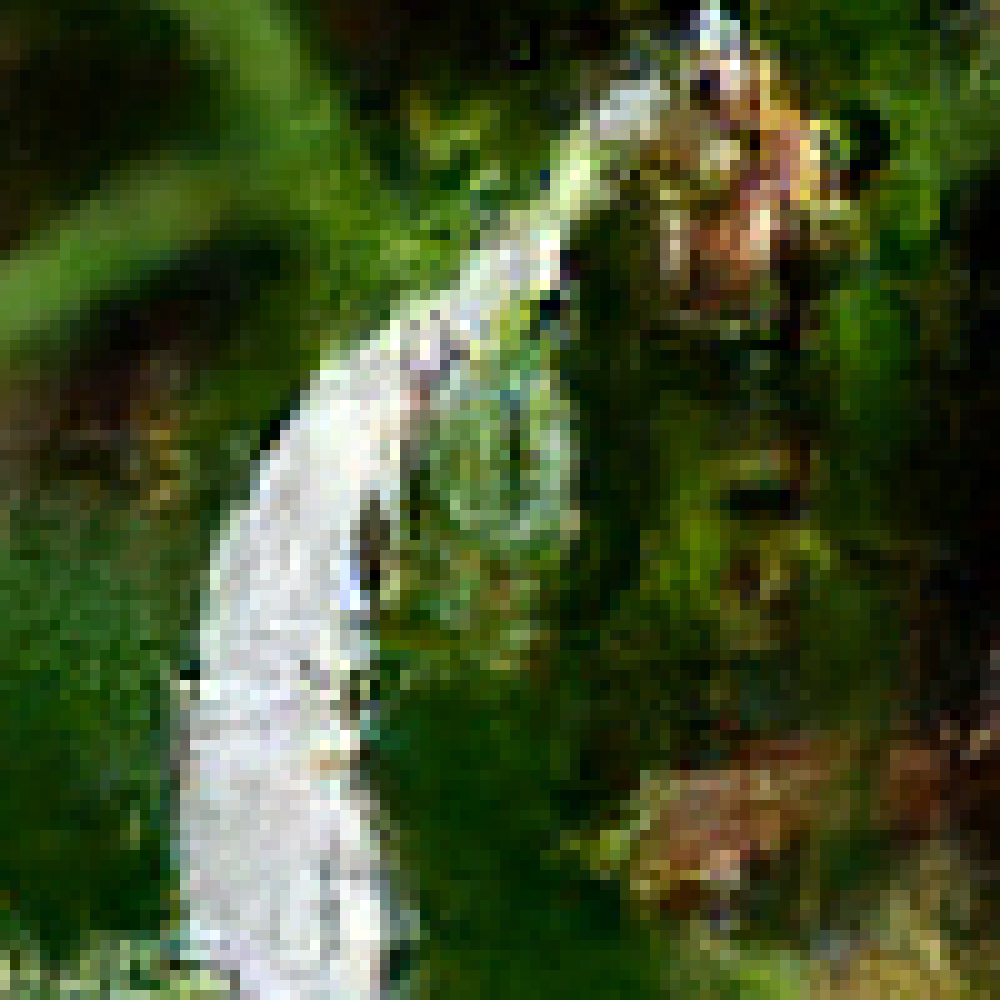}}&
\frame{\includegraphics[height=0.12\textwidth]{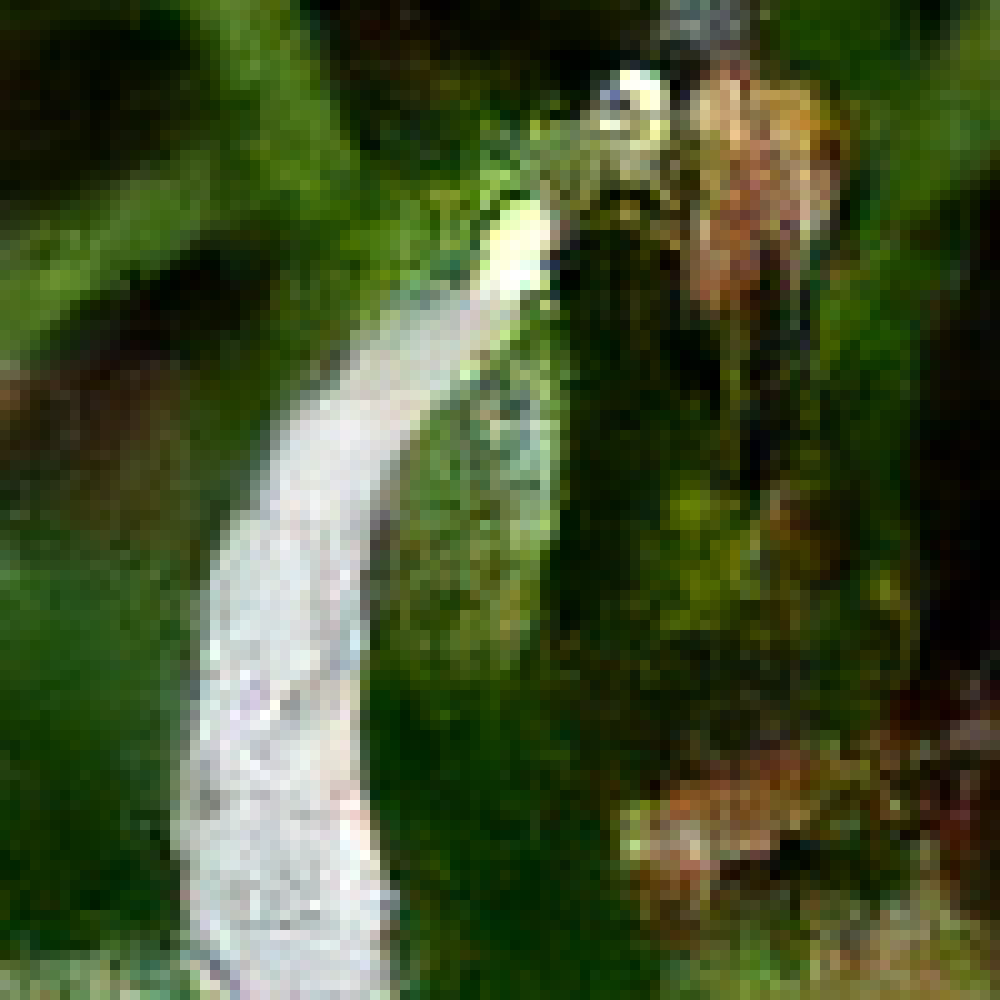}}\\
\frame{\includegraphics[height=0.12\textwidth]{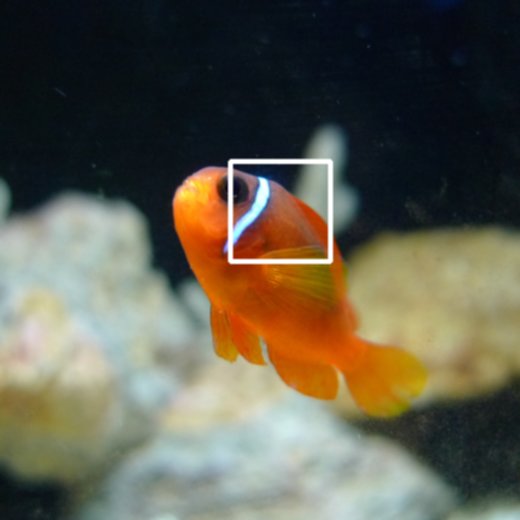}}&
\frame{\includegraphics[height=0.12\textwidth]{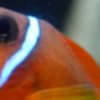}}&
\frame{\includegraphics[height=0.12\textwidth]{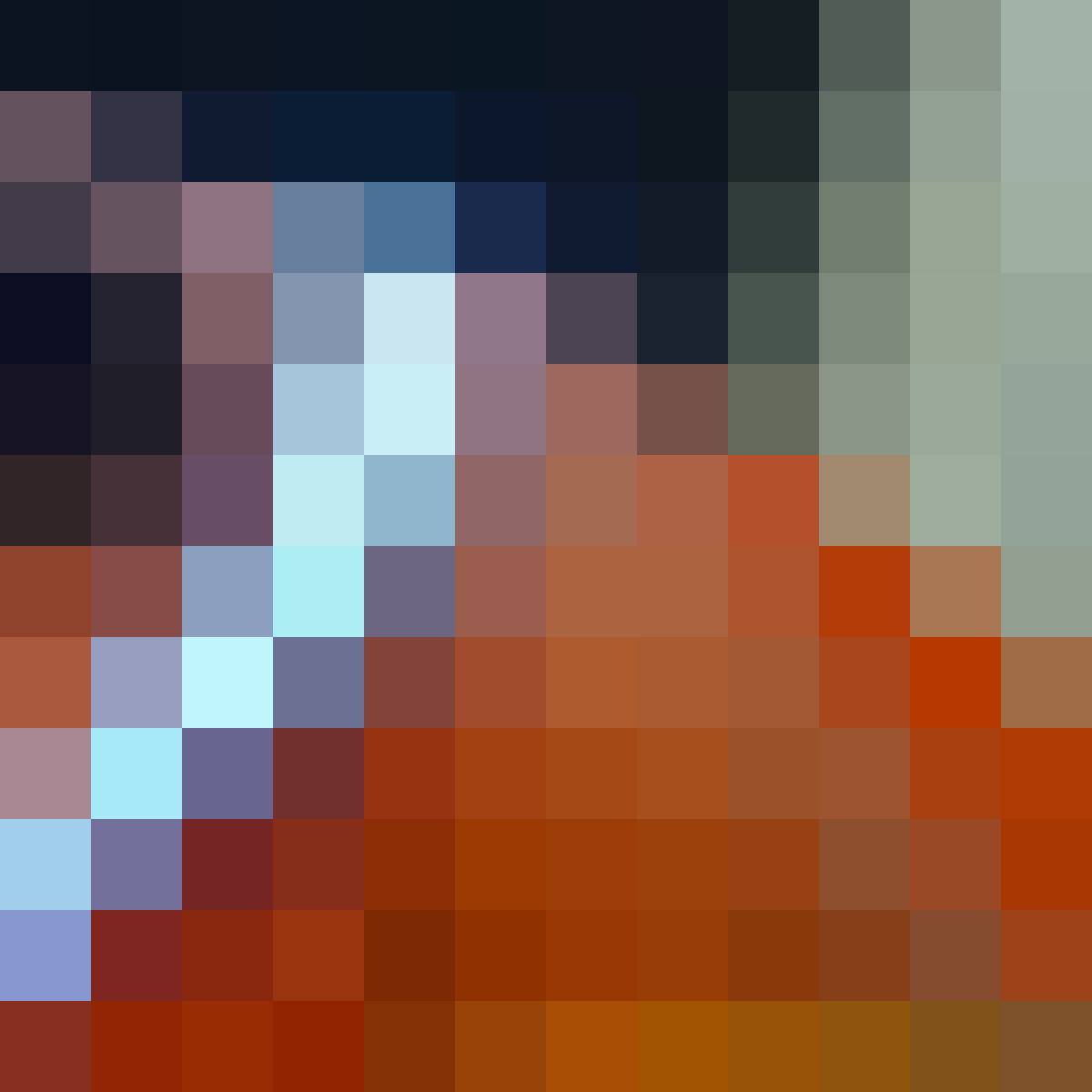}}&
\frame{\includegraphics[height=0.12\textwidth]{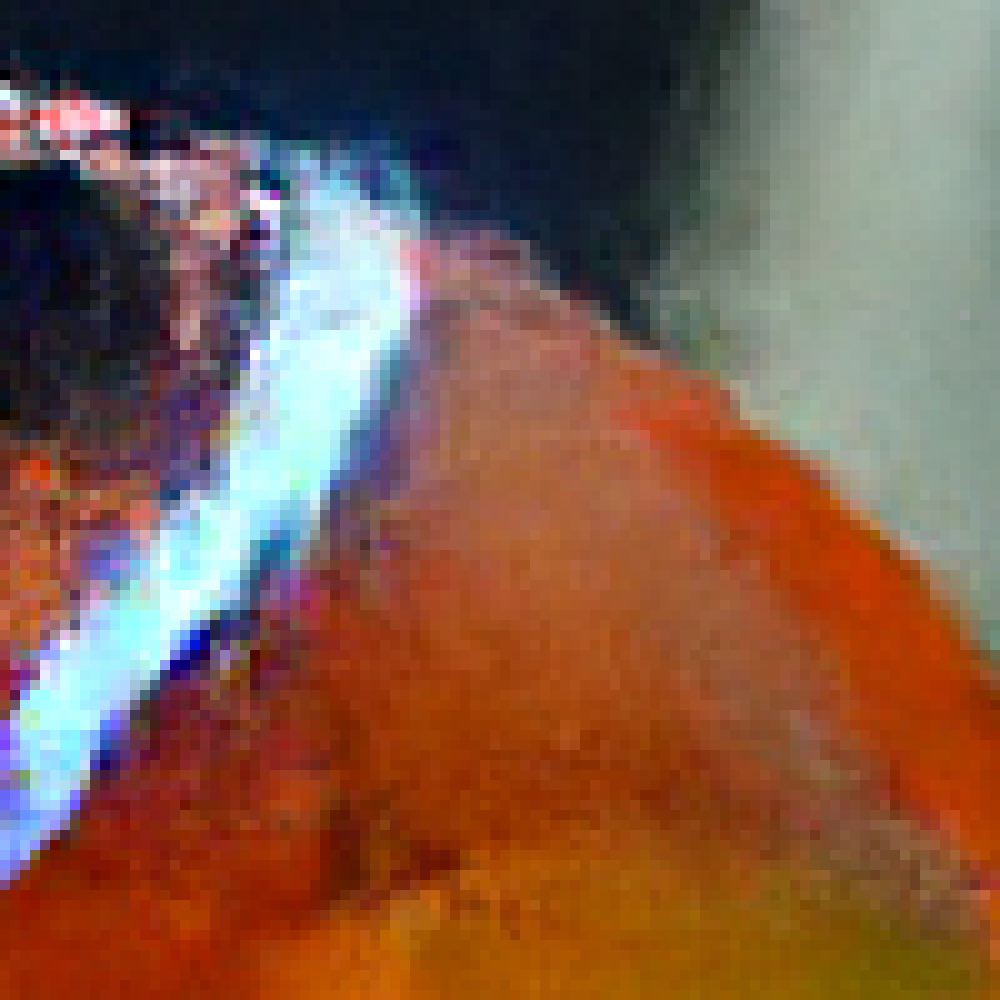}}&
\frame{\includegraphics[height=0.12\textwidth]{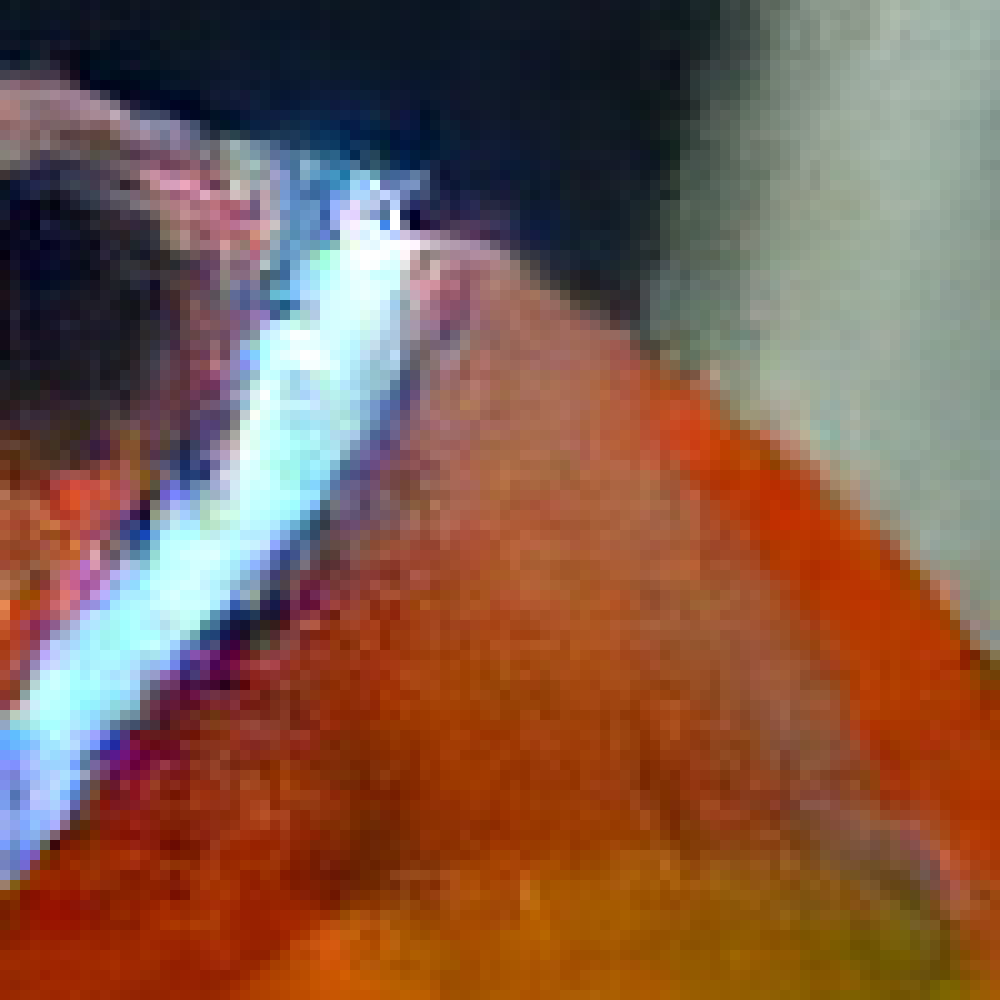}}&
\frame{\includegraphics[height=0.12\textwidth]{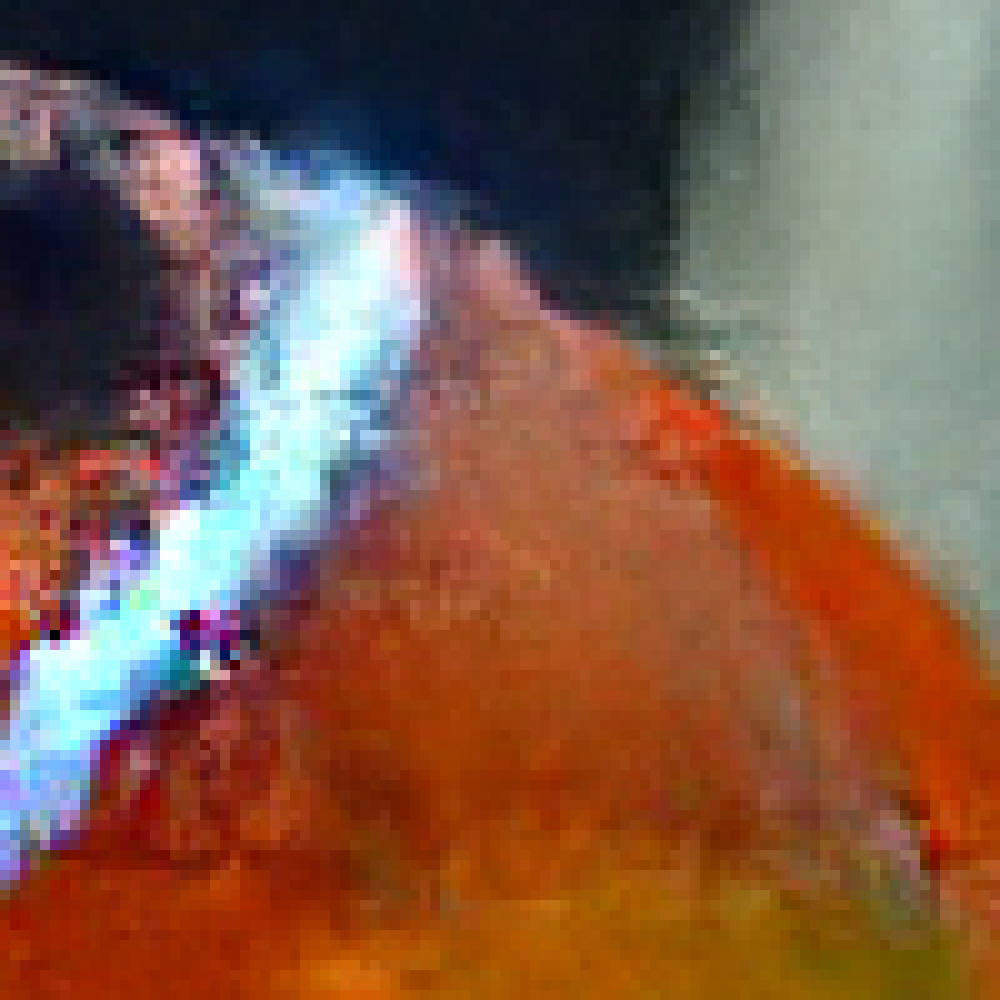}}&
\frame{\includegraphics[height=0.12\textwidth]{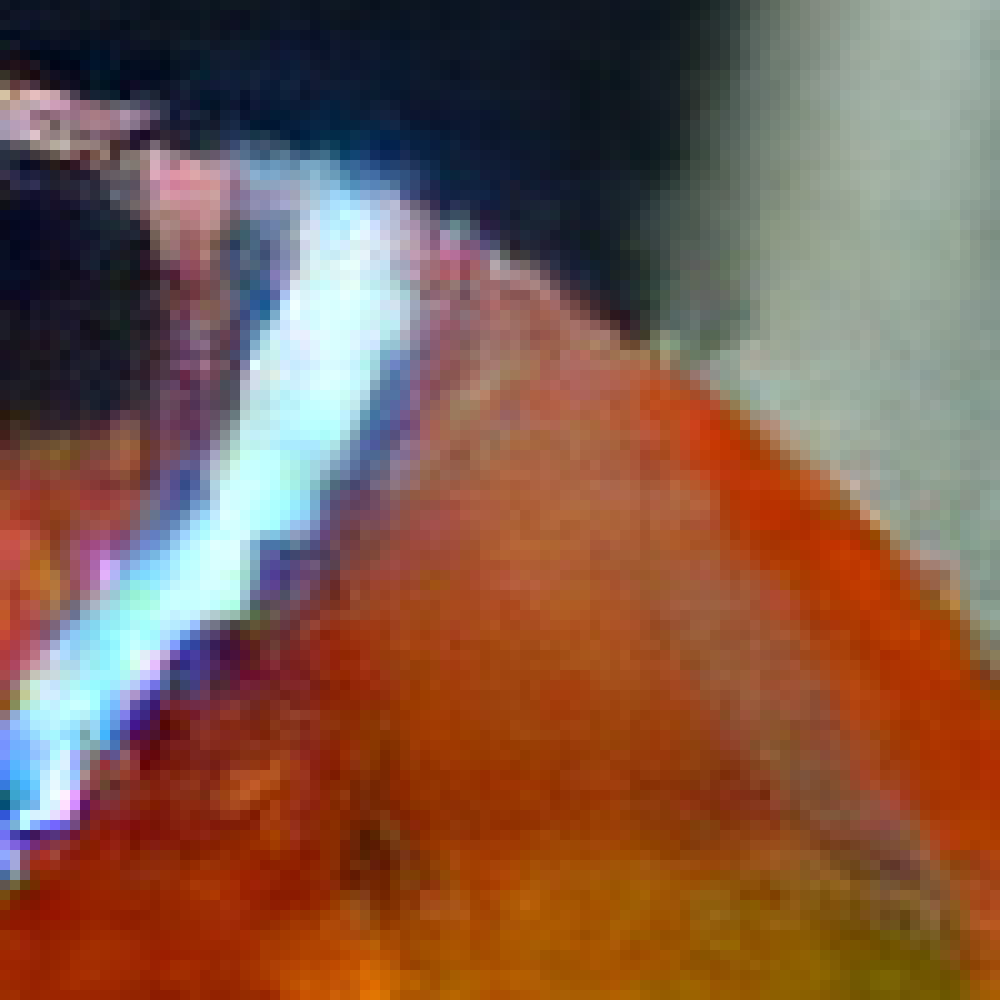}}&
\frame{\includegraphics[height=0.12\textwidth]{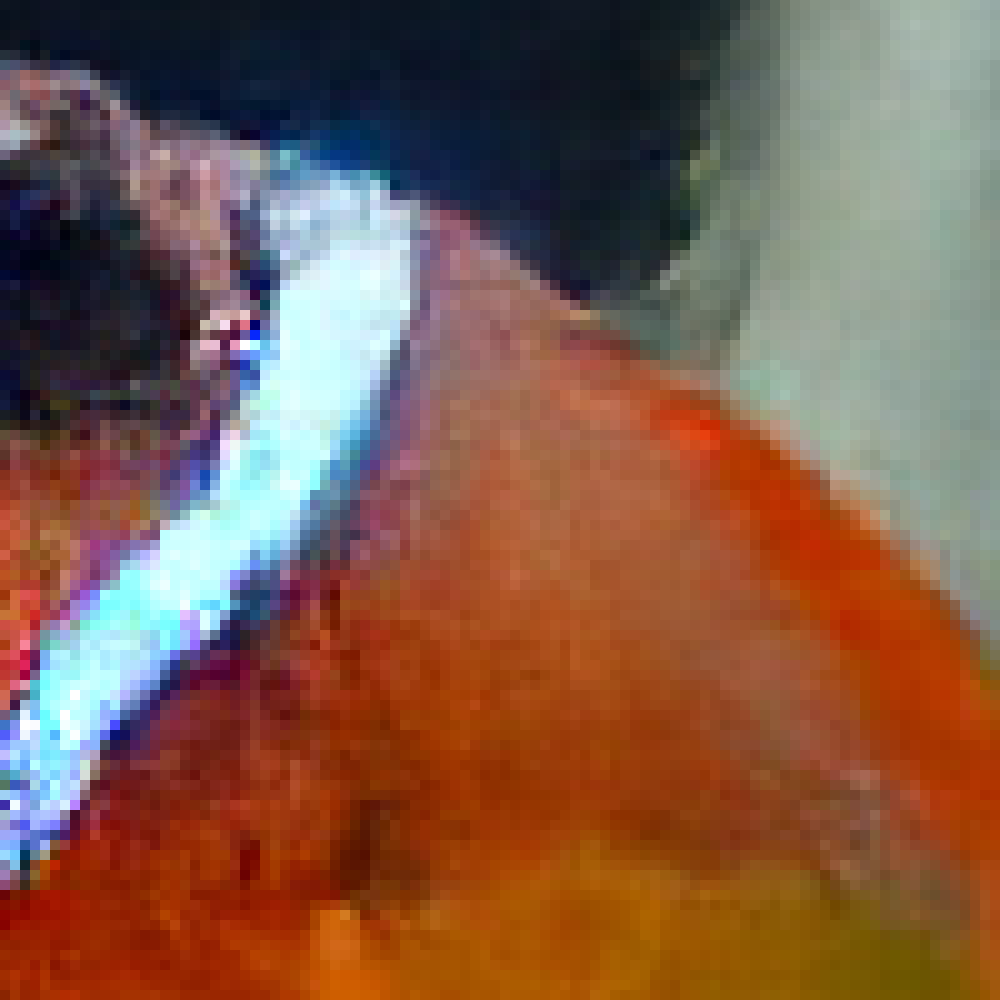}}\\
\frame{\includegraphics[height=0.12\textwidth]{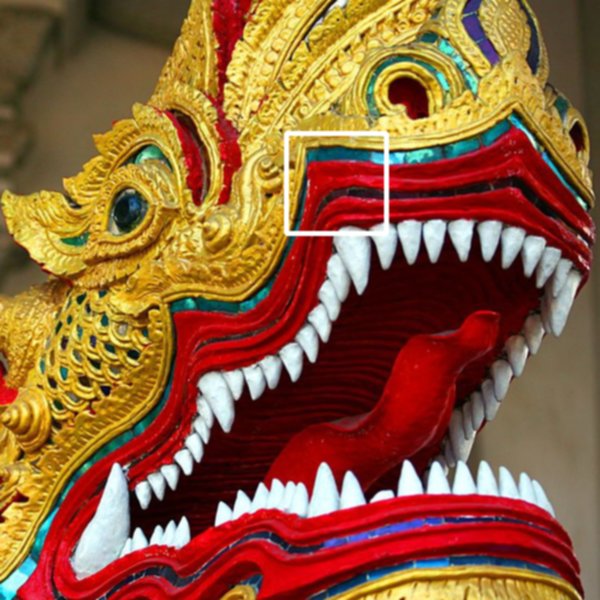}}&
\frame{\includegraphics[height=0.12\textwidth]{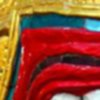}}&
\frame{\includegraphics[height=0.12\textwidth]{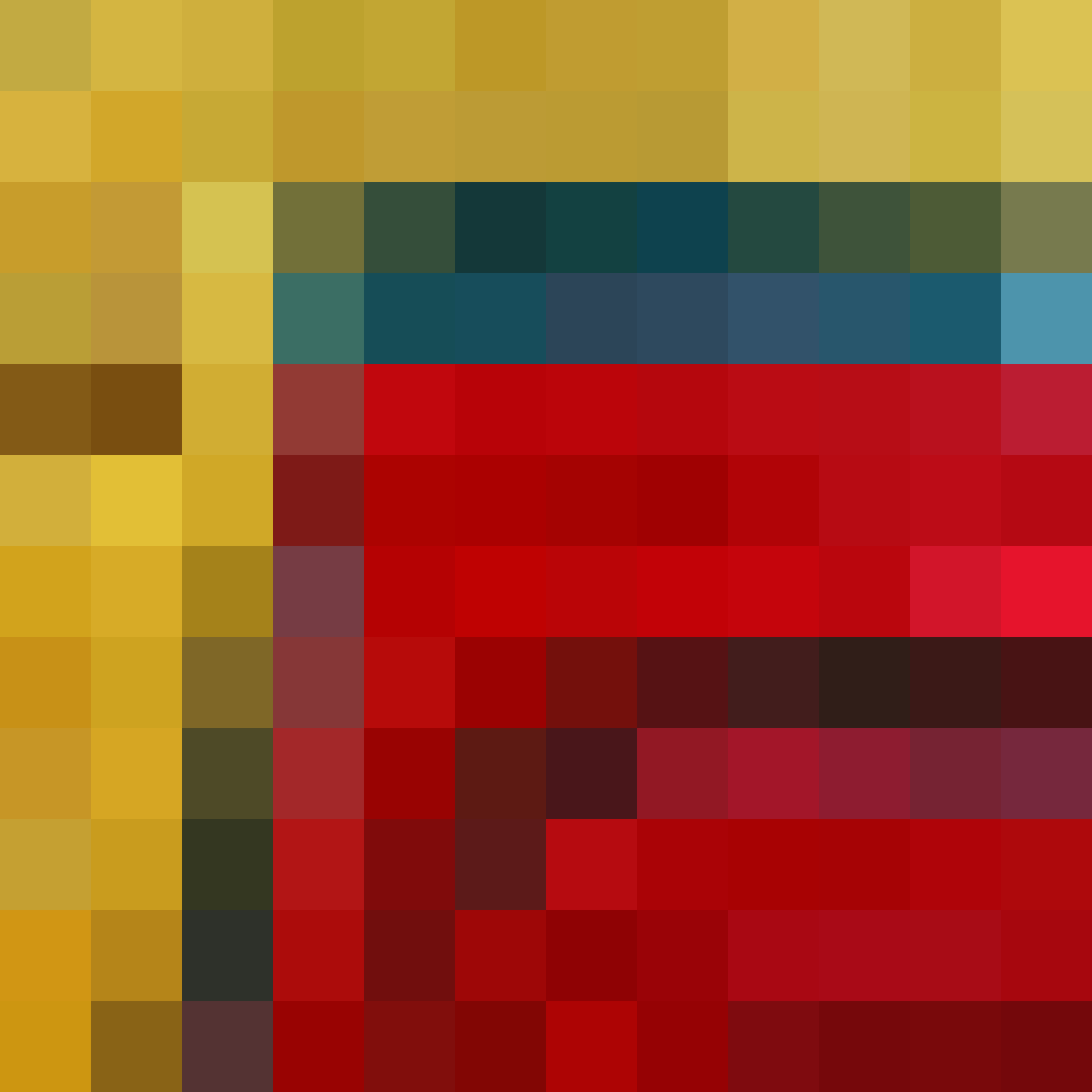}}&
\frame{\includegraphics[height=0.12\textwidth]{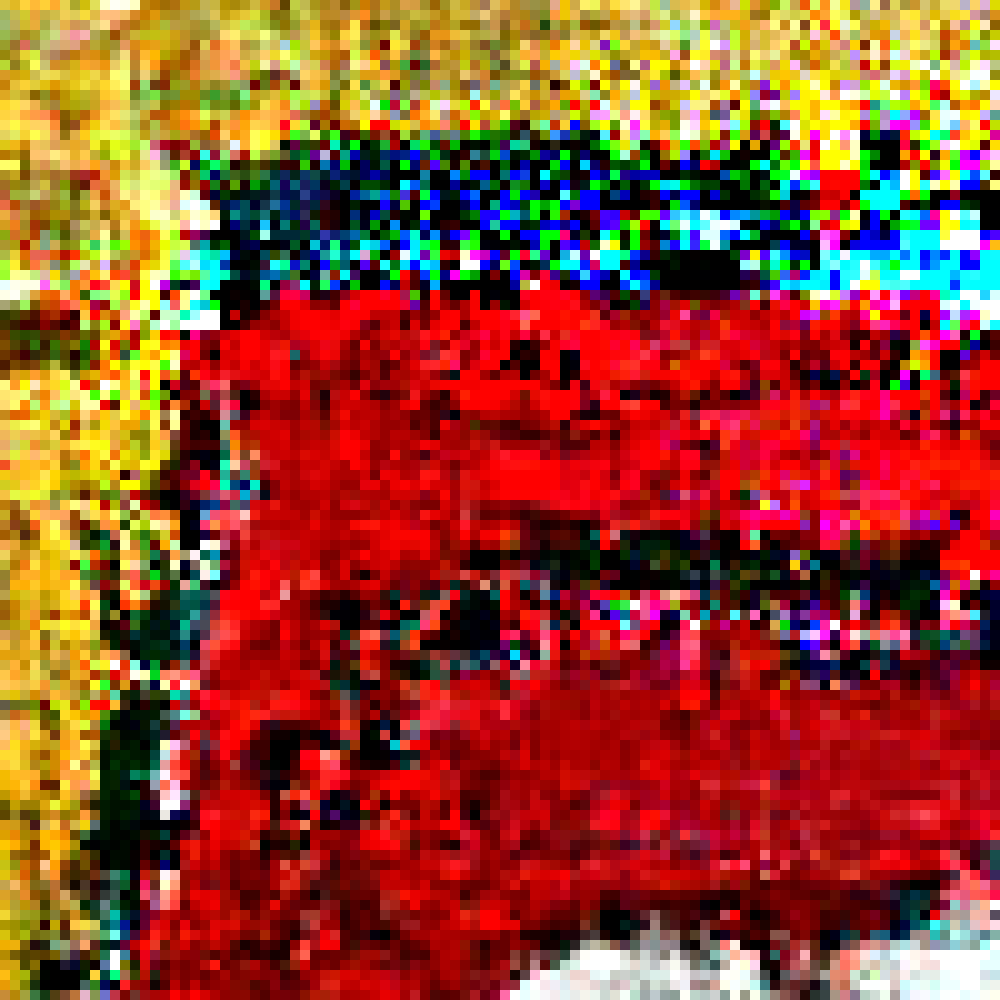}}&
\frame{\includegraphics[height=0.12\textwidth]{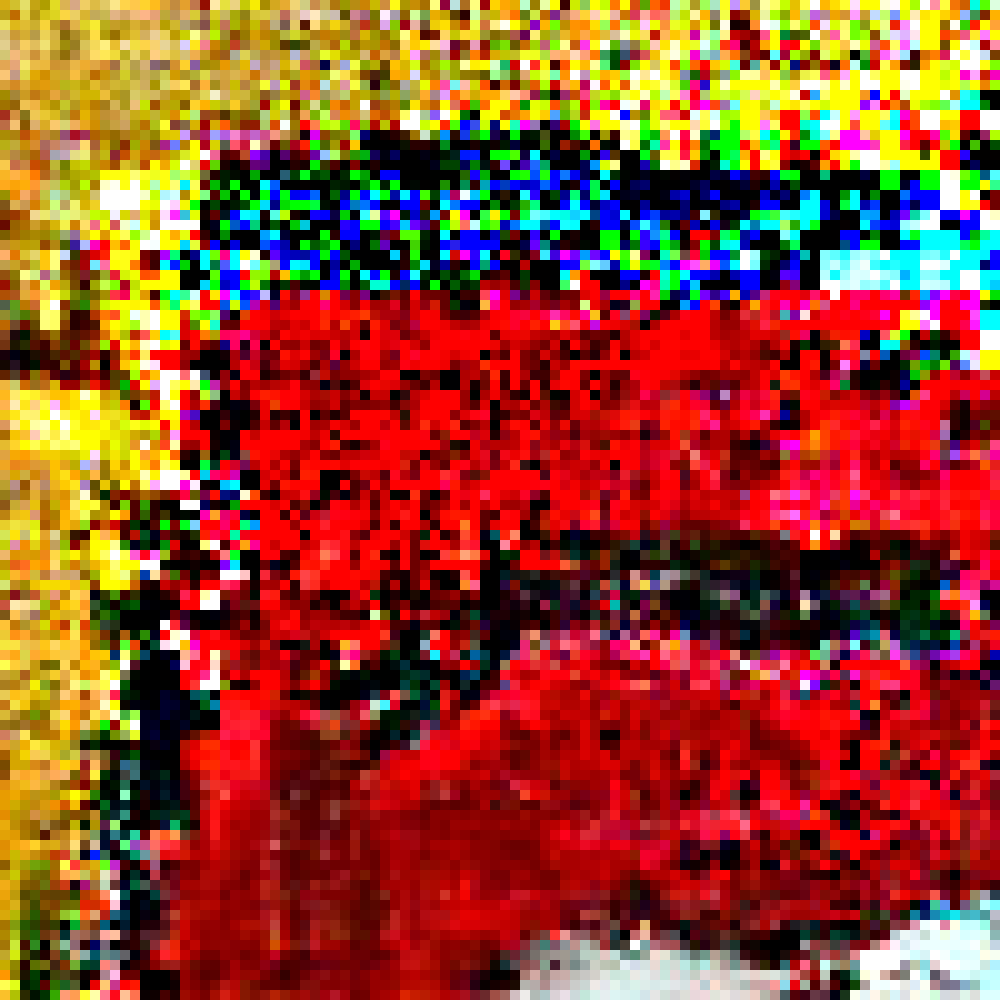}}&
\frame{\includegraphics[height=0.12\textwidth]{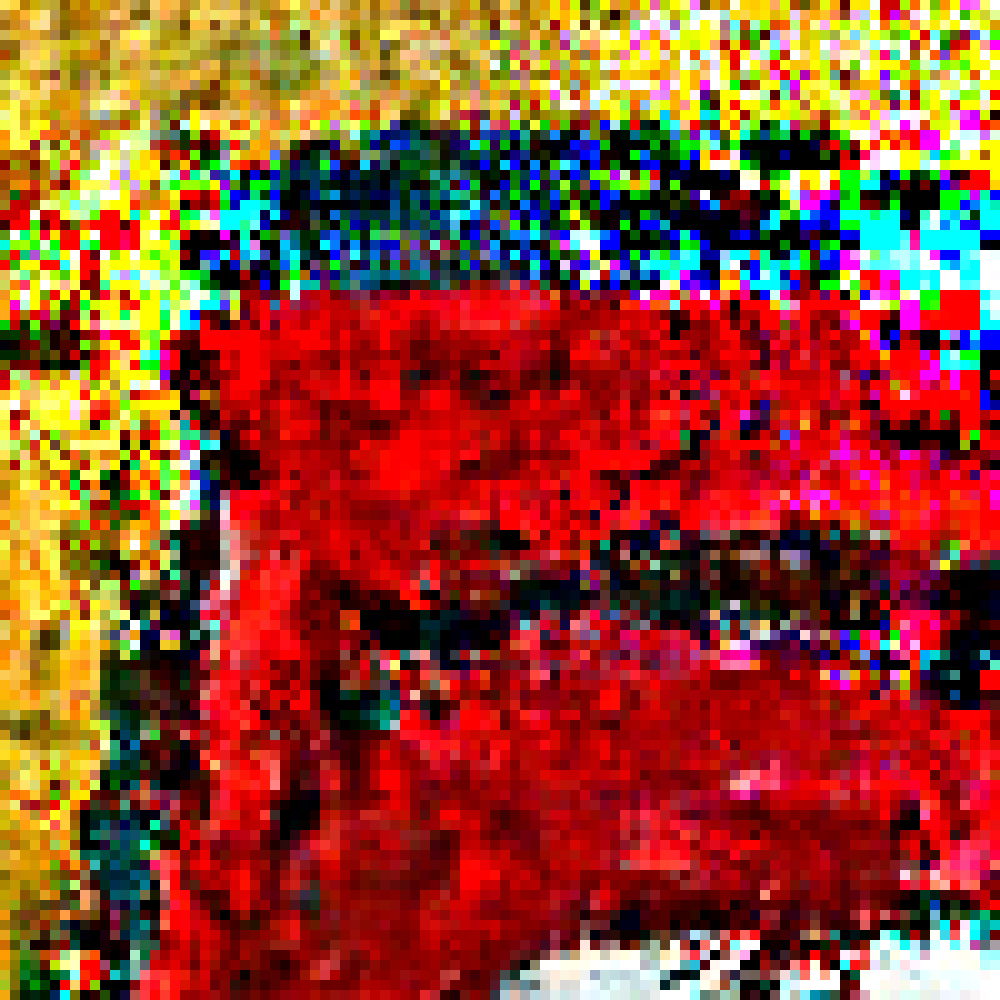}}&
\frame{\includegraphics[height=0.12\textwidth]{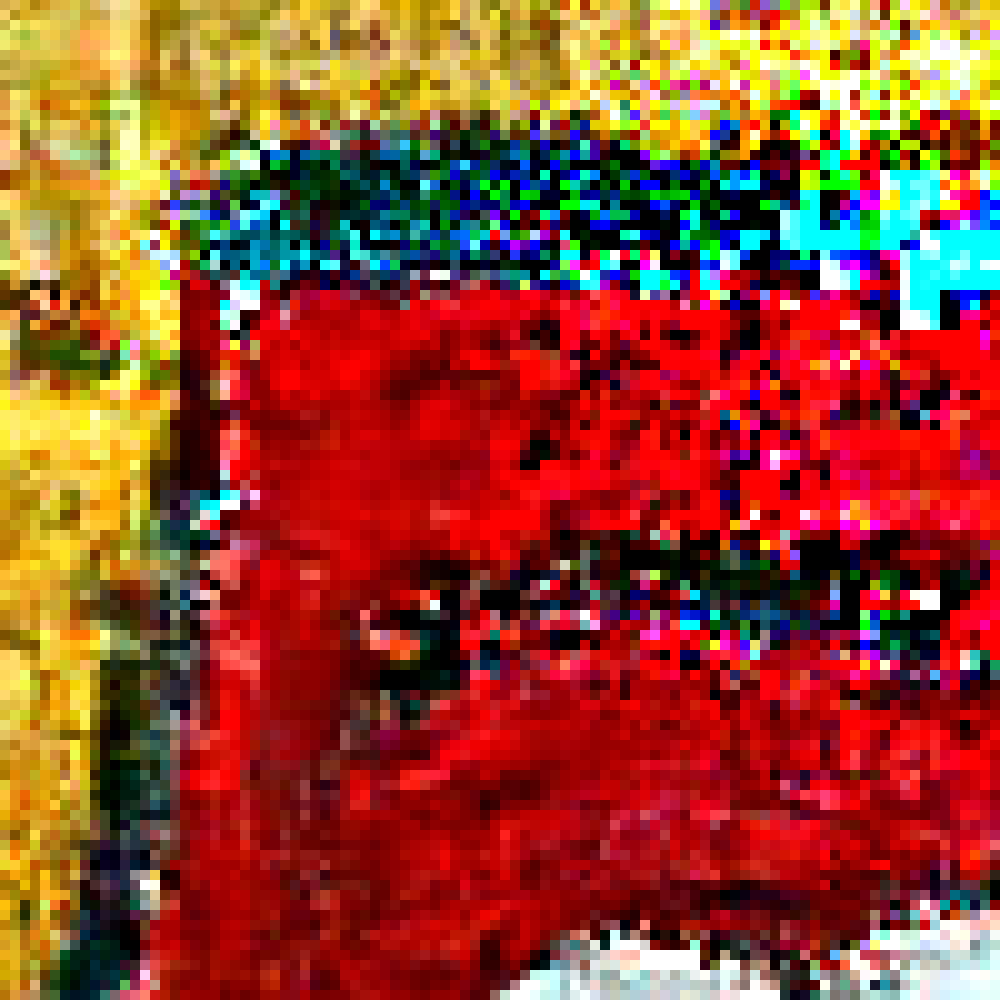}}&
\frame{\includegraphics[height=0.12\textwidth]{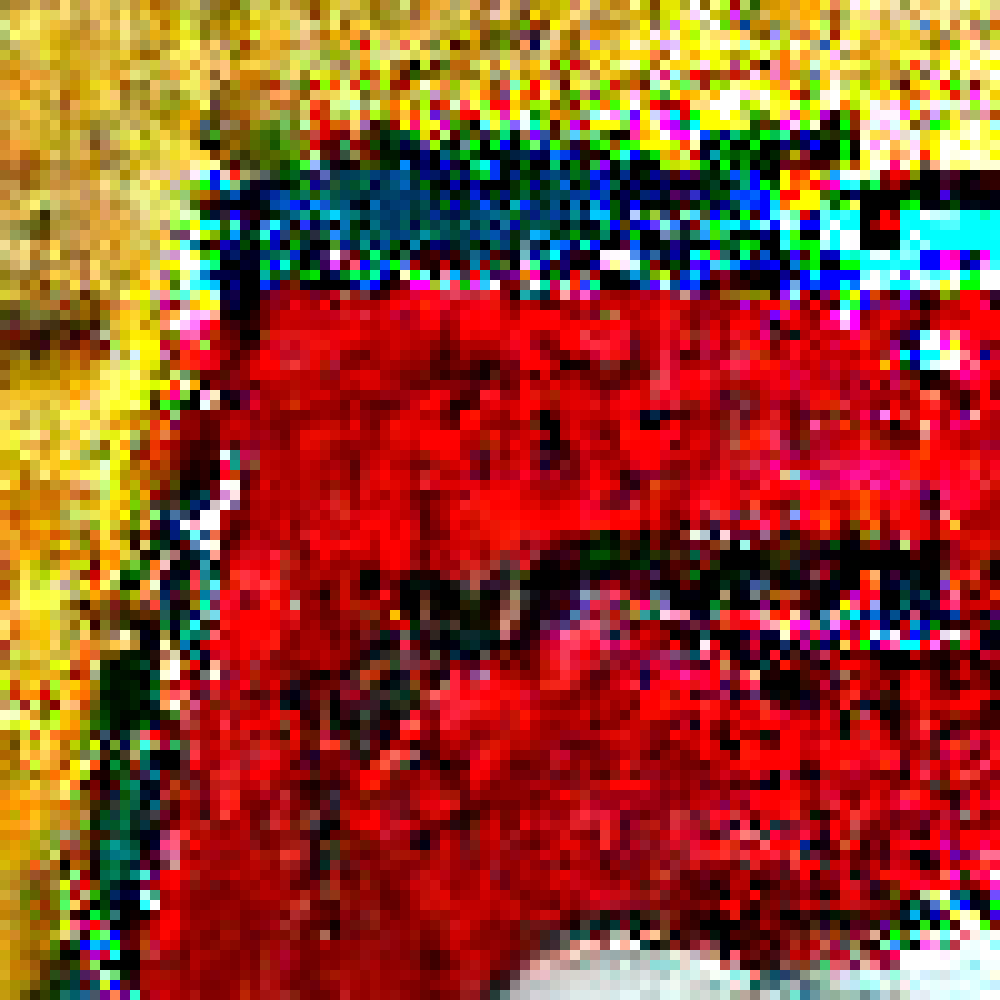}}\\
\frame{\includegraphics[height=0.12\textwidth]{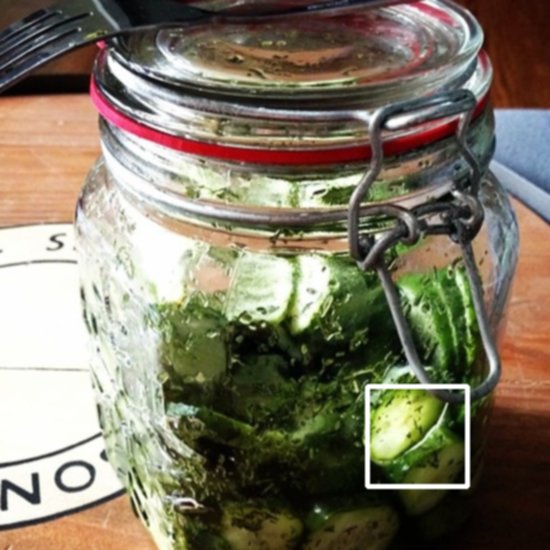}}&
\frame{\includegraphics[height=0.12\textwidth]{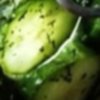}}&
\frame{\includegraphics[height=0.12\textwidth]{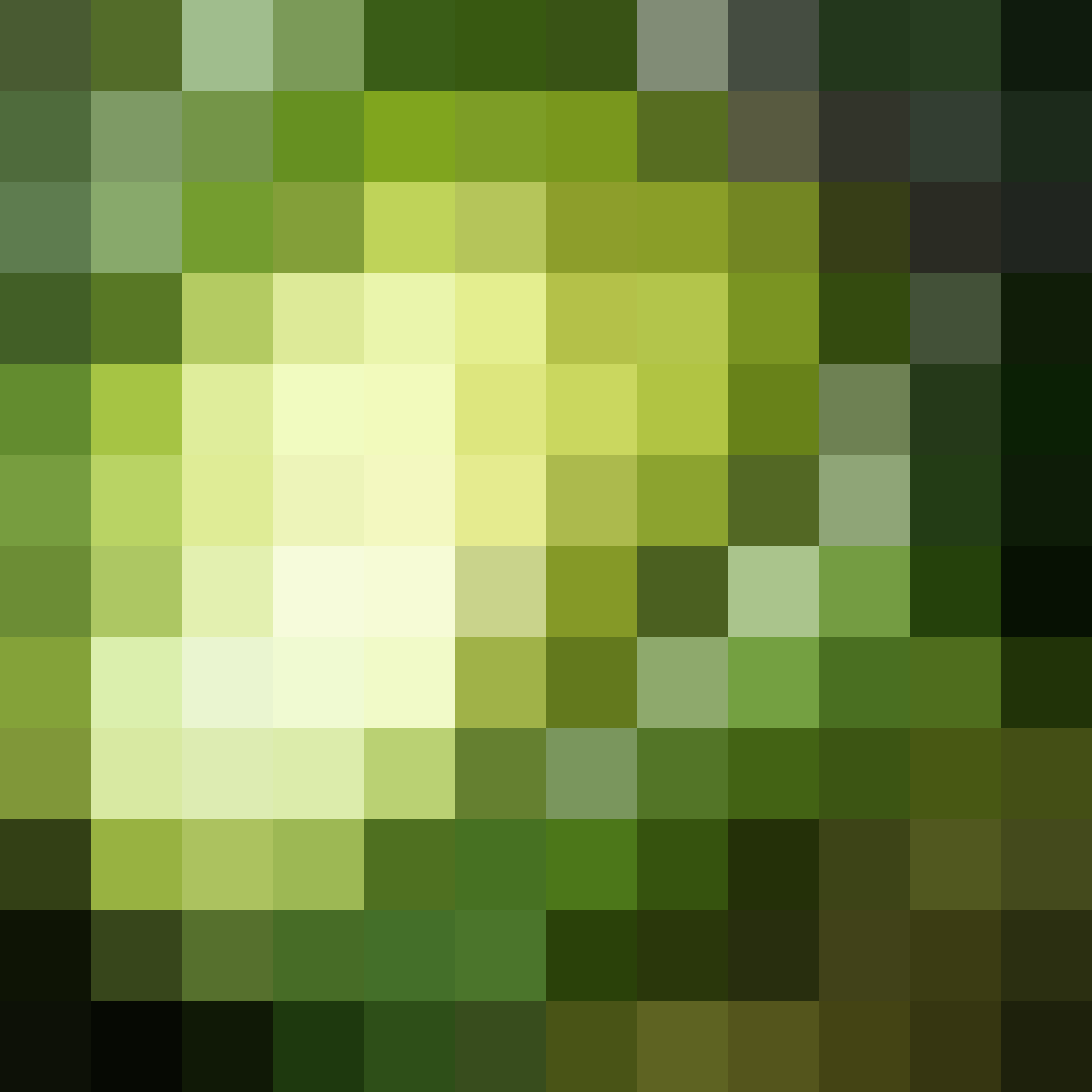}}&
\frame{\includegraphics[height=0.12\textwidth]{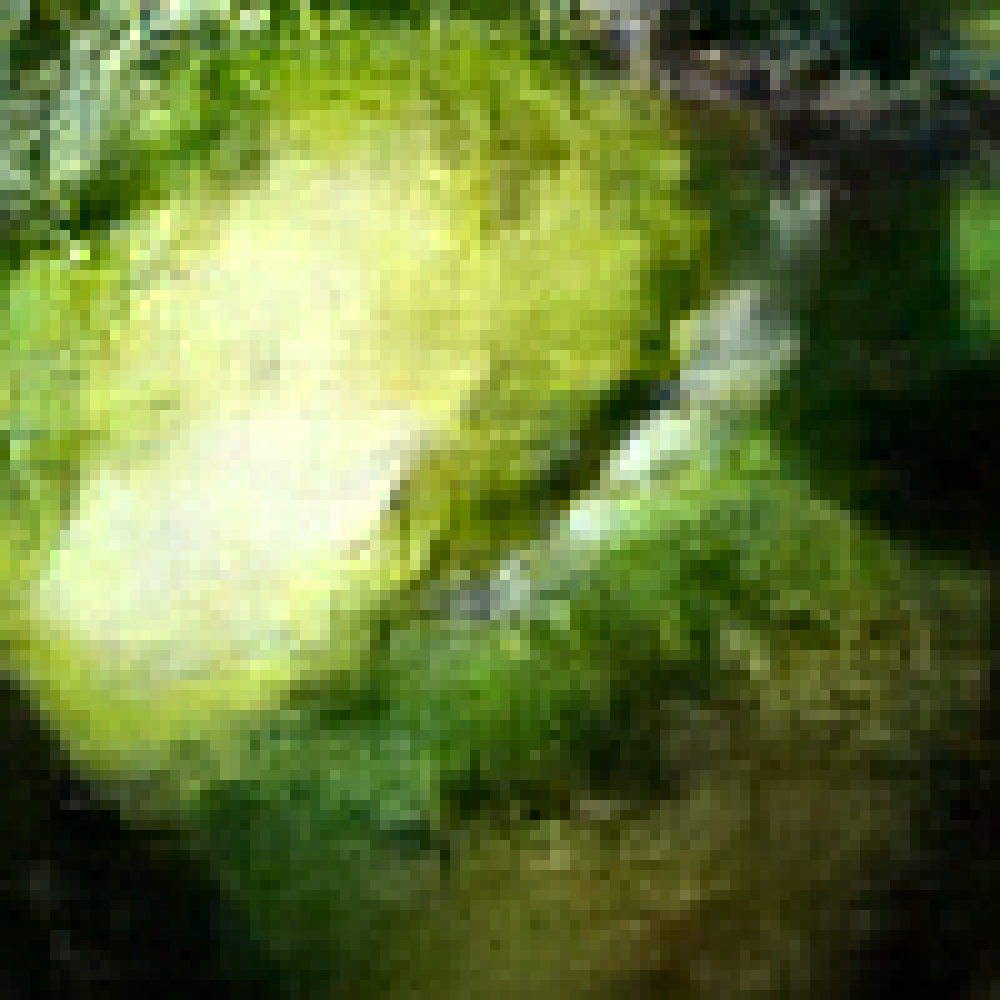}}&
\frame{\includegraphics[height=0.12\textwidth]{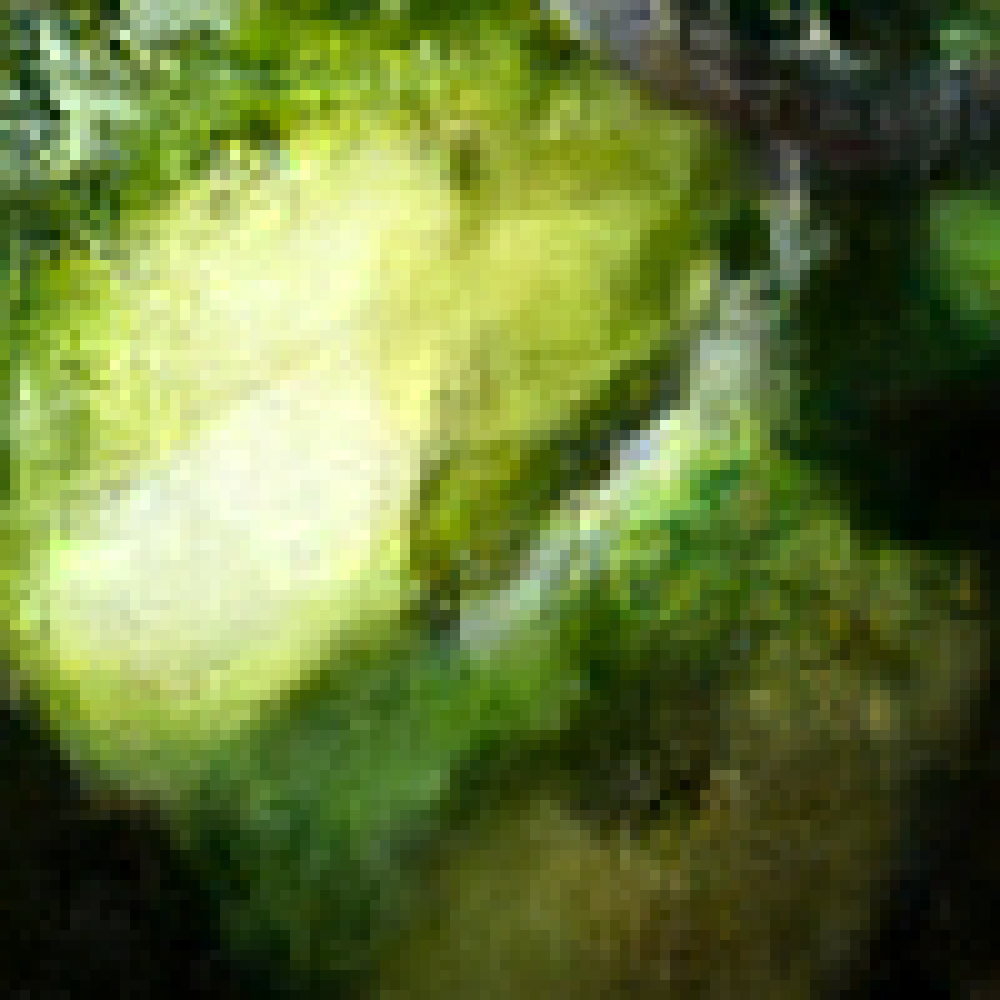}}&
\frame{\includegraphics[height=0.12\textwidth]{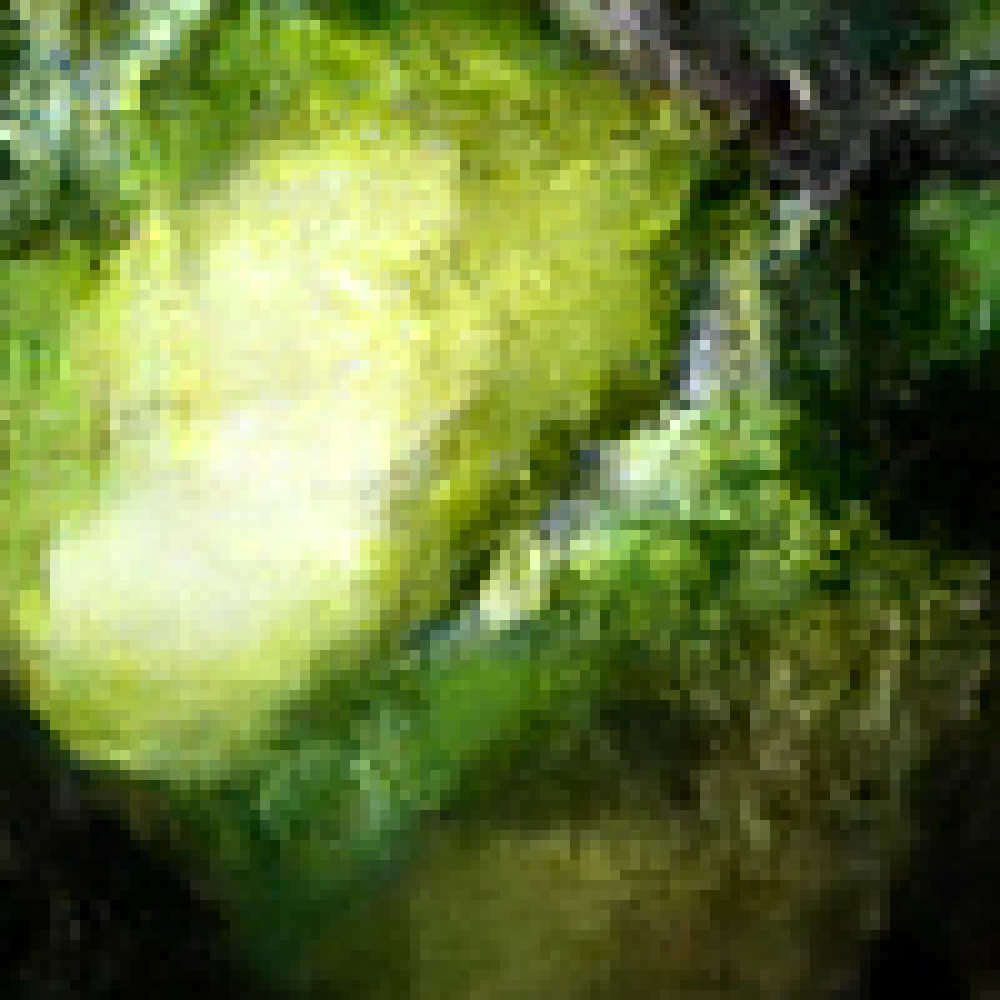}}&
\frame{\includegraphics[height=0.12\textwidth]{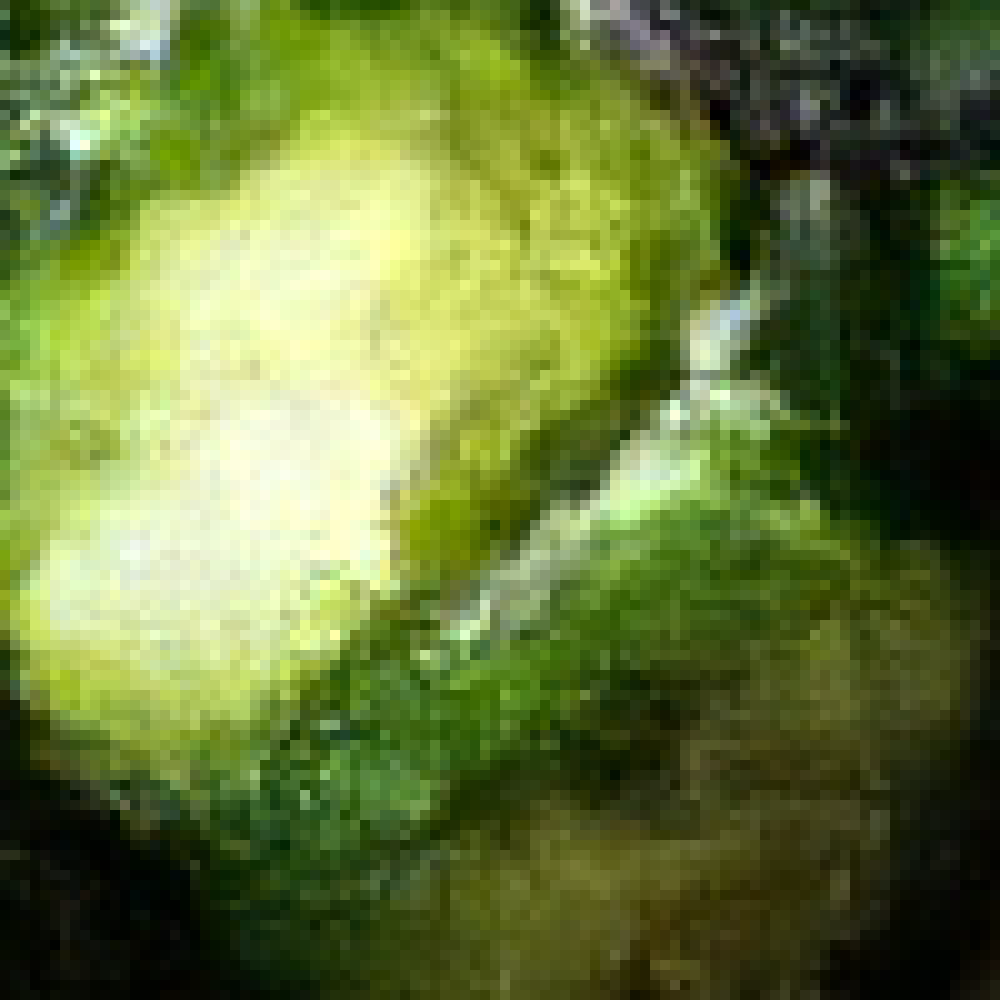}}&
\frame{\includegraphics[height=0.12\textwidth]{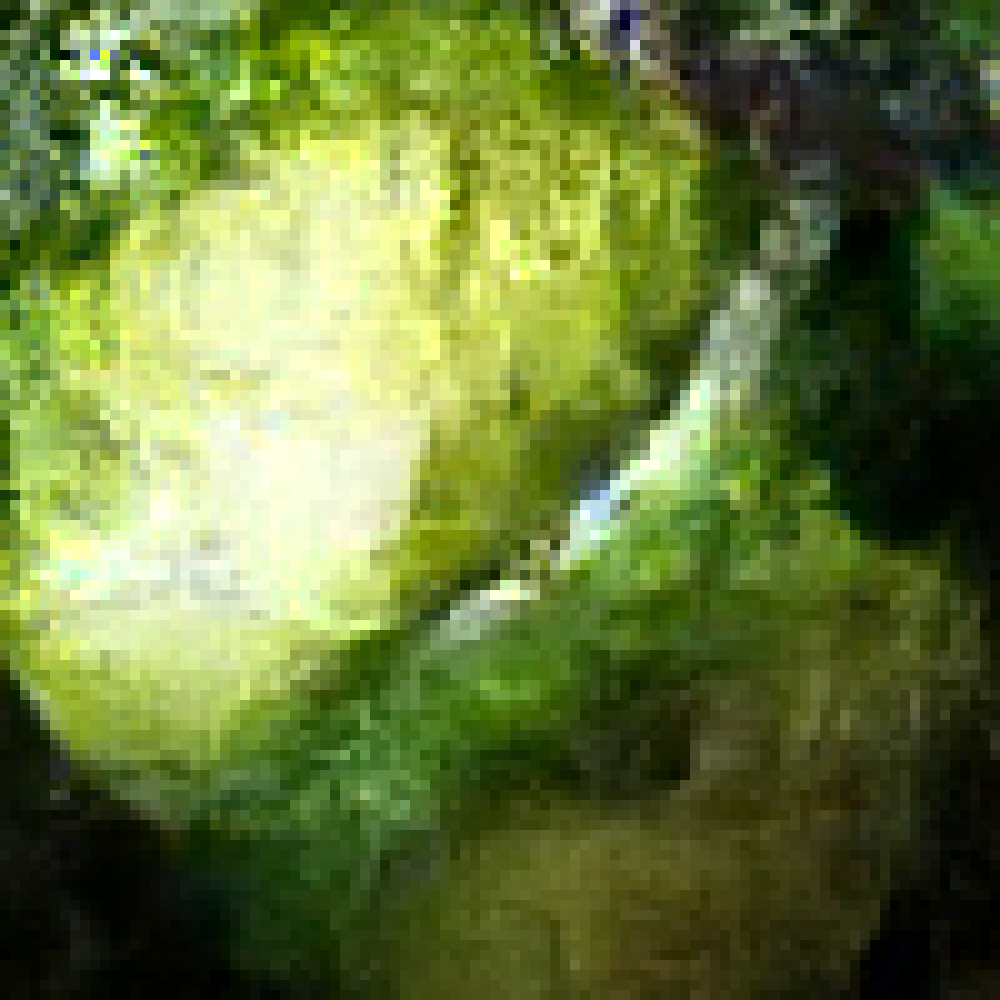}}\\
\frame{\includegraphics[height=0.12\textwidth]{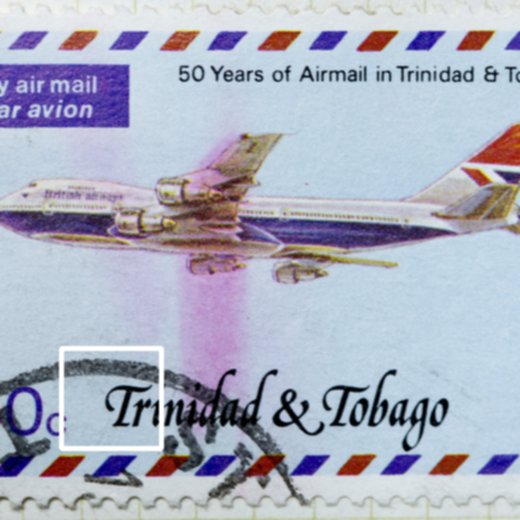}}&
\frame{\includegraphics[height=0.12\textwidth]{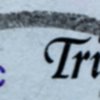}}&
\frame{\includegraphics[height=0.12\textwidth]{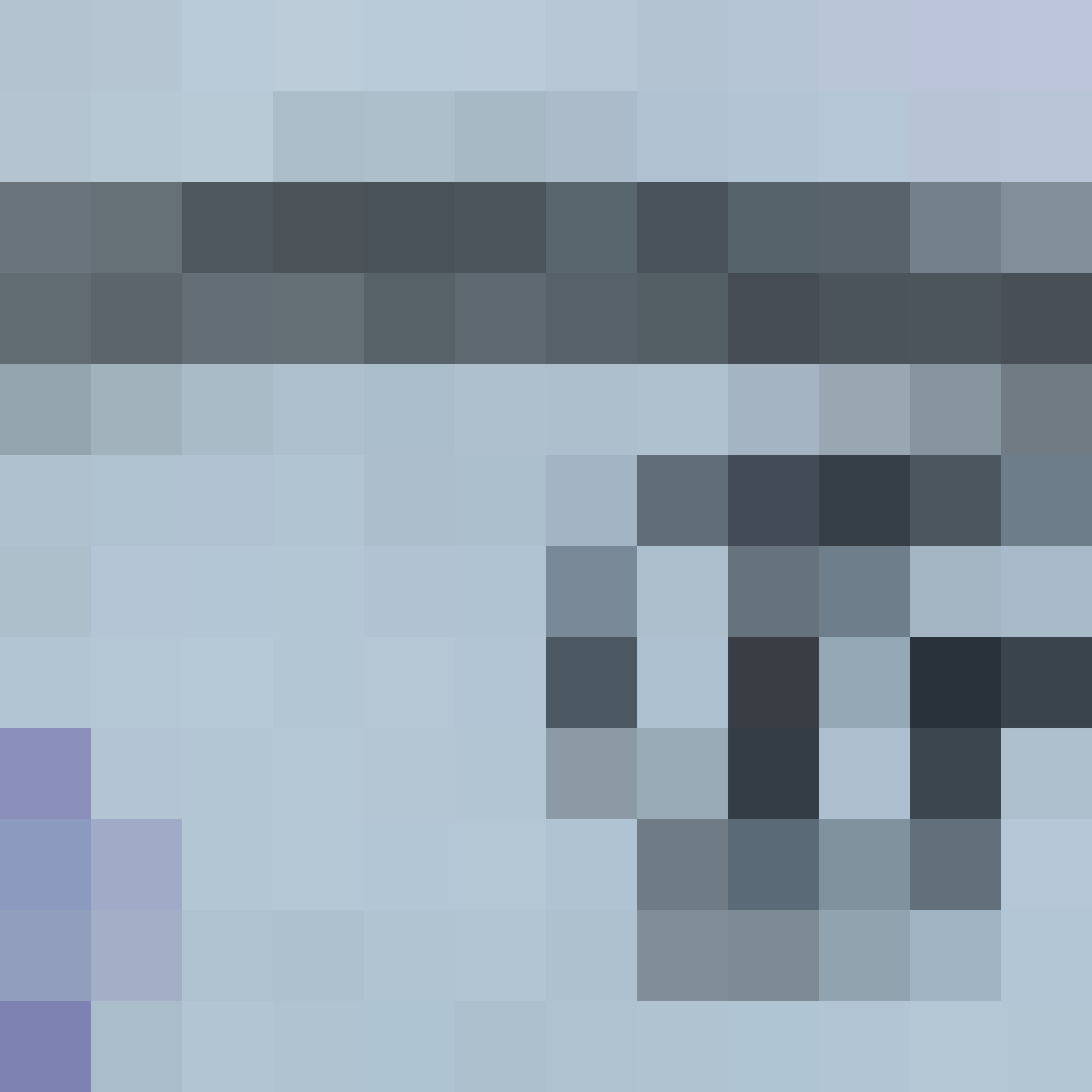}}&
\frame{\includegraphics[height=0.12\textwidth]{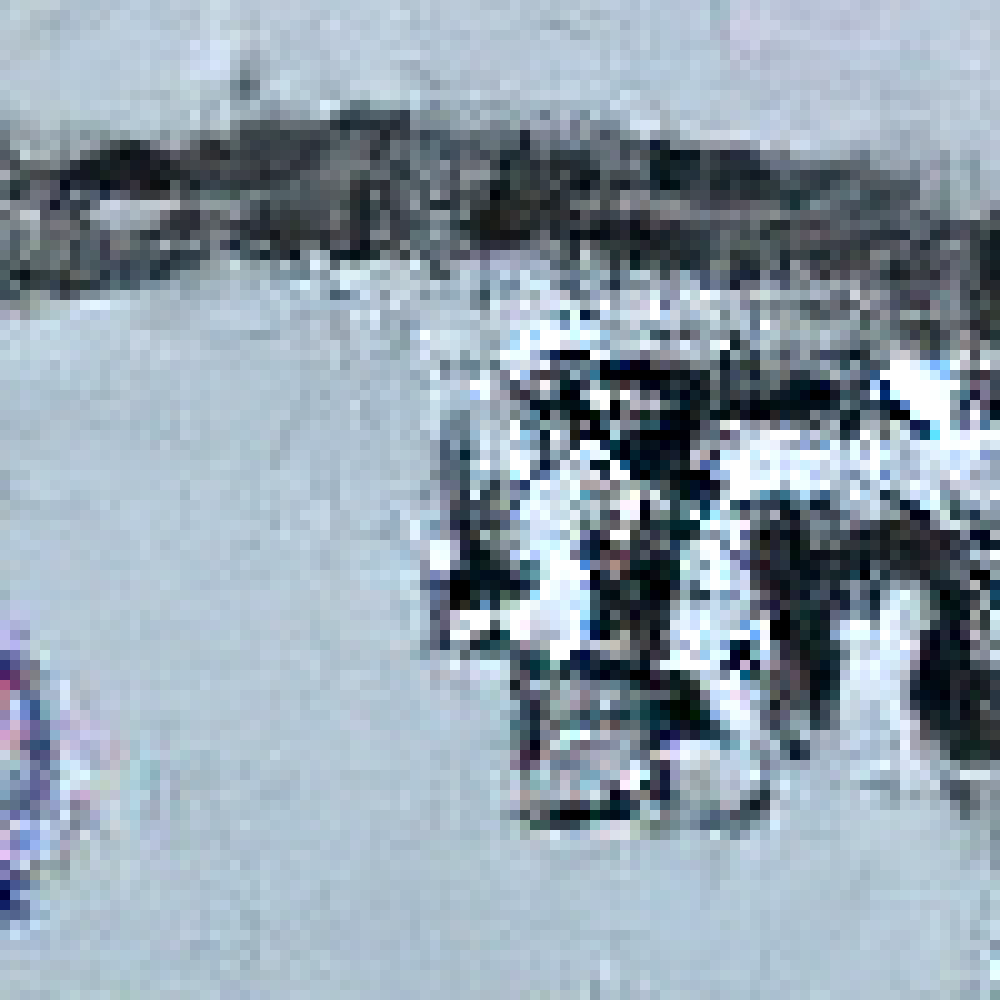}}&
\frame{\includegraphics[height=0.12\textwidth]{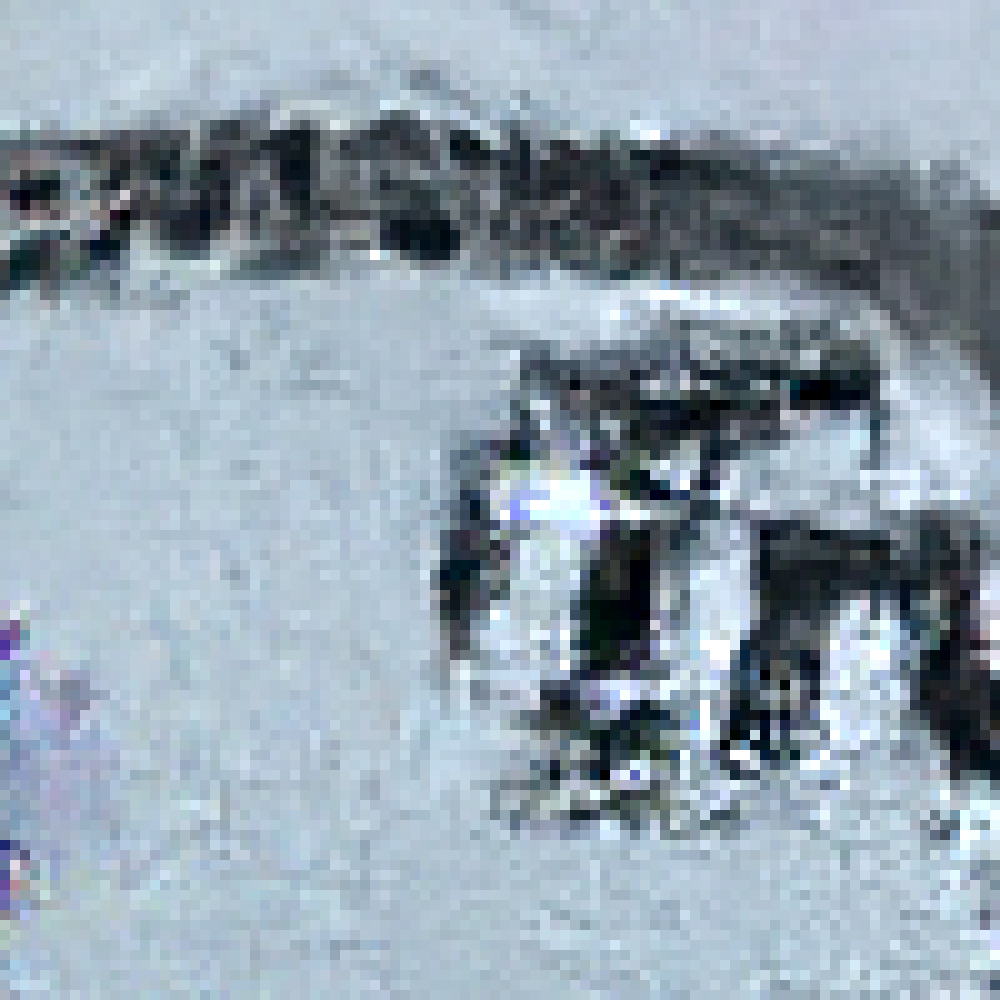}}&
\frame{\includegraphics[height=0.12\textwidth]{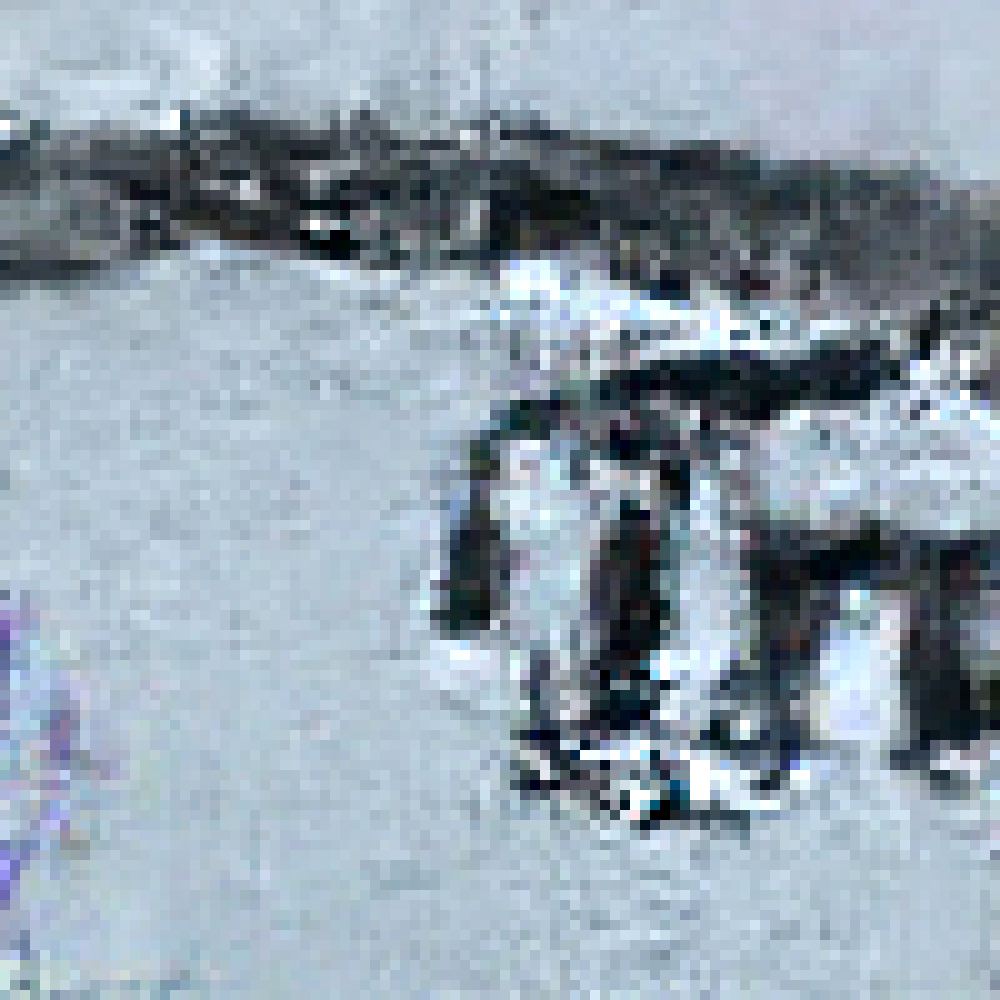}}&
\frame{\includegraphics[height=0.12\textwidth]{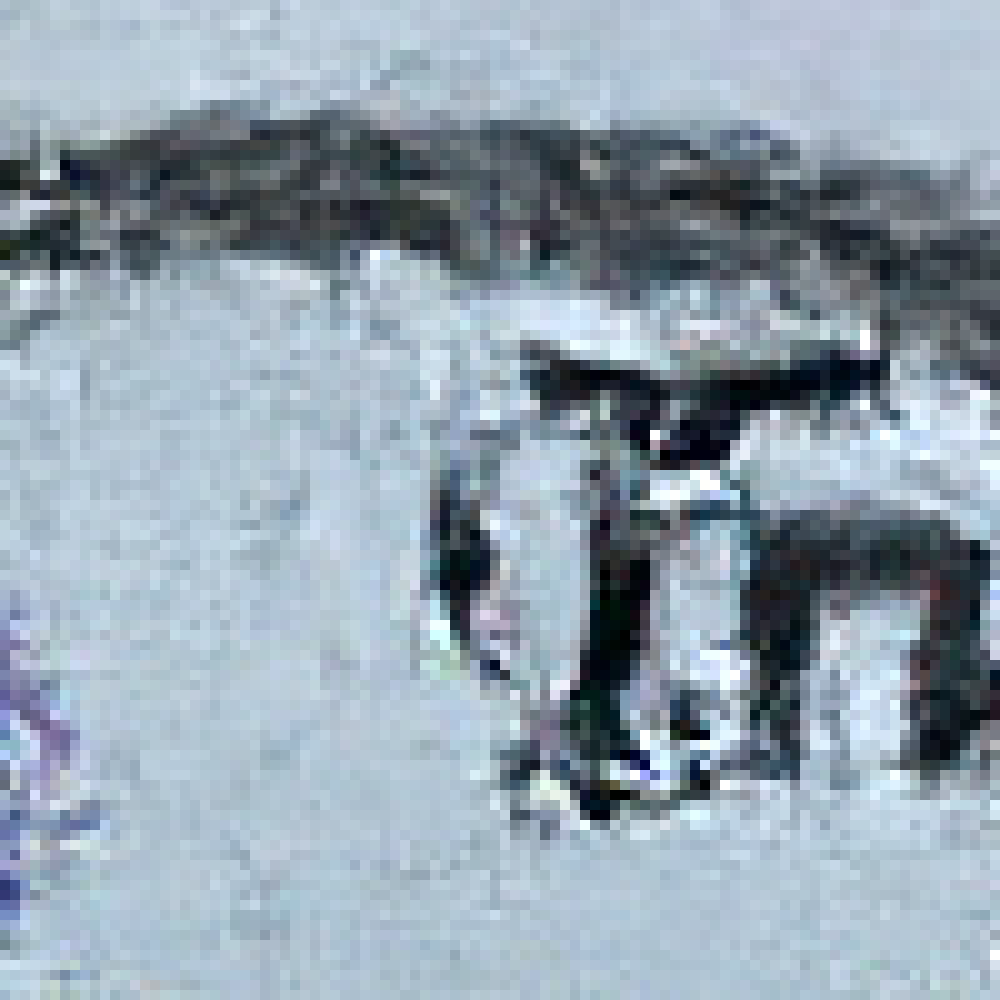}}&
\frame{\includegraphics[height=0.12\textwidth]{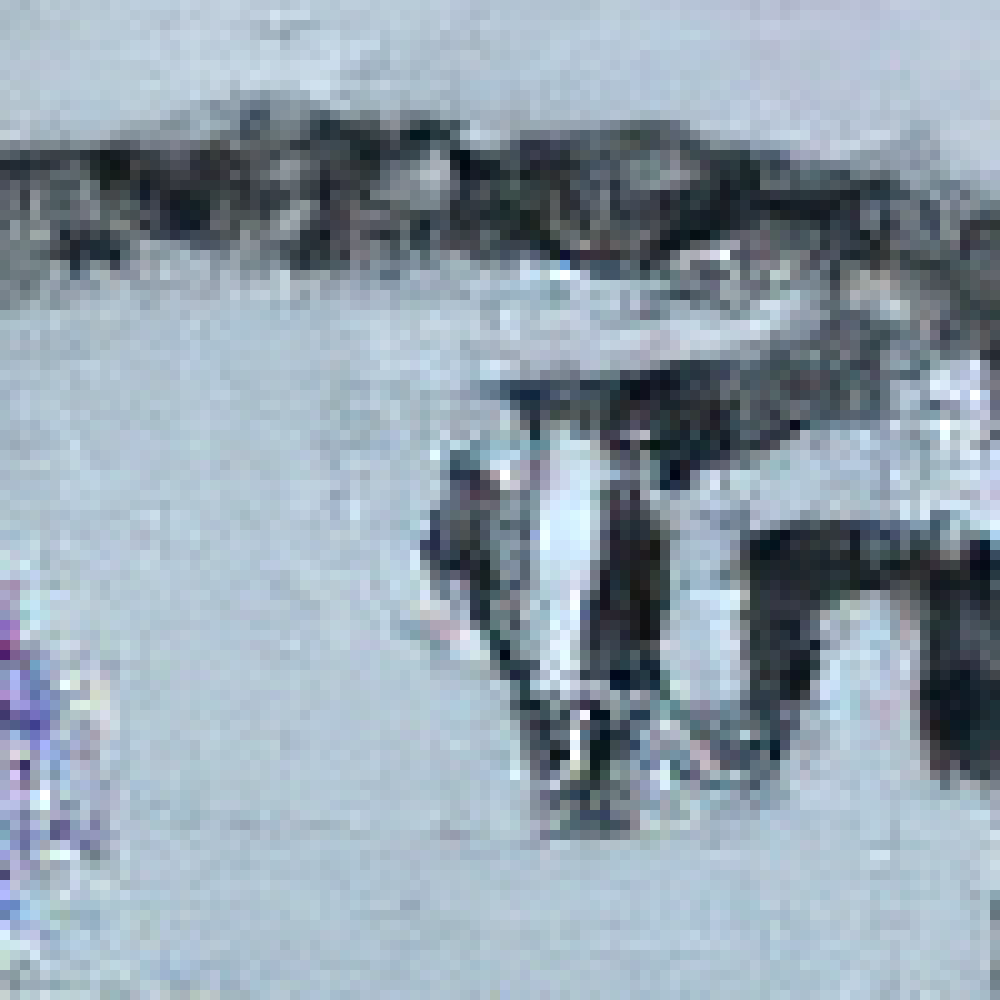}}\\
\frame{\includegraphics[height=0.12\textwidth]{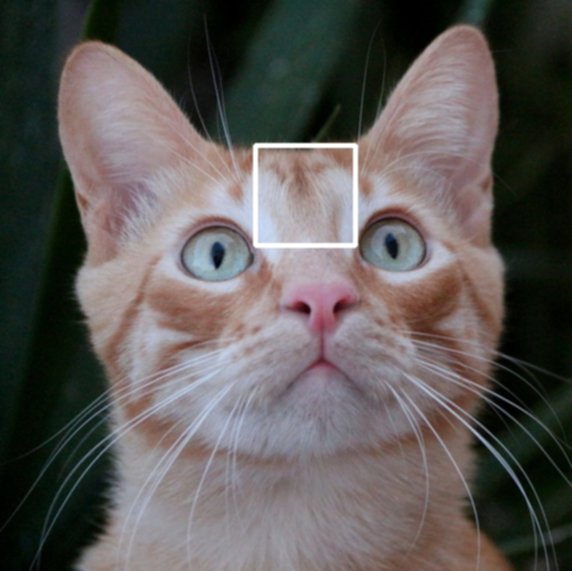}}&
\frame{\includegraphics[height=0.12\textwidth]{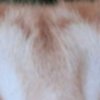}}&
\frame{\includegraphics[height=0.12\textwidth]{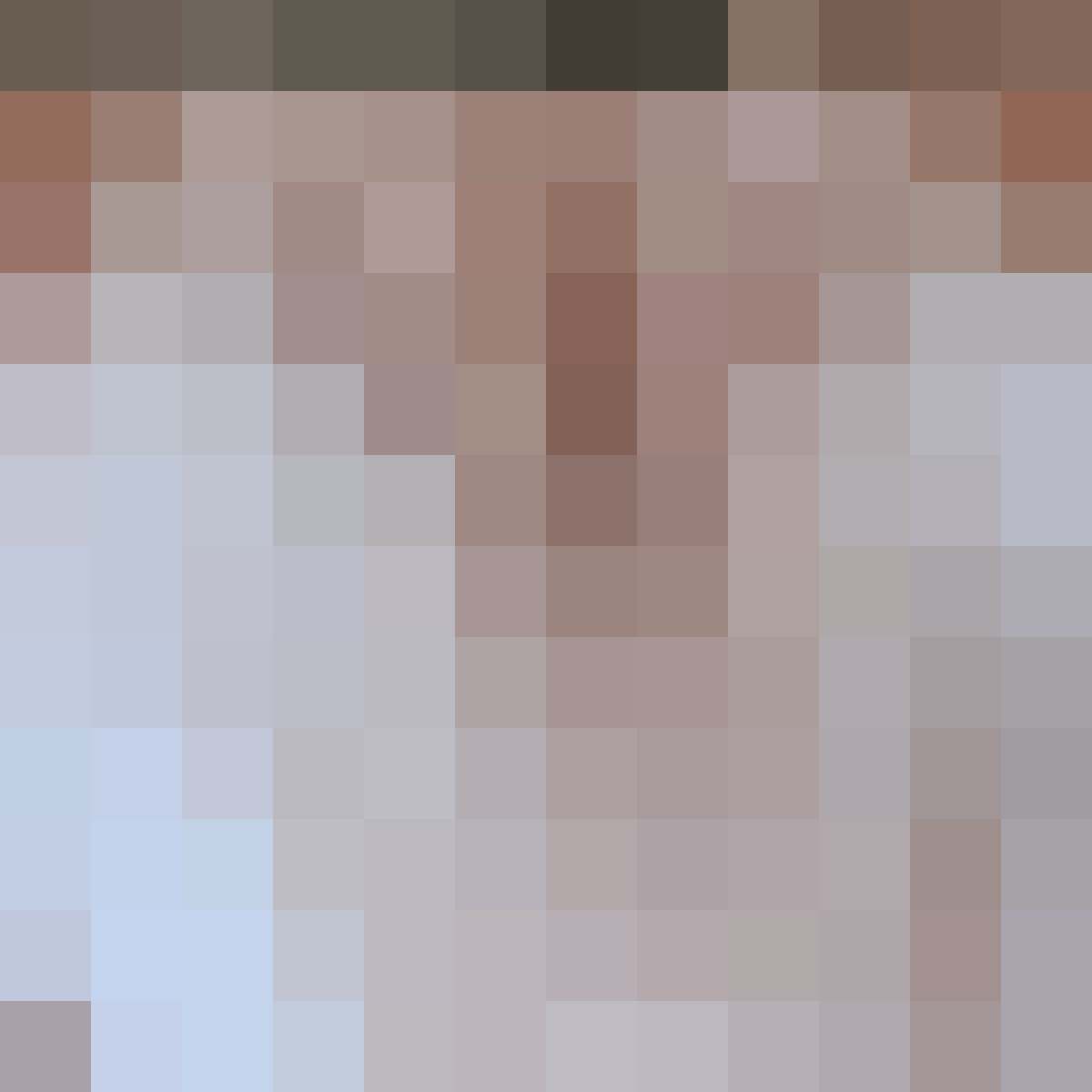}}&
\frame{\includegraphics[height=0.12\textwidth]{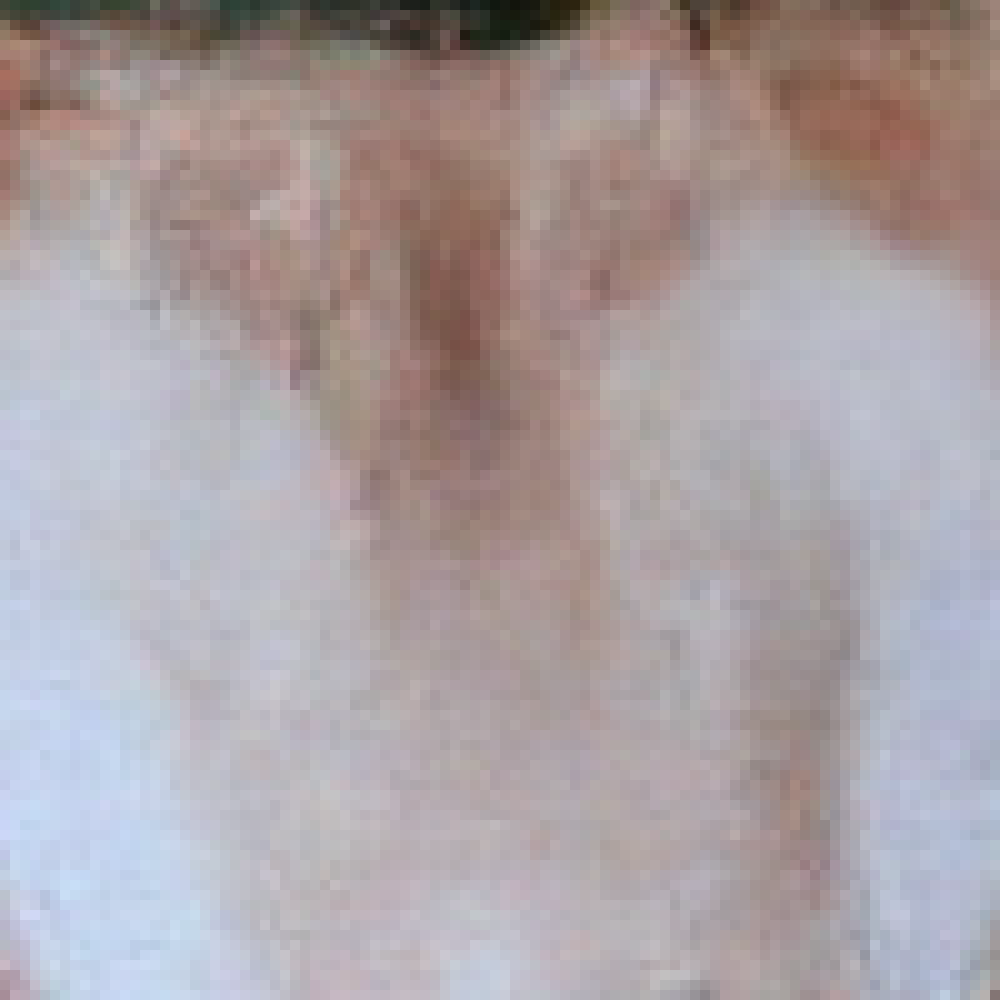}}&
\frame{\includegraphics[height=0.12\textwidth]{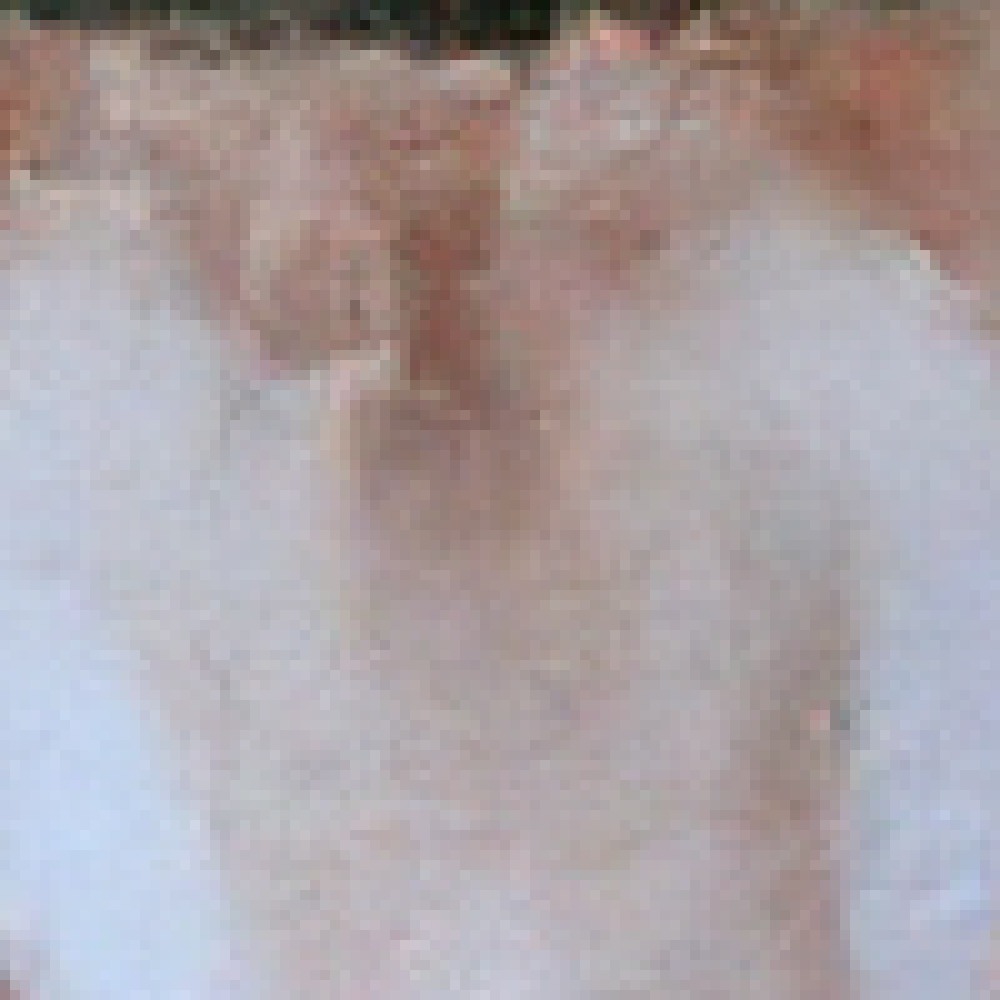}}&
\frame{\includegraphics[height=0.12\textwidth]{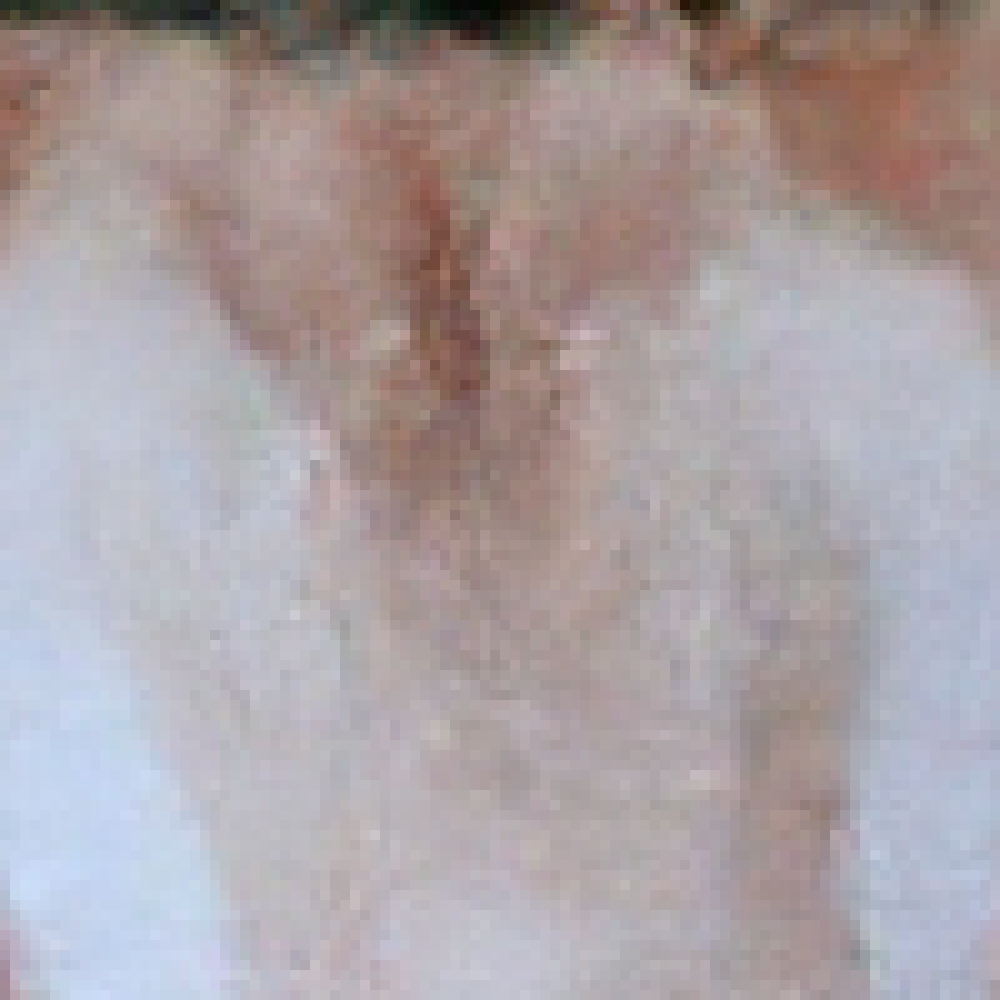}}&
\frame{\includegraphics[height=0.12\textwidth]{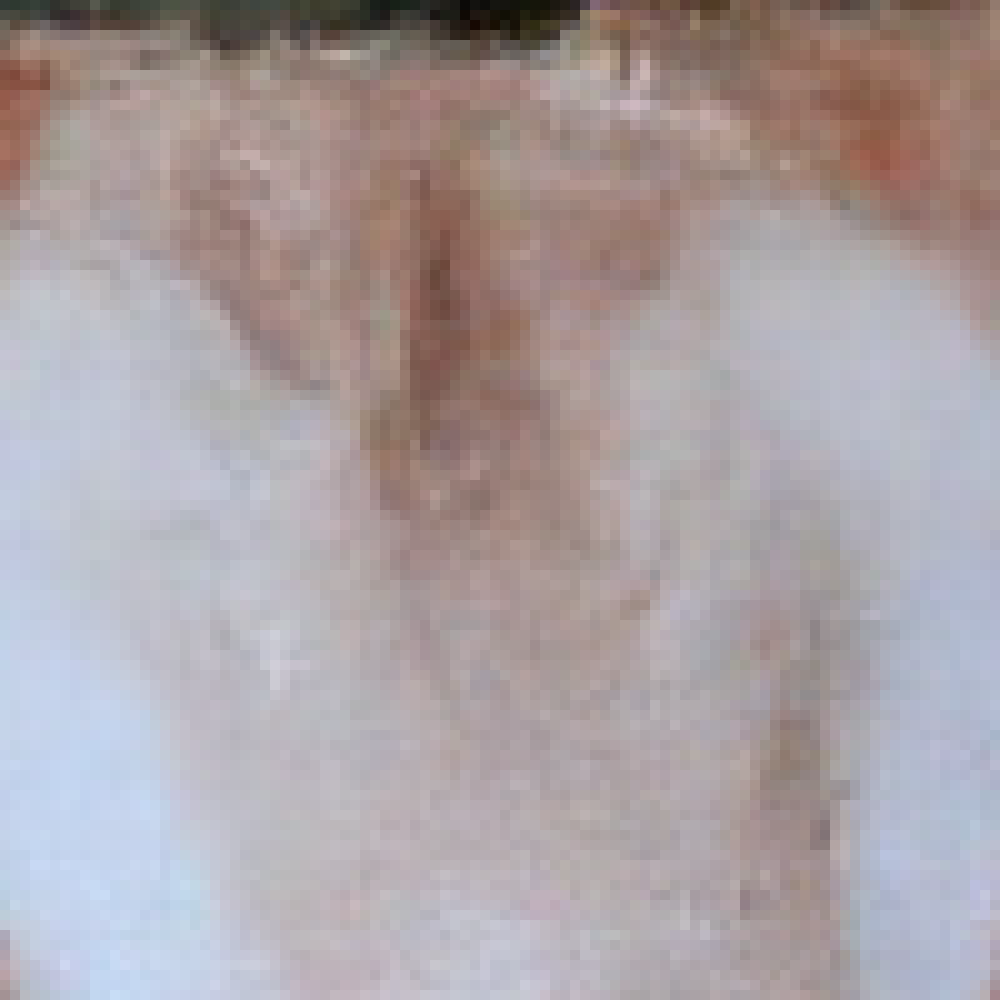}}&
\frame{\includegraphics[height=0.12\textwidth]{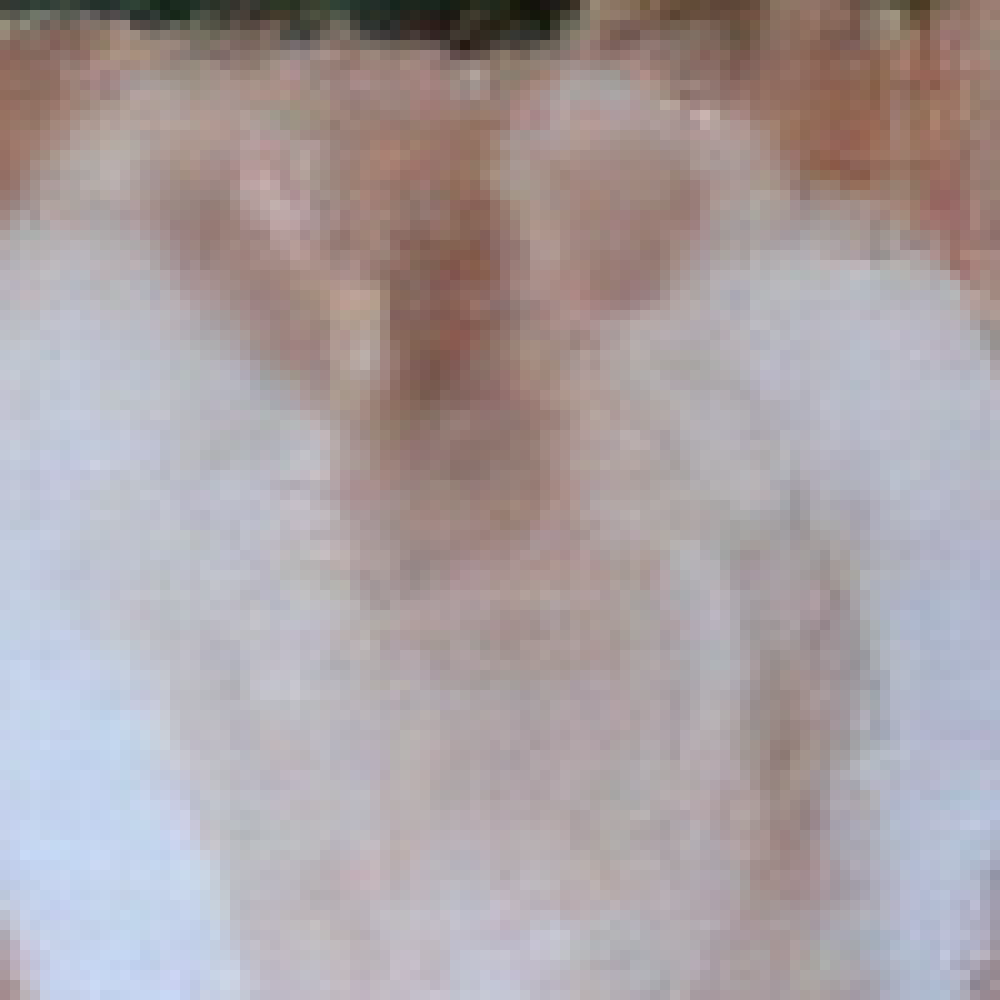}}\vspace{-2mm}
  \end{tabular}
  \caption{\textbf{Super-resolution distribution visualization.}
  We sample images from our network to visualize what it learns.
  From left to right:
  original image $x^{(0)}$, $x^{(0)}$ (zoomed in), downsampled image $x^{(3)}$, and samples.
  We sample images from full $x^{(3)}$, but just present the zoomed-in 100$^2$-pixel view for clear presentation.
   (Best viewed on screen.)
   \vspace{-1mm}}
  \label{fig:sampled}
\end{figure}

\subsection{Qualitative Analysis}
\lblsec{qualitative}
\paragraph{Super-Resolution Distribution Visualization.}
Our network learns a distribution over possible super-resolutions.
In \figref{sampled}, we present images sampled from the distribution (conditioned on $x^{(3)}$) to visualize what the network learns.
We see that our model learns a wide range of possible super-resolutions.\footnote{
Artifacts are present in some cases, in part due to that the network is not optimized for image generation.
The good bpsp suggests that the learned distribution does model the image distribution well.
}
We conjecture that the diversity is crucial for generalization to unseen images,
contributing to the improved compression rate.

\begin{figure}[t]
\setlength{\tabcolsep}{0.1em}
\center
\resizebox{\columnwidth}{!}{\begin{tabular}{@{}cccccccc@{}}
1.24\tiny/1.55/0.85 & 1.34\tiny/1.78/1.16 & 1.38\tiny/1.80/1.24 & 1.39\tiny/1.93/1.24 & 1.45\tiny/1.93/1.42 & 1.46\tiny/1.96/1.25 & 1.48\tiny/2.16/1.33 & 1.51\tiny/2.10/1.51\\
\frame{\includegraphics[height=0.12\textwidth]{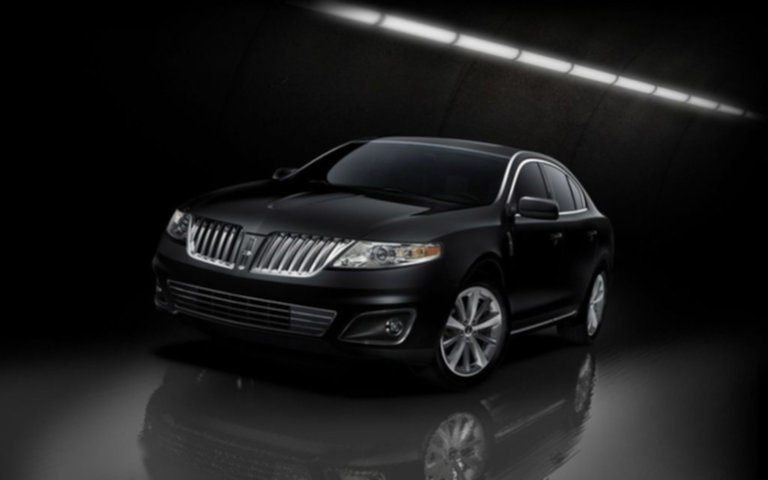}}&
\frame{\includegraphics[height=0.12\textwidth]{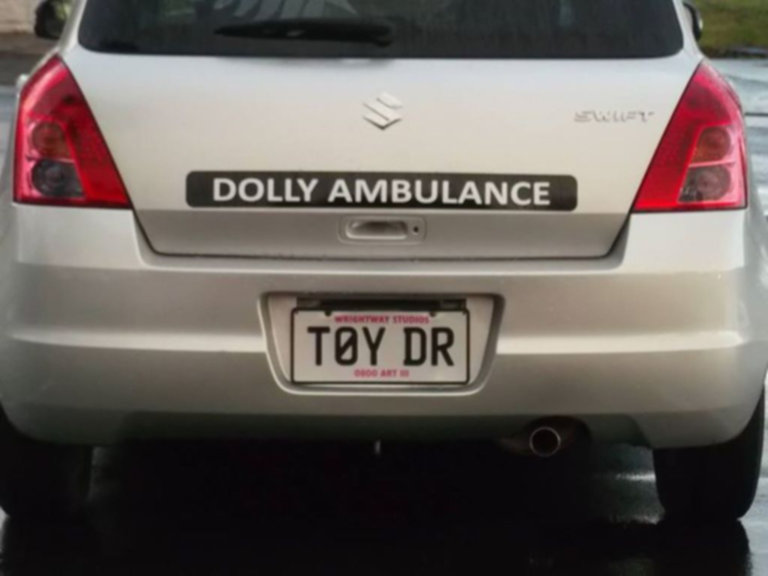}}&
\frame{\includegraphics[height=0.12\textwidth]{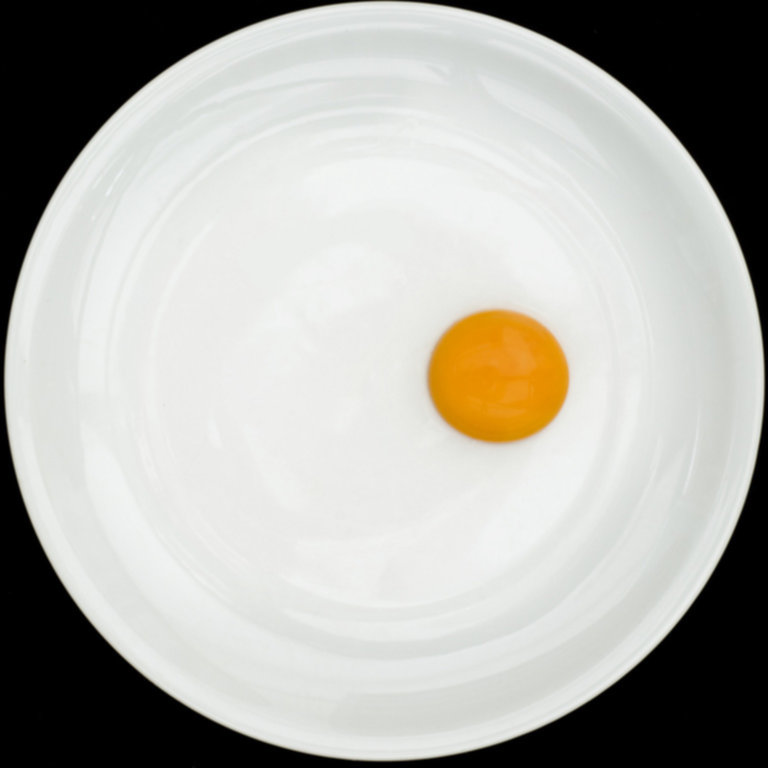}}&
\frame{\includegraphics[height=0.12\textwidth]{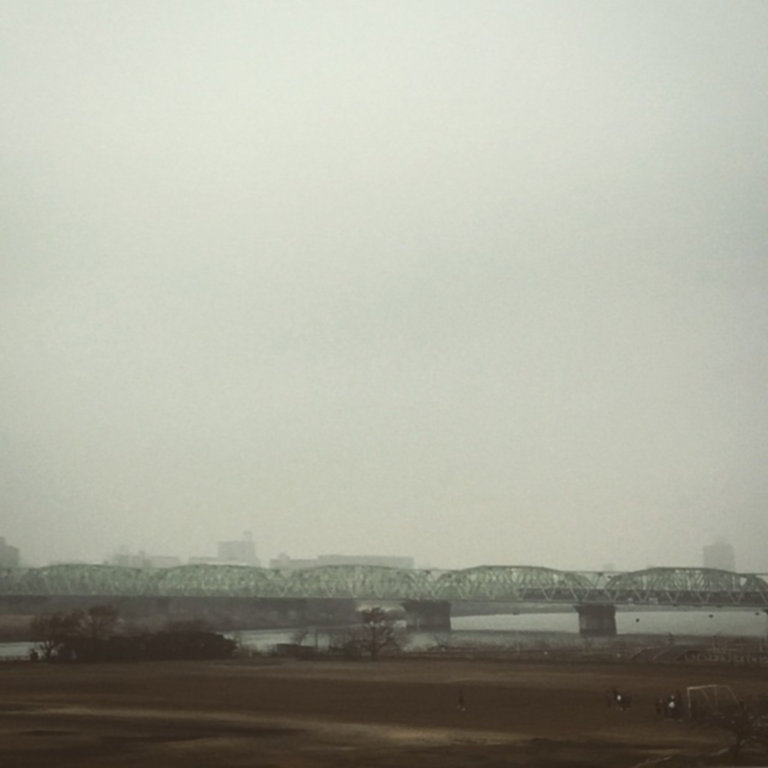}}&
\frame{\includegraphics[height=0.12\textwidth]{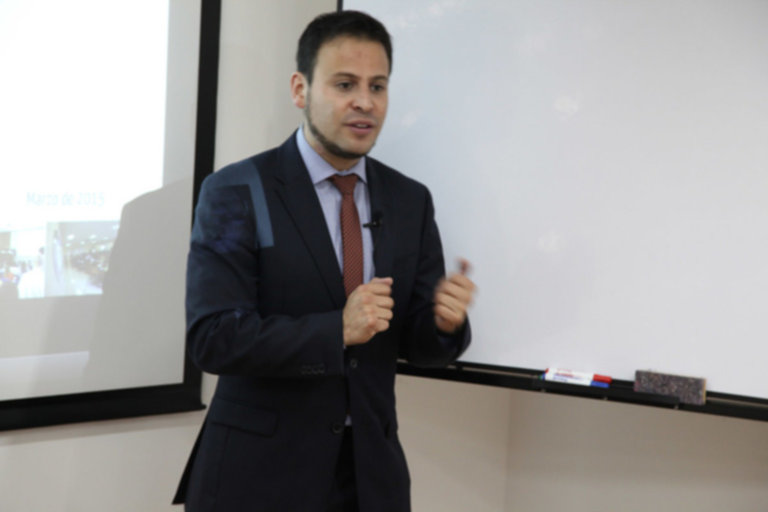}}&
\frame{\includegraphics[height=0.12\textwidth]{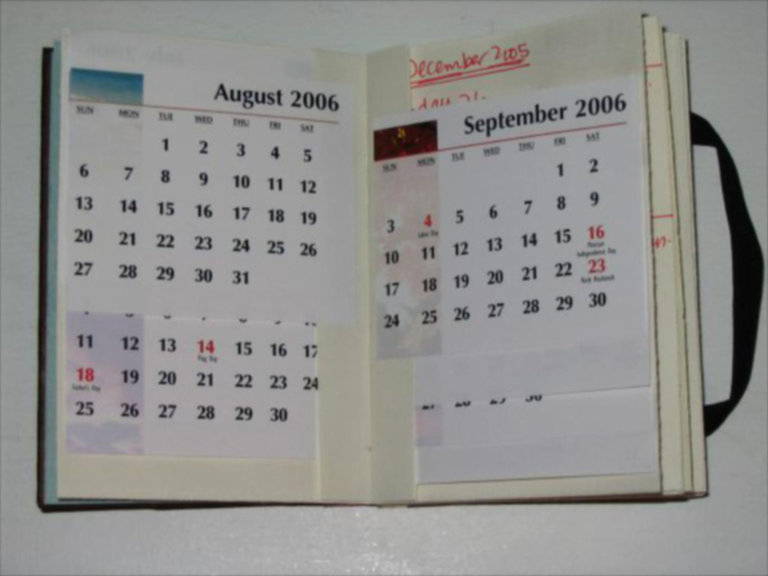}}&
\frame{\includegraphics[height=0.12\textwidth]{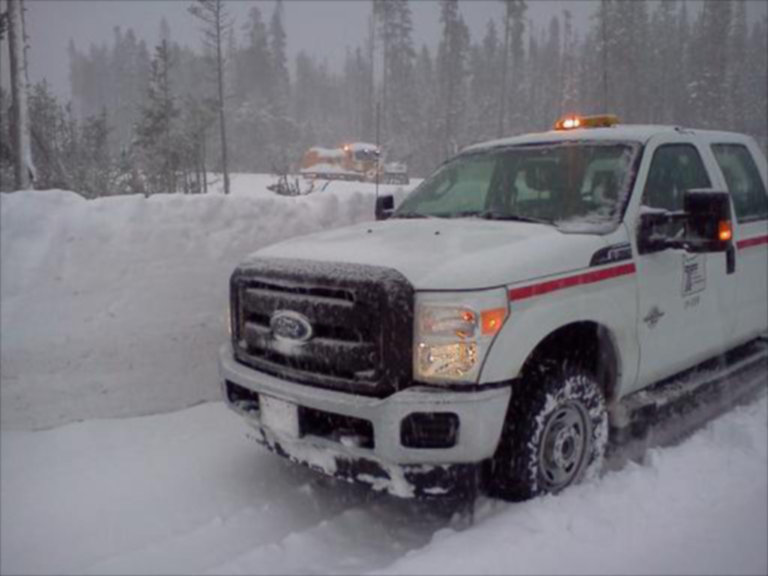}}&
\frame{\includegraphics[height=0.12\textwidth]{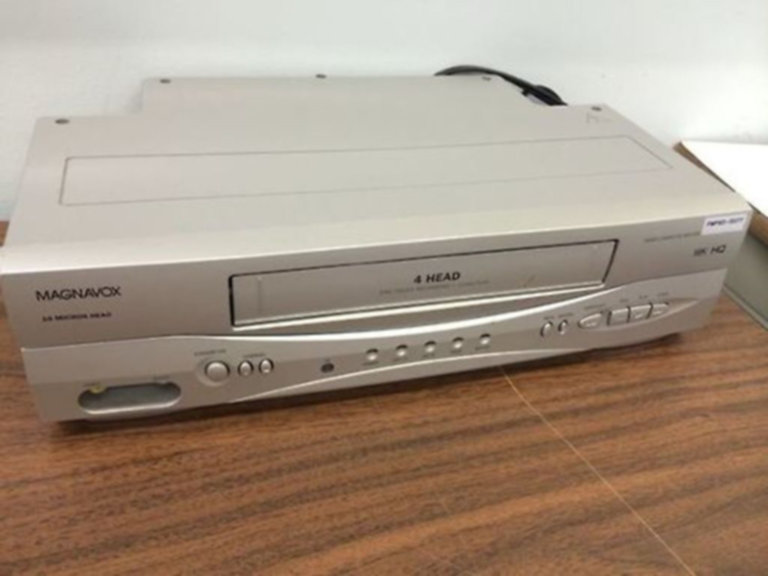}}
  \end{tabular}}
\resizebox{\columnwidth}{!}{\begin{tabular}{@{}cccccccc@{}}
1.52\tiny/2.12/1.44 & 1.53\tiny/2.52/1.59 & 1.55\tiny/2.33/1.63 & 1.56\tiny/2.25/1.54 & 1.59\tiny/2.32/1.66 & 1.62\tiny/2.44/1.67 & 1.63\tiny/2.33/1.62 & 1.63\tiny/2.46/1.82\\
\frame{\includegraphics[height=0.12\textwidth]{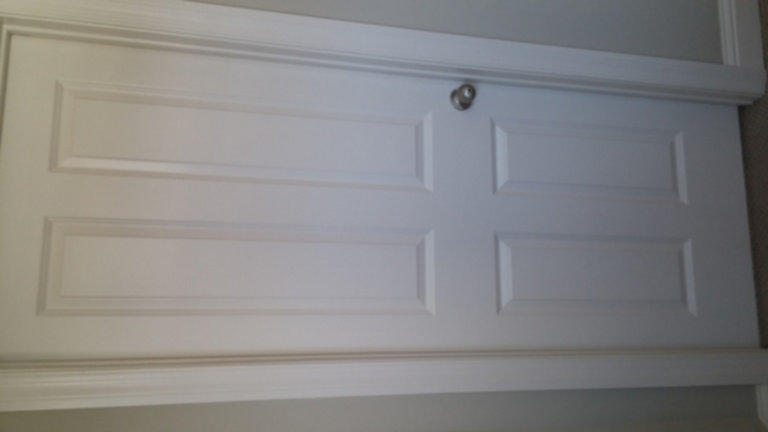}}&
\frame{\includegraphics[height=0.12\textwidth]{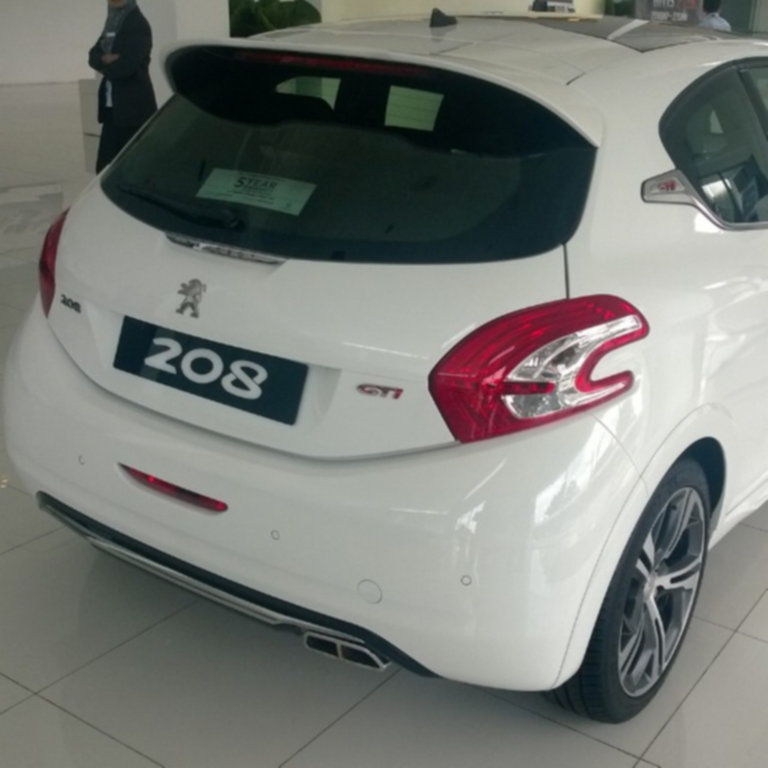}}&
\frame{\includegraphics[height=0.12\textwidth]{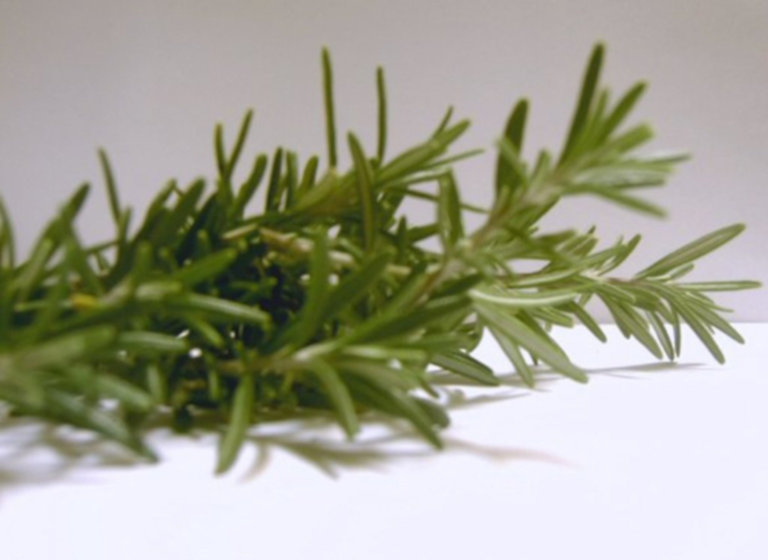}}&
\frame{\includegraphics[height=0.12\textwidth]{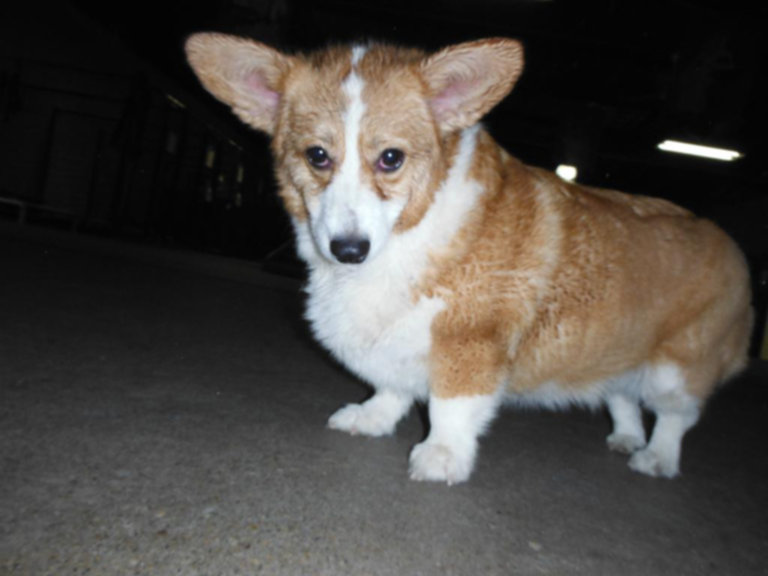}}&
\frame{\includegraphics[height=0.12\textwidth]{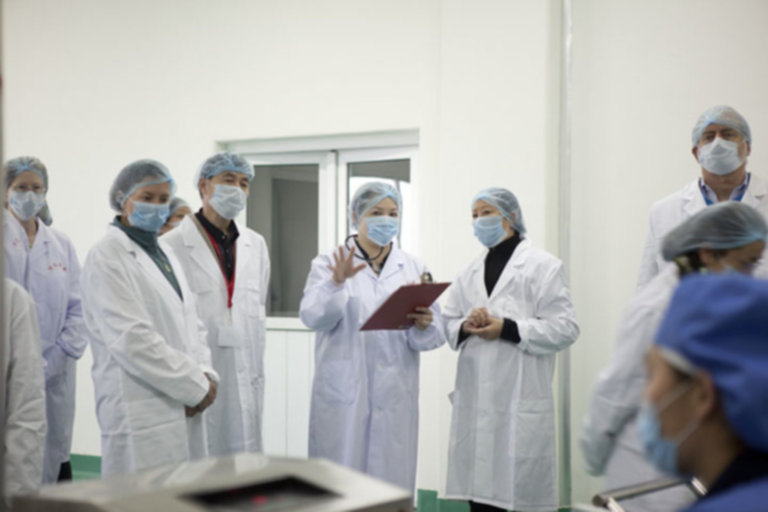}}&
\frame{\includegraphics[height=0.12\textwidth]{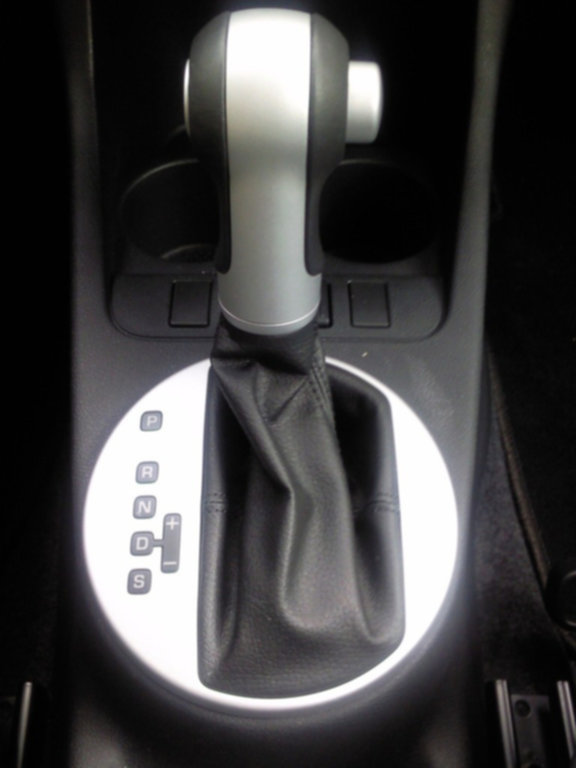}}&
\frame{\includegraphics[height=0.12\textwidth]{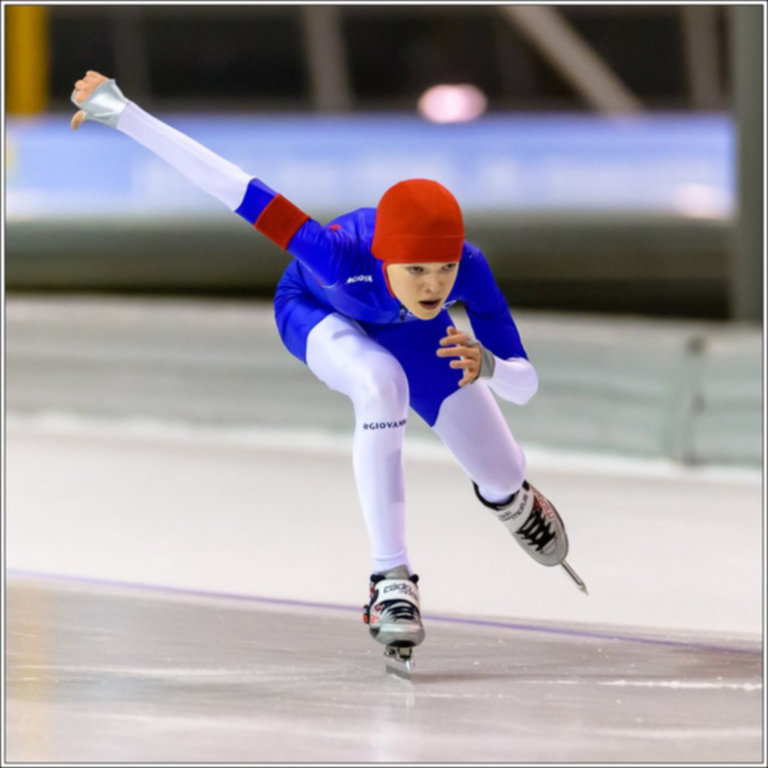}}&
\frame{\includegraphics[height=0.12\textwidth]{figs/ranking/015_1_6313152313232422.jpg}}
  \end{tabular}}
\resizebox{\columnwidth}{!}{\begin{tabular}{@{}cccccccc@{}}
1.63\tiny/2.33/1.67 & 1.69\tiny/2.74/1.99 & 1.70\tiny/2.54/1.60 & 1.71\tiny/2.48/1.90 & 1.71\tiny/2.65/1.85 & 1.73\tiny/2.57/1.99 & 1.75\tiny/2.67/1.86 & 1.75\tiny/2.59/1.83\\
\frame{\includegraphics[height=0.12\textwidth]{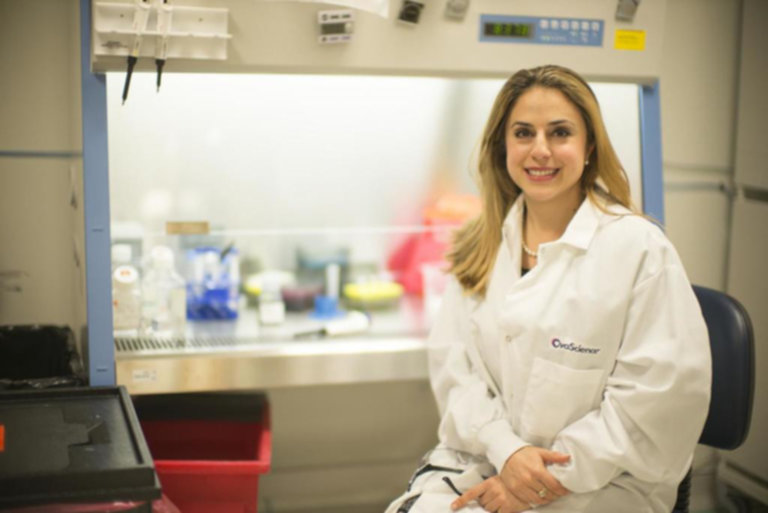}}&
\frame{\includegraphics[height=0.12\textwidth]{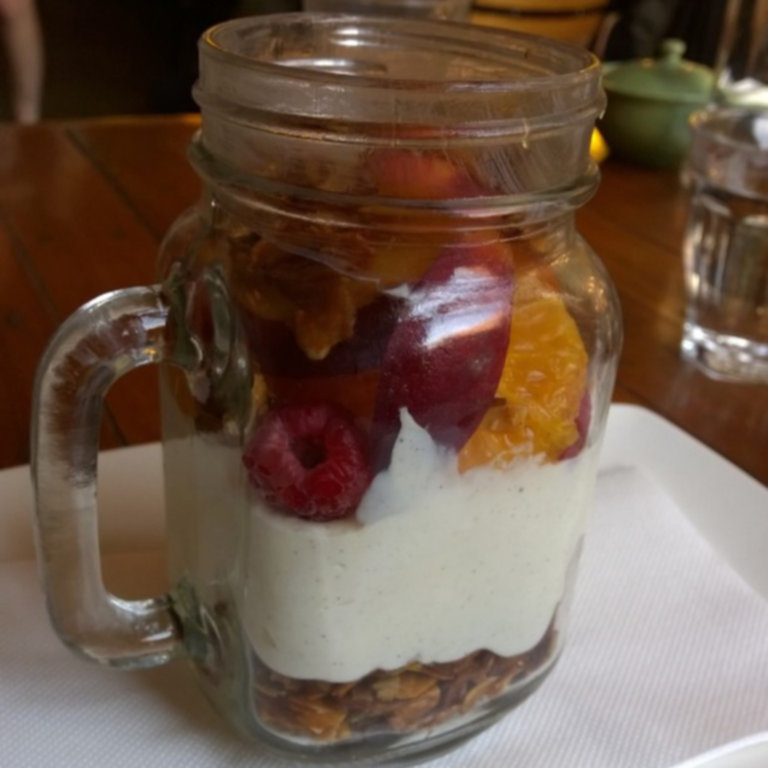}}&
\frame{\includegraphics[height=0.12\textwidth]{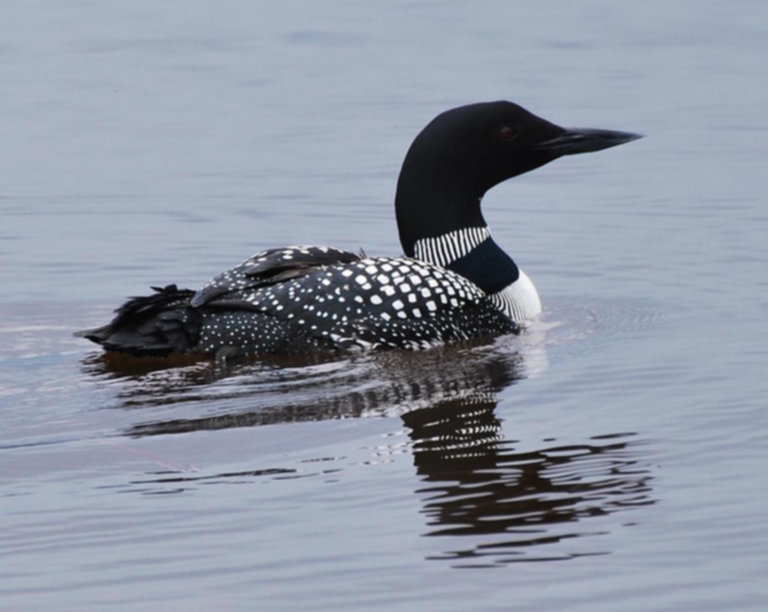}}&
\frame{\includegraphics[height=0.12\textwidth]{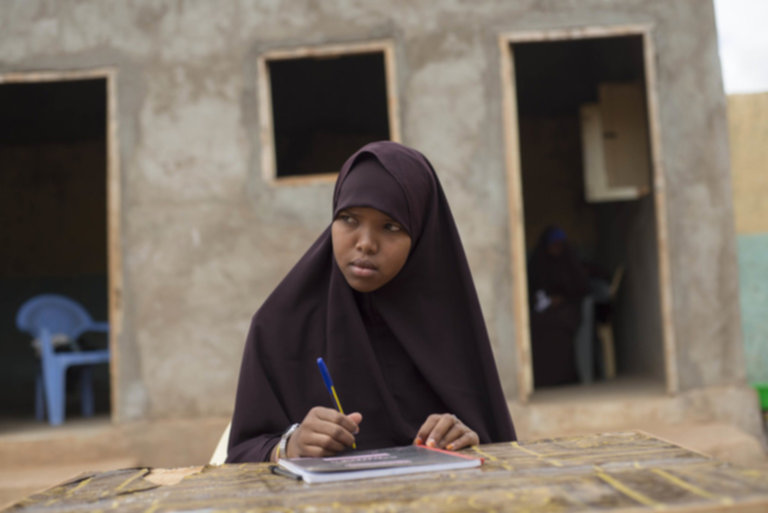}}&
\frame{\includegraphics[height=0.12\textwidth]{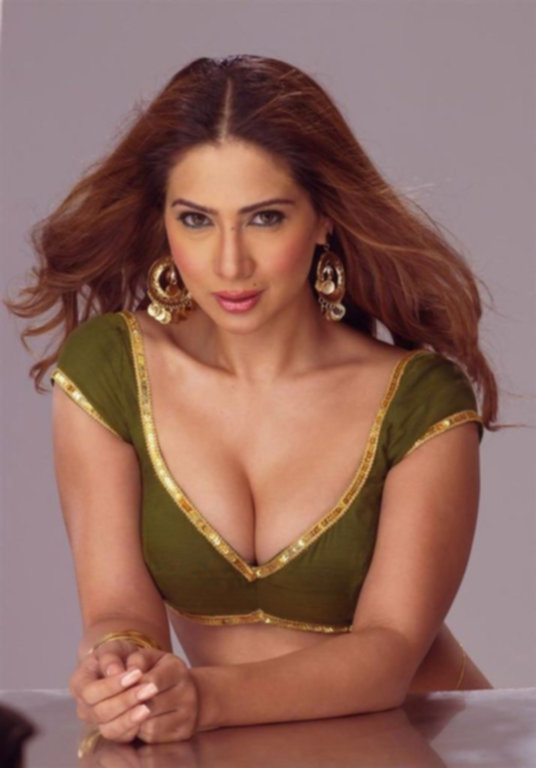}}&
\frame{\includegraphics[height=0.12\textwidth]{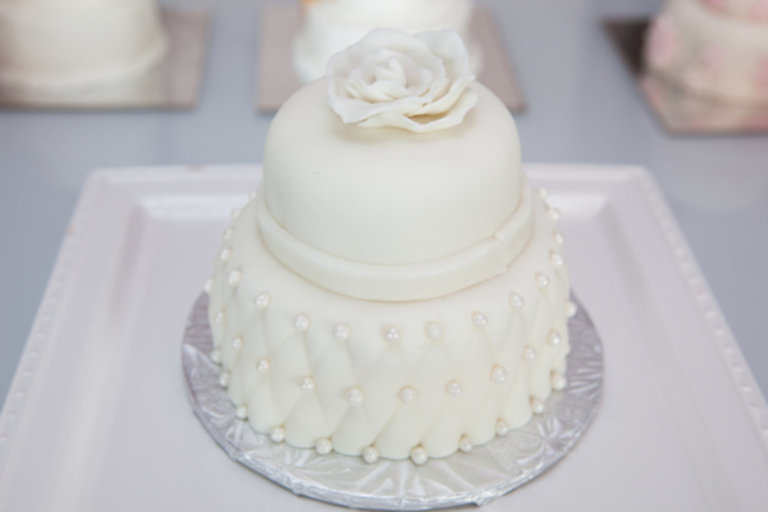}}&
\frame{\includegraphics[height=0.12\textwidth]{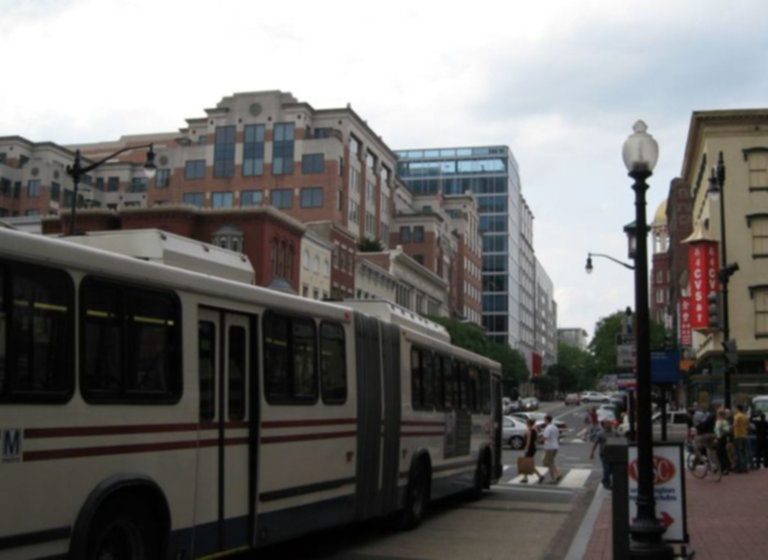}}&
\frame{\includegraphics[height=0.12\textwidth]{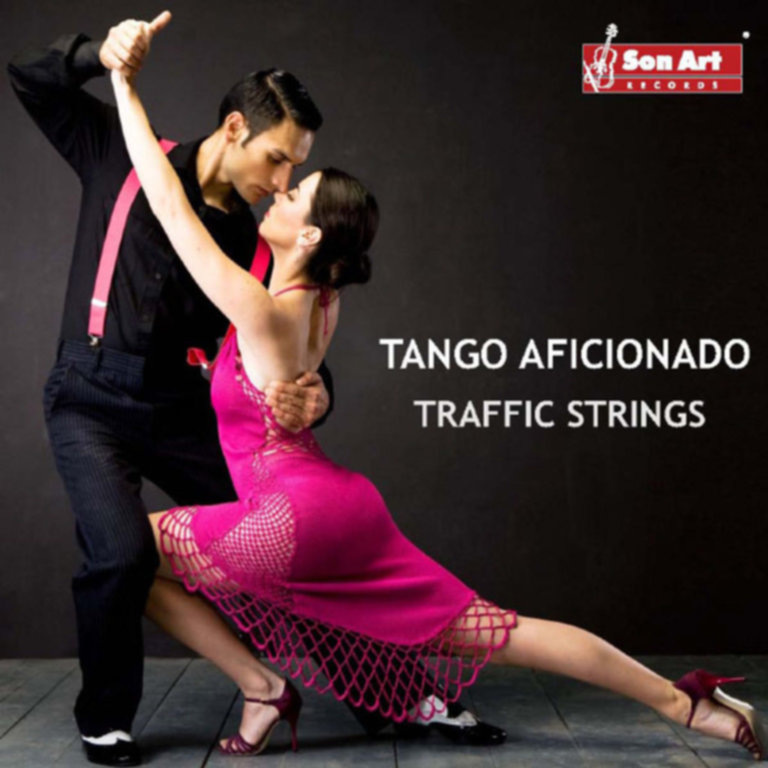}}\\
  \end{tabular}}
\resizebox{\columnwidth}{!}{\begin{tabular}{@{}ccccccc@{}}
{4.07}\tiny/5.18/4.25 & {4.08}\tiny/5.10/4.56 & {4.08}\tiny/5.92/4.61 & {4.12}\tiny/5.42/4.36 & {4.14}\tiny/5.29/4.48 & {4.16}\tiny/5.38/4.43 &
{4.17}\tiny/5.47/4.52\\
\frame{\includegraphics[height=0.12\textwidth]{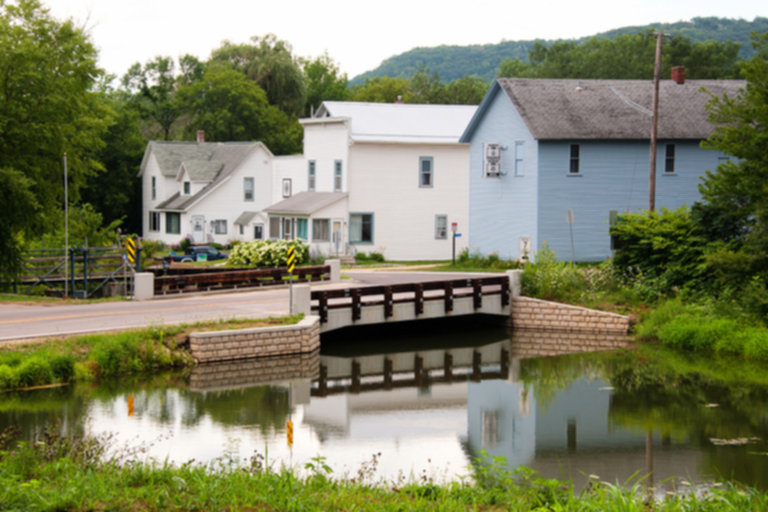}}&
\frame{\includegraphics[height=0.12\textwidth]{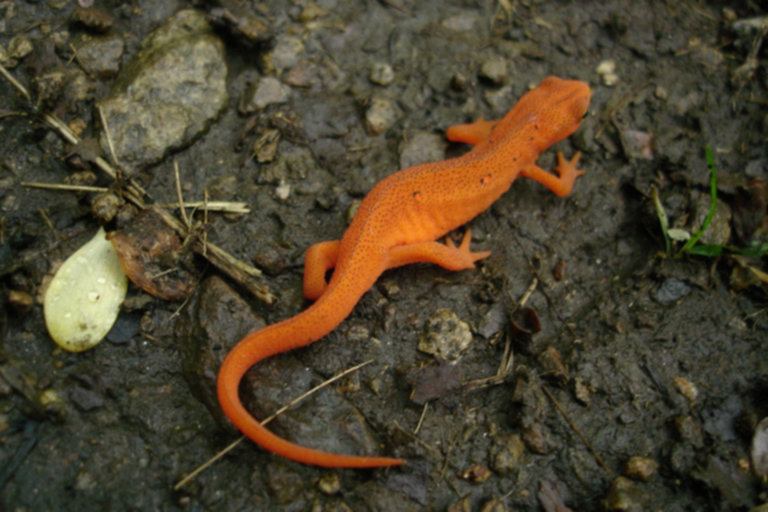}}&
\frame{\includegraphics[height=0.12\textwidth]{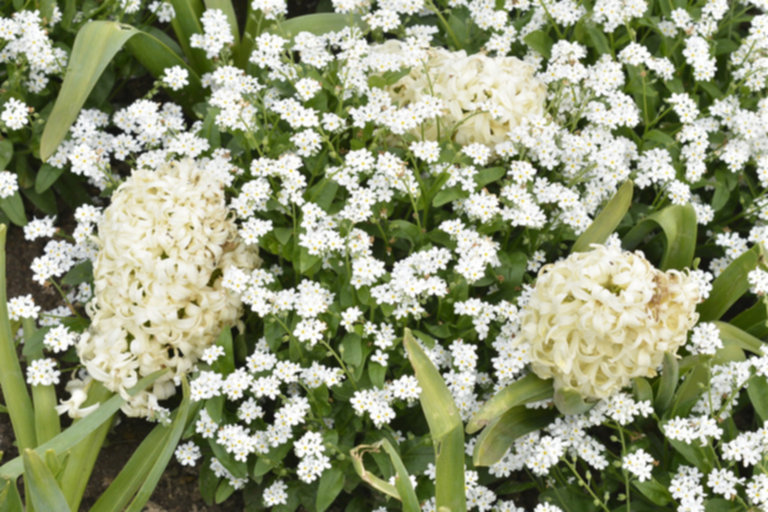}}&
\frame{\includegraphics[height=0.12\textwidth]{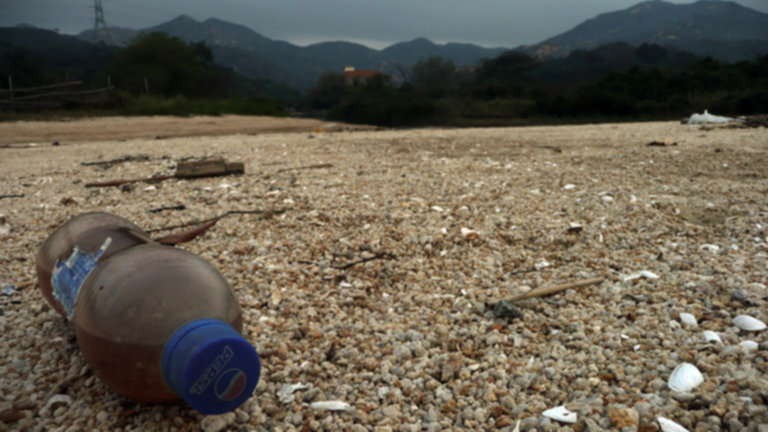}}&
\frame{\includegraphics[height=0.12\textwidth]{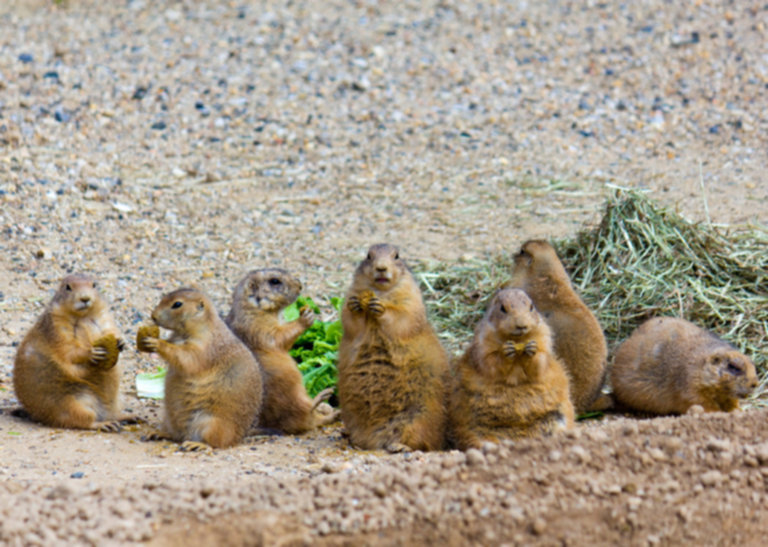}}&
\frame{\includegraphics[height=0.12\textwidth]{figs/ranking/483_4_164795398712158.jpg}}&
\frame{\includegraphics[height=0.12\textwidth]{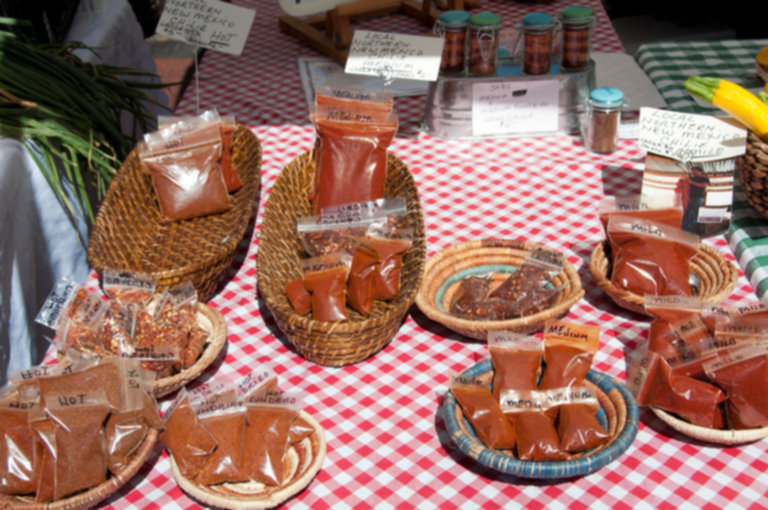}}
  \end{tabular}}
\resizebox{\columnwidth}{!}{\begin{tabular}{@{}cccccccc@{}}
{4.18}\tiny/6.10/4.62 & {4.23}\tiny/5.20/4.34 & {4.31}\tiny/5.86/4.69 & {4.41}\tiny/6.01/4.85 & {4.43}\tiny/5.59/5.03 & {4.46}\tiny/6.09/4.93 & {4.47}\tiny/6.39/5.01 & {4.63}\tiny/5.83/4.91\\
\frame{\includegraphics[height=0.12\textwidth]{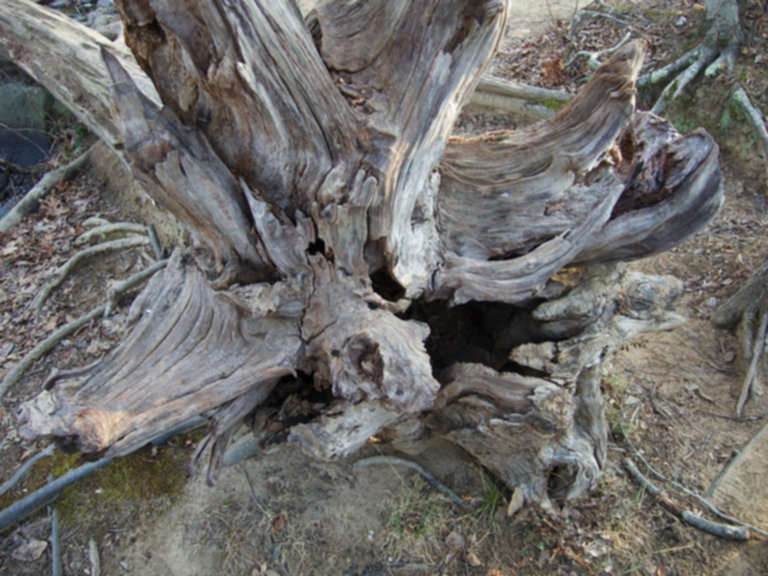}}&
\frame{\includegraphics[height=0.12\textwidth]{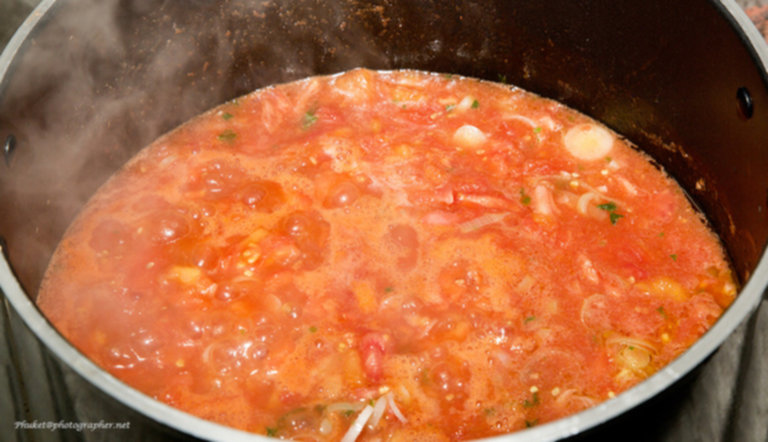}}&
\frame{\includegraphics[height=0.12\textwidth]{figs/ranking/487_4_311310768127441.jpg}}&
\frame{\includegraphics[height=0.12\textwidth]{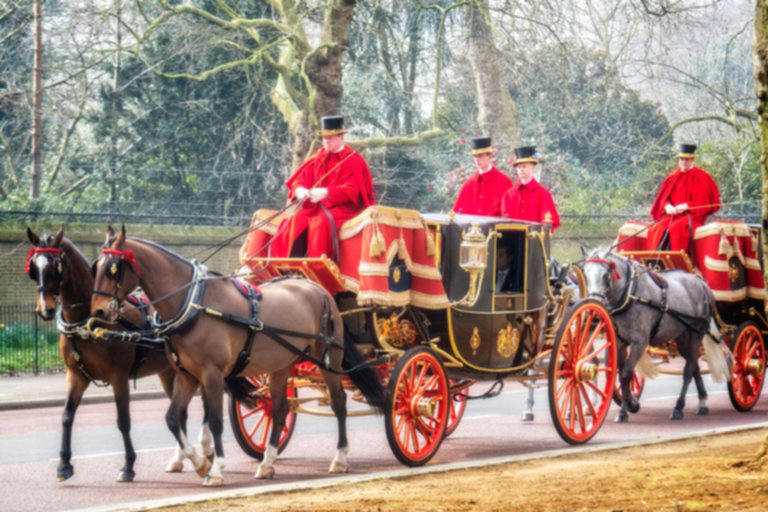}}&
\frame{\includegraphics[height=0.12\textwidth]{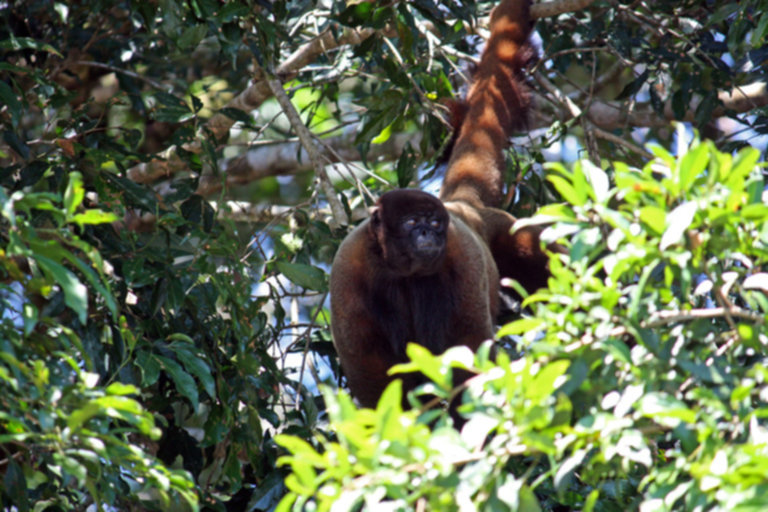}}&
\frame{\includegraphics[height=0.12\textwidth]{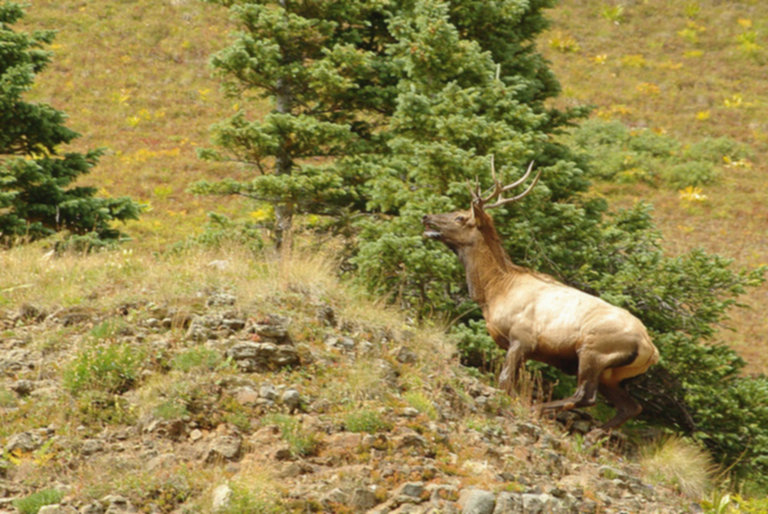}}&
\frame{\includegraphics[height=0.12\textwidth]{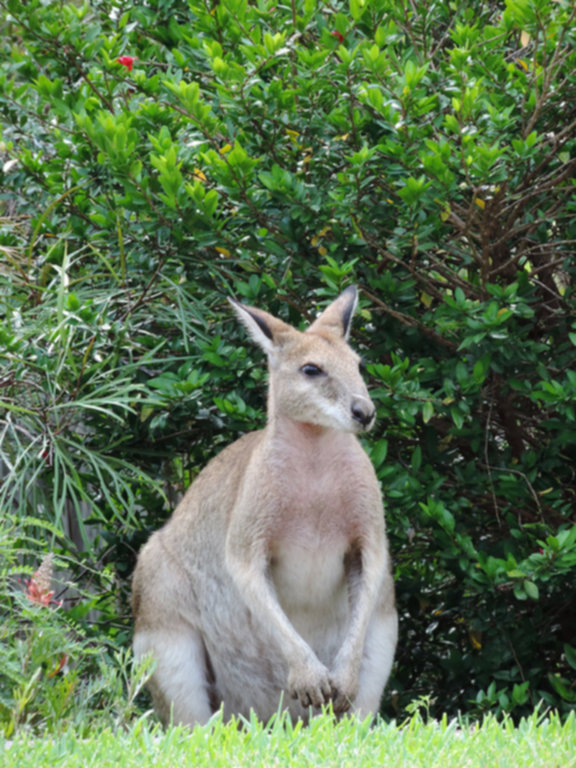}}&
\frame{\includegraphics[height=0.12\textwidth]{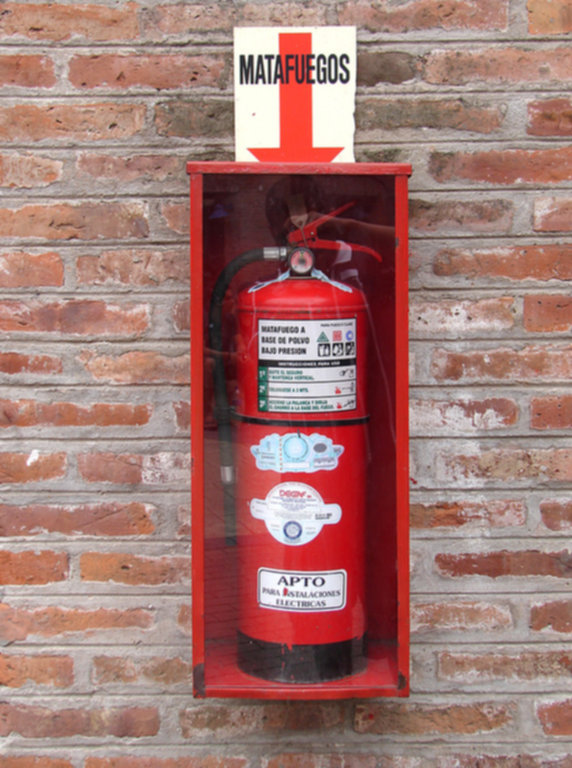}}
  \end{tabular}}
\resizebox{\columnwidth}{!}{\begin{tabular}{@{}ccccccc@{}}
{4.66}\tiny/6.62/5.36 & {4.73}\tiny/6.66/5.07 & {4.73}\tiny/5.81/5.10 & {4.79}\tiny/6.59/5.02 & {5.04}\tiny/6.43/5.60 & {5.39}\tiny/6.61/5.71 & {5.46}\tiny/6.60/5.75\\
\frame{\includegraphics[height=0.12\textwidth]{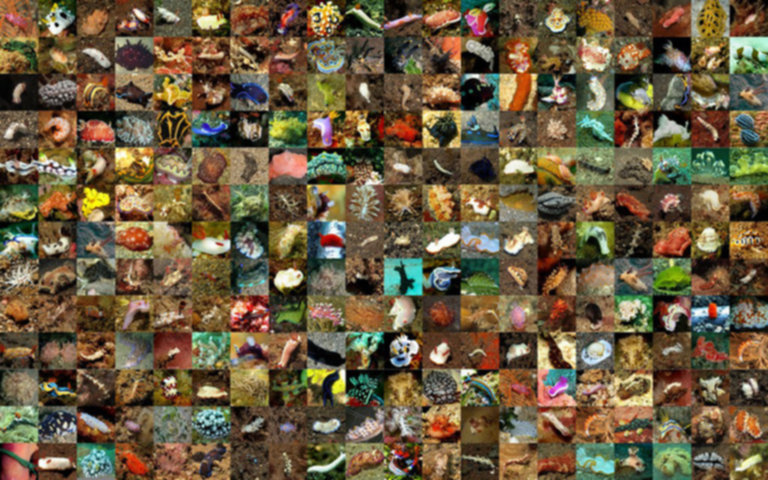}}&
\frame{\includegraphics[height=0.12\textwidth]{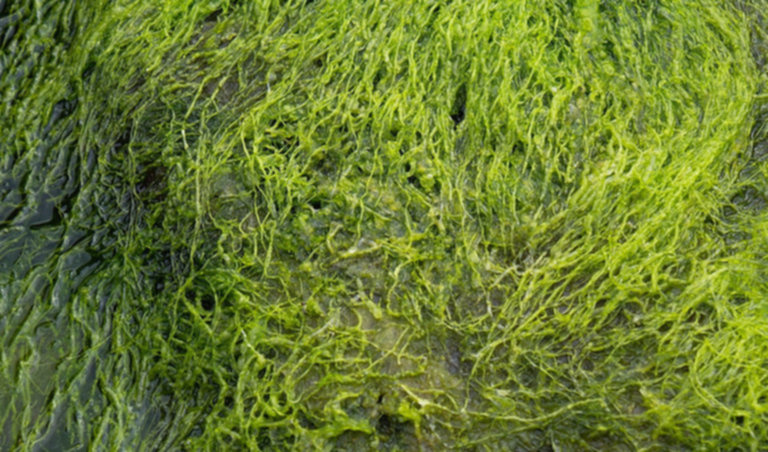}}&
\frame{\includegraphics[height=0.12\textwidth]{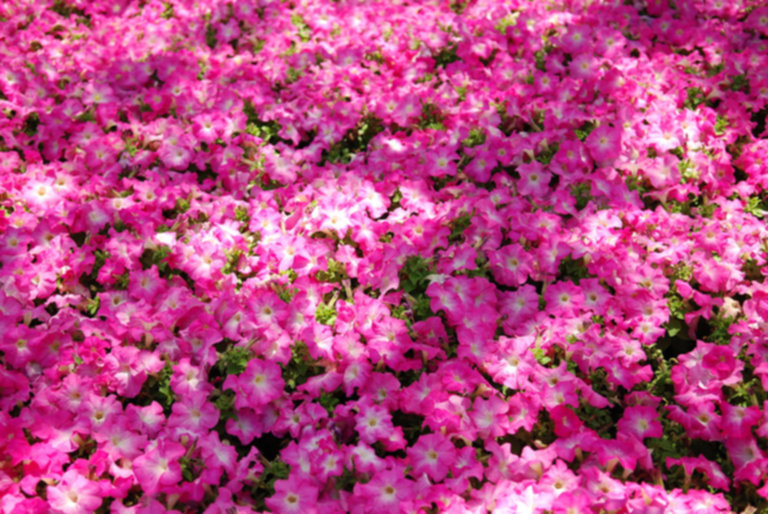}}&
\frame{\includegraphics[height=0.12\textwidth]{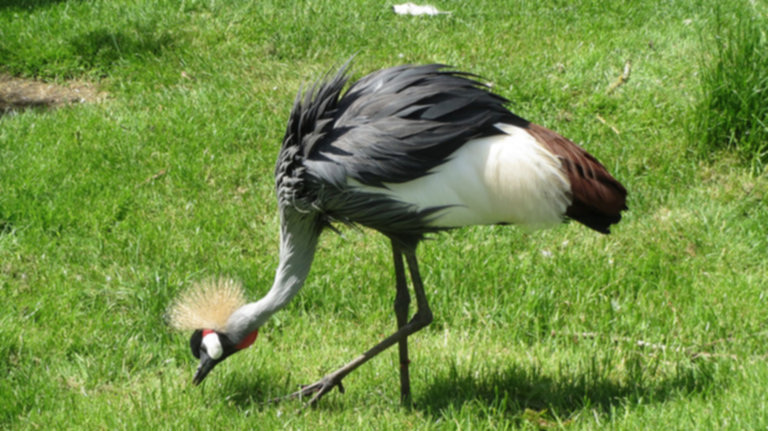}}&
\frame{\includegraphics[height=0.12\textwidth]{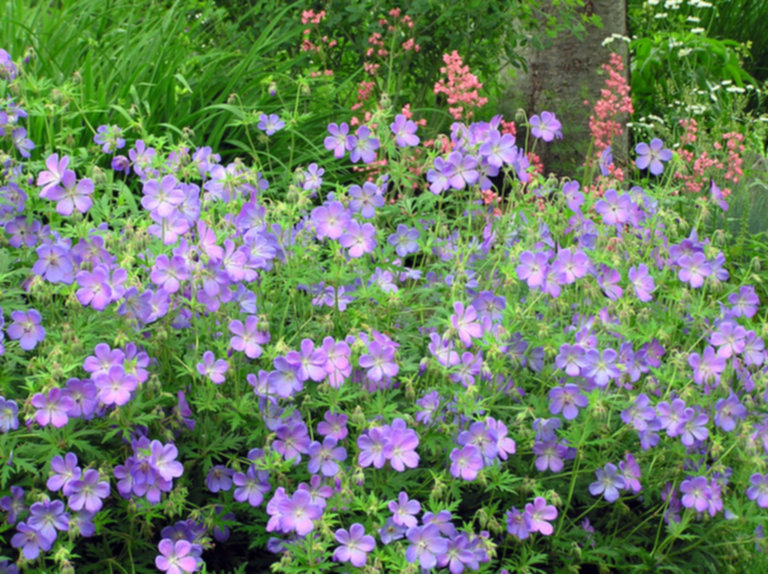}}&
\frame{\includegraphics[height=0.12\textwidth]{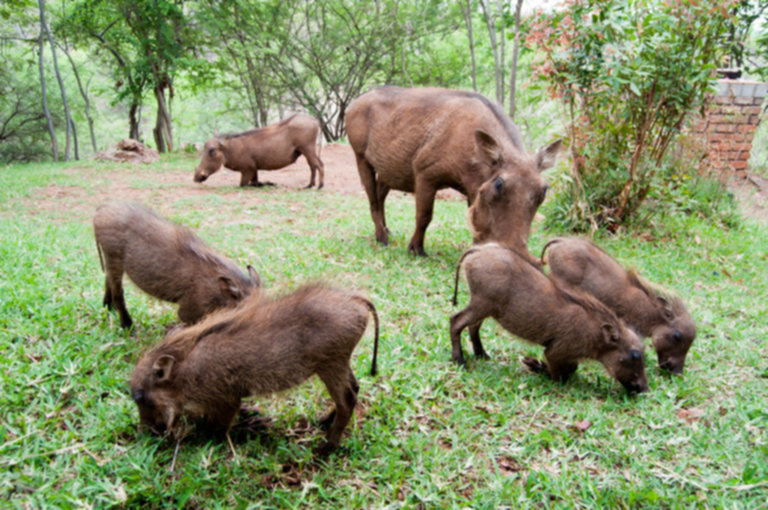}}&
\frame{\includegraphics[height=0.12\textwidth]{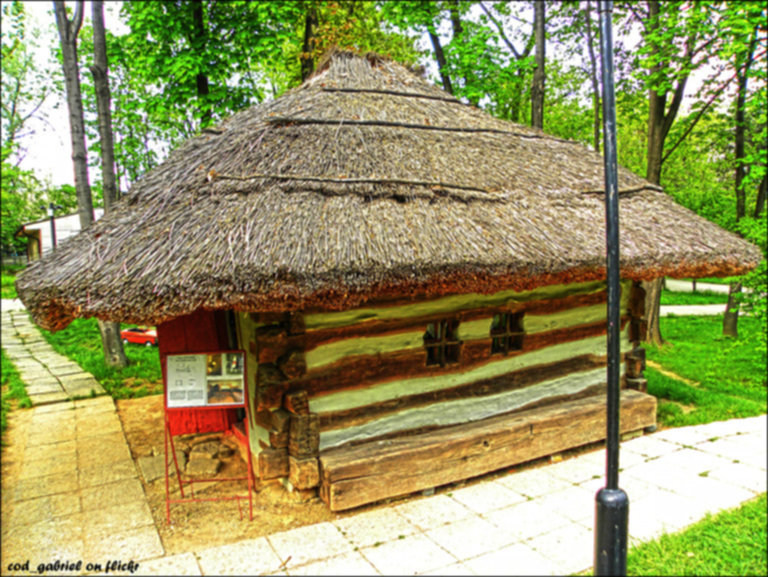}}
  \end{tabular}}
  \caption{\textbf{Case Study.}
  We present the most and the least compressible images (measured by SReC bpsp) in the validation set of Open Images~\cite{openimages}.
  The numbers above each image is the bpsp of ``SReC{\tiny/PNG/WebP}" respectively.
  Compared to traditional methods, SReC obtains larger performance gain on more challenging cases (lower part of the table).
  This shows that our network effectively models complicated patterns that hand-engineered methods fail to exploit.
   (Best viewed on screen.)}
\vspace{-1em}
  \label{fig:sampled}
\end{figure}

\paragraph{Case Study.}
We analyze what images SReC and other methods can compress more (or less) in \figref{sampled}.
Not surprisingly, images with consistent colors or simple patterns are easier to compress.
Images with high frequency changes or fine details are more challenging to compress.
Compared to traditional methods, we note that SReC obtains larger performance gain on more challenging (less compressible) cases.
This suggests that our network effectively models complicated image patterns that
hand-engineered methods struggle at.

\subsection{Comparison to Prior Work}
\lblsec{compare}

In \tabref{priorwork}, we compare SReC with engineered codecs, PNG~\cite{png}, WebP~\cite{webp}, and FLIF~\cite{flif},
as well as deep learning based methods L3C~\cite{l3c} and IDF~\cite{idf}.
We are not able to train IDF on full resolution due to GPU memory constraints and thus tiled the model over 64$\x$64 crops on Open Images~\cite{openimages}.
On Open Images~\cite{openimages},
SReC outperforms all prior work, while being efficient --- $\app$55$\x$ faster than
the second best performing method, IDF~\cite{idf}.
Engineered codecs, such as PNG or FLIF, are more efficient than deep-network based methods.
However, they fail to achieve good compression rates.
On 64$\x$64 small images (ImageNet64~\cite{imagenet64}), SReC again demonstrates a strong compression rate.
It outperforms all methods, except for IDF, which is 30$\x$ slower to encode and 14$\x$ slower to decode.
We also list performance of an efficient PixelCNN variant of Reed~\etal~\cite{mspixelcnn} purely for reference.
Reed~\etal~\cite{mspixelcnn} achieves a good compression rate, but its runtime is not practical (66$\x$ slower than SReC).
Our method is also parameter efficient --- smaller than other deep-learning based methods.
For example, SReC is $\app$20$\x$ smaller than than the second-best-performing method.

\begin{table}[t]
\resizebox{\columnwidth}{!}{\tablestyle{3pt}{1.12}\begin{tabular}{@{\extracolsep{2pt}}lx{32}x{39}x{39}x{25}x{39}x{39}x{25}x{25}@{}}
&&\multicolumn{3}{c}{ImageNet64} & \multicolumn{4}{c}{Open Images}\\
\cline{3-5} \cline{6-9}
 & \#params (10$^6$) & encode time (s) & decode time (s) & bpsp & encode time (s) & decode time (s) & bpsp (JPEG)\footnotemark & bpsp \\
\shline
\demph{Reed~\etal~\cite{mspixelcnn}\footnotemark} &&\demph{-}&\demph{$\app$4.68}&\demph{3.70} & \demph{-} & \demph{-} & \demph{-} & \demph{-}\\
PNG~\cite{png} &-& \textbf{1.3$\cdot$10$^{-3}$} & \textbf{8.0$\cdot$10$^{-5}$} & 5.74 & \textbf{0.17} & \textbf{9.8$\cdot$10$^{-5}$} & 3.78 & 4.03 \\
WebP~\cite{webp} &-& 0.021 & 2.1$\cdot$10$^{-4}$ & 4.64 & 0.40 & 7.0$\cdot$10$^{-4}$ & 2.67 & 3.03\\
FLIF~\cite{flif} &-& 0.022 & 0.010 & 4.54 & 1.23 & 0.30 & 2.47 & 2.87\\
L3C~\cite{l3c} & 5.01 & 0.031 & 0.023 & 4.42 & 1.33 & 1.13 & 2.58 & 2.99\\
IDF~\cite{idf} & 84.33 & 1.33 & 1.02 & \textbf{3.90} & 57.31 & 62.33 & \underline{2.34} & \underline{2.76} \\
\textbf{SReC} &\textbf{4.20} & {0.044} & {0.071} & \underline{4.29} & {0.99} & {1.15} &\textbf{2.29} & \textbf{2.70}\\
\end{tabular}}
\vspace{2mm}
\caption{\textbf{Comparison to prior work.}
We compare compression performance of SReC vs. other methods on ImageNet64~\cite{imagenet64} and Open Images~\cite{openimages} in bpsp, runtime, and number of parameters.
We additionally list an efficient PixelCNN variant of Reed~\etal~\cite{mspixelcnn} purely for reference.
It is not practical for lossless compression yet due to its long runtime.
SReC outperforms all practical algorithms in terms of bpsp, while being efficient and small in size.
}
\vspace{-2em}
\label{tab:priorwork}
\end{table}
\footnotetext{Bpsp results are taken from Reed~\etal~\cite{mspixelcnn}.
Timing is extrapolated from 32$\x$32 runtime following~\cite{l3c}.}

\section{Conclusions}
We propose Super-Resolution based Compression (SReC),
which relies on multiple levels of lossless super-resolution.
We show that lossless super-resolution operators are efficient to store,
due to the natural constraints induced by the super-resolution setting.
On multiple datasets, we show state-of-the-art compression rates.
Overall, our method is simple and efficient.
Its model size is small,
and its runtime is on par with or faster than other deep-network based compression methods.

\bibliographystyle{splncs04}
\bibliography{srec}

\clearpage

\begin{subappendices}
\renewcommand{\thesection}{\Alph{section}}%
\section{Architecture Details}
Each level $l$ of super-resolution uses a network to predict a distribution of $x^{(l)}$ given $y^{(l+1)}$.
Network weights are not shared across levels.
Levels $l=0, 1$ additionally take the output activation $z^{(l+1)}$ from the previous level as the second input.
See left hand side of \figref{architecture} for a schematic illustration of the network (of one level).
The network for each level predicts the first three pixels in each 4-pixel block in an autoregressive fashion.

The right hand side of \figref{architecture} illustrates the exact architecture.
All convolutional layers use a kernel size of 3$\x$3, stride of 1$\x$1, dilation of 1$\x$1, padding of 1$\x$1, and 64 output channels except for the first convolution in each CNN, which differs in using kernel size of 1$\x$1.
Since each image has 3 channels, the inputs, $y^{(l+1)}$, $[y^{(l+1)}, x^{(l)}\topleft]$, and $[y^{(l+1)}, x^{(l)}\toptwo]$
has the number of input channels $C_{in}=$ 3, 6, and 9, respectively.\footnote{Square brackets denote channel concatenation.}

\paragraph{Residual block} (denoted `Res' in \figref{architecture})
consists of two convolutional layers with a leaky ReLU~\cite{maas2013rectifier} in between.
A skip connection~\cite{resnet} is used to sum the input and the output of this two-convolutional-layer block.

\paragraph{Logistic Mixture Output.}
The convolutional layer before the discrete logistic mixture output~\cite{pixelcnnpp} is a stacked atrous convolution operator~\cite{chen2017rethinking}, following the same design in L3C~\cite{l3c}.
Following Salimans~\etal~\cite{pixelcnnpp}, we use 10 mixtures, with each discrete logistic being parameterized by 12 parameters.

\paragraph{Upsampling layer} (denoted `Up' in \figref{architecture}), is implemented as a PixelShuffle operator~\cite{shi2016real}.
It doubles the width and height of the input, while simultaneously shrinks the channel size by a factor of 4.
The convolution before PixelShuffle has 256 output channels, such that after applying PixelShuffle, the channel size remains 64.

\paragraph{Additional Training Details.}
We apply random 128$\x$128 cropping while training on Open Images~\cite{openimages}.
We keep ImageNet64~\cite{imagenet64} images as 64$\x$64.
We apply random horizontal flipping during training.
\end{subappendices}

\end{document}